\newcommand\aap{{A}\&{A}}
\newcommand\aaps{{A}\&{AS}}
\newcommand\aapr{{A}\&{ARv}}
\newcommand\aj{{AJ}}
\newcommand\apj{{ApJ}}

\newcommand\apjs{{ ApJS}}

\newcommand\mnras{{MNRAS}}

\newcommand\pasp{{PASP}}
\newcommand\pasj{{PASJ}}
\newcommand\science{{Science}}

\documentclass[useAMS,usenatbib]{mn2e}
\usepackage{graphicx}
\usepackage{epsf}
\usepackage{bm}
\usepackage{color}
\usepackage{times}
\usepackage{supertabular}
\usepackage{hyperref}
\usepackage{amsmath}
\usepackage{longtable}
\usepackage{tabularx}
\usepackage{booktabs}
\usepackage{array}
\usepackage{lscape}
\usepackage{amssymb}
\usepackage{threeparttable}
\usepackage{booktabs}
\title[$T_{\rm eff}$-colour-metallicity relations]{Empirical metallicity-dependent calibrations of effective temperature against colours for dwarfs and giants based on interferometric data}

\author[Huang et al.] {Y. Huang$^{1}$\thanks{E-mails: yanghuang@pku.edu.cn (YH); x.liu@pku.edu.cn (XWL)},  
                 X.-W. Liu$^{1,2}$\footnotemark[1], H.-B. Yuan$^{2}$\thanks{LAMOST Fellow}, M.-S. Xiang$^{1}$, B.-Q. Chen$^{1}$\footnotemark[2], H.-W. Zhang$^{1,2}$\\        
$^{1}$Department of Astronomy, Peking University, Beijing 100871, People's Republic of China\\
$^{2}$Kavli Institute for Astronomy and Astrophysics, Peking University, Beijing 100871, People's Republic of China}

\begin{document}

\date{}

\pagerange{\pageref{firstpage}--\pageref{lastpage}} \pubyear{2015}

\maketitle

\begin{abstract}
We present empirical metallicity-dependent calibrations of effective temperature against colours for dwarfs of luminosity classes IV and V and for giants of luminosity classes II and III, based on a collection from the literature of about two hundred nearby stars with direct effective temperature measurements of better than 2.5 per cent.
The calibrations are valid for an effective temperature range 3,100 -- 10,000\,K for dwarfs of spectral types M5 to A0 and 3,100 -- 5,700\,K for giants of spectral types K5 to G5.
A total of twenty-one colours  for dwarfs and eighteen colours  for giants of bands of four photometric systems, i.e. the Johnson ($UBVR_{\rm J}I_{\rm J}JHK$), the Cousins ($R_{\rm C}I_{\rm C}$), the Sloan Digital Sky Survey (SDSS, $gr$) and the Two Micron All Sky Survey (2MASS, $JHK_{\rm s}$), have been calibrated.
Restricted by the metallicity range of the current sample, the calibrations are mainly applicable for disk stars ([Fe/H]$\,\gtrsim\,-1.0$).
The normalized percentage residuals of the calibrations are typically 2.0 and 1.5 per cent for dwarfs and giants, respectively.
Some systematic discrepancies at various levels are found between the current scales and those  available in the literature (e.g. those based on the infrared flux method IRFM or spectroscopy).
Based on the current calibrations, we have re-determined the colours of the Sun. 
We have also investigated the systematic errors in effective temperatures yielded by the current on-going large scale low- to intermediate-resolution stellar spectroscopic surveys. 
We show that the calibration of colour ($g-K_{\rm s}$) presented in the current work provides an invaluable tool for the estimation of stellar effective temperature for those on-going or upcoming surveys.
\end{abstract}


\vspace{-100pt}
\section{Introduction}
The determination of effective temperatures of stars is of upmost importance in stellar astrophysics.
Accurate stellar effective temperatures are essential for reliable estimates of stellar metallicity by spectroscopy.
The distribution of stellar metallicities provide vital clues for the understanding of the Galactic chemical evolution.
For high resolution spectroscopy, the effective temperature is required as an input parameter for the stellar abundance analysis and thus its uncertainties is propagated to the resultant elemental abundances.
For low- and intermediate-resolution spectroscopy, the basic atmospheric parameters  (i.e. effective temperature $T_{\rm eff}$, surface gravity log $g$ and metallicity [Fe/H]) are often determined simultaneously from the measured spectrum, say, for example, by spectral template matching (e.g. Lee et al. 2008; Luo et 2015; Xiang et al. 2015) and there may exist some degeneracies amongst the stellar parameters thus deduced, in particular between $T_{\rm eff}$ and [Fe/H].
Therefore, accurate and model-independent estimation of $T_{\rm eff}$ is a critical step for stellar abundance determinations, either to minimize the uncertainties in the case of  high resolution spectroscopy or to break the degeneracy in the case of low- and intermediate-resolution spectroscopy.
As a basic stellar parameter, $T_{\rm eff}$, is also directly related to the observables of a star, such as its photometric colours, and the relations between $T_{\rm eff}$ and photometric colours are widely employed in a variety of astrophysical studies.
For example, the relations are used to convert a stellar isochrone of a given (simple) stellar population, i.e. the locus of stellar luminosity $L$ as a function of $T_{\rm eff}$, yielded by theoretical stellar evolution models, to a relation in observable space, such as colour-magnitude diagrams (CMDs; Demarque \& Larson 1964; Yi et al. 2001; Girardi et al. 2002, 2004; Dotter et al. 2008). 
Such CMDs have been widely applied to interpret photometric measurements of stellar clusters to determine their basic properties (e.g. age, distance and metallicity). 

A variety of methods have been developed to estimate $T_{\rm eff}$. 
For high resolution spectroscopy, techniques such as fitting the profiles of Balmer lines (Gray 1992; Heiter et al. 2002) or adjusting the ``excitation balance'' of Fe lines (Takeda et al. 2002; Santos et al. 2004) are often used. 
For low- and intermediate-resolution spectrophotometry, fitting the stellar continuum flux (over a wavelength range as wide as possible) with model atmosphere synthetic fluxes (e.g. Smalley 2005; Norris et al. 2013) or the strength of Balmer jump (Sokolov 1995) is generally used.
All these indirect methods of $T_{\rm eff}$ estimation require however  reliable stellar model atmospheres, and, as a consequence, lead to some model-dependent of the results.

The most popular and least model-dependent (or semi-direct) method for $T_{\rm eff}$ determination is the infrared flux method (IRFM), which is first developed by Blackwell \& Shallis (1977).
Since then, tremendous amount of work by many authors have used the IRFM to determine effective temperatures of stars (Alonso et al. 1996, 1999; Ram\'ires \& Mel\'endez 2005; Casagrande et al. 2006; Gonz\'alez Hern\'andes \& Bonifacio 2009; Casagrande et al. 2010).
However, the lack of absolute zero-point calibration of the IRFM could induce some systematic errors, by as much as $\sim$ 100 K, among the different studies (cf., e.g., the discussion of Casagrande et al. 2010).

Reliable estimates of $T_{\rm eff}$ can also be obtained from photometric colours using accurately calibrated metallicity-dependent $T_{\rm eff}$--colour relations\footnote{With accurate enough photometric data, one can determine $T_{\rm eff}$ without the knowledge of (spectroscopic) metallicity if suitable colour  measurements are available (e.g. $U-B$ for early, hot OB stars, $V-K$ for late, cool KM stars). In those cases, $T_{\rm eff}$ can be determined by iteration, assuming, for example, an initial value of metallicity such as that of the Sun.}.
The method allows one to accurately determine values of $T_{\rm eff}$ for large numbers of stars based on (relatively ``cheap'') photometric measurements alone. 
This is extremely important in the current era of large scale surveys, with a number of large scale stellar spectroscopic surveys on-going, including the Sloan Extension for Galactic Understanding and Exploration (SEGUE; Yanny et al. 2009), the Radial Velocity Experiment (RAVE; Steinmetz et al. 2006), the Apache Point Observatory Galactic Evolution Experiment (APOGEE; Eisenstein et al. 2011),  the LAMOST Spectroscopic Survey of the Galactic Anti-center (LSS-GAC; Liu et al. 2014, Yuan et al. 2014) and Gaia (Perryman et al. 2001).
Those surveys are generating a huge amount of data that will dramatically improve our understanding of the formation and evolution of the Milky Way.

The key to a robust metallicity-dependent $T_{\rm eff}$--colour relation is accurate, model-independent measurements of $T_{\rm eff}$.
For the moment, both theoretical model atmospheres (e.g. Bessell, Castelli \& Plez 1998; Houdashelt, Bell \& Sweigart 2000) and empirical measurements (e.g. Gratton, Carretta, \& Castelli 1996; Weiss \& Salaris 1999;  Vandenberg \& Clem 2003) are used to calibrate the relations.
However, the synthetic broad-band colours derived from model atmospheres may deviate the real values due to some defects, for example, in the choice of microturbulence velocity, especially for cool stars (i.e. $T_{\rm eff}\,\lesssim 4000\,{\rm K}$; Castelli 1999; Ku\v cinskas et al. 2005; Casagrande \& VandenBerg 2014) or in the treatment of convection of late-type giants (Ku\v cinskas et al. 2005).
As for the empirical approach, the source of error is mainly contributed by uncertainties in the estimates of $T_{\rm eff}$, which may result, for example, as a consequence of adopting inappropriate stellar models in the case of indirect methods or the usage of an  incorrect zero-point in the case of IRFM method.
In order to construct a robust empirical metallicity-dependent $T_{\rm eff}$--colour relation, accurate direct measurements of $T_{\rm eff}$  are indispensable, preferably for a number of stars coverage wide ranges of stellar parameters, including the metallicity. 

Direct measurements of stellar effective temperature are now available for a large number of stars for which their angular diameters directly measured using long-baseline optical/infrared interferometry (LBOI; e.g. Code et al. 1976).
With angular diameters measurement from LBOI, linear radii can be derived when the geometric parallaxes (e.g. from the Hipparcos; van Leeuwen 2007) are also available.
Then, the liner radii combined with bolometric flux measurements yield directly the stellar effective temperatures through the Stefan-Boltzmann equation (Mozurkewich et al. 2003, hereafter M03; Baines et al. 2009; Boyajian et al. 2009, 2012a,b, 2013, hereafter B12a, B12b and B13, respectively; Creevey et al. 2012; White et al. 2013).
Due to the progress in observations in the past few years, stellar angular diameters can now be measured to an accuracy of better than 5.0 per cent, implying $T_{\rm eff}$ measurements of better than 2.5 per cent.
Those { data have however not been fully utilized to calibrate the aforementioned, very important, empirical metallicity-dependent $T_{\rm eff}$--colours relations.
Quite recently, B12b derive the metallicity-dependent $T_{\rm eff}$--colour relations in the Johnson $BVJHK$ bands based on a small  sample of 40 G/K/M dwarfs with direct angular diameter measurement of better than 5 per cent, while B13 do so based on a sample of over one hundred A to M dwarfs with direct $T_{\rm eff}$ measurements.
In the latter study, the effects of metallicity are considered for colour $(B-V)$ only.
Both calibrations are restricted to population I (i.e. [Fe/H] $> -1.0$) dwarfs of luminosity classes IV and V.

In this paper, we improve, on the basis of work by B12a,b and B13, the metallicity-dependent empirical $T_{\rm eff}$--colour relations for dwarfs by adding more metal-poor stars with direct $T_{\rm eff}$ measurements to the calibration sample. 
Accurate estimates of the interstellar extinction are also made for all calibration stars.
In addition, we present similar calibration for giants based on available direct $T_{\rm eff}$ measurements for those types of star.
The paper is organized as follows.
In Section 2 we introduce the calibration sample compiled from the literature and the extinction determinations for the sample stars.
The empirical calibrations of $T_{\rm eff}$ versus colours and [Fe/H] are presented in Section 3.
Comparisons with previous work and applications of the newly derived relations are given in Sections 4 to 6, respectively. 
Finally, we conclude in Section 7.
\vspace{-10pt}
\section{Data}

In B13, direct measurements of $T_{\rm eff}$ for 125 dwarf stars (luminosity classes: IV/V) are collected from their surveys of stellar diameters and effective temperatures (B12a, B12b and B13) as well as from 24 other publications.
We take 121 stars from the B13 sample, discarding two stars with $T_{\rm eff}$ errors greater than 2.5 per cent and another two with incommensurate measurements of  $T_{\rm eff}$ amongst the different work.
On top of this, we collect another 13 dwarf stars with direct $T_{\rm eff}$ measurements from 6 publications.
Thus the sample of dwarfs employed in the current work consists of a total of 134 stars with direct $T_{\rm eff}$ measurements of better than 2.5 per cent.
The sample stars span spectral types from M5 to A0, with $T_{\rm eff}$ ranging from $\sim 3,100$ to 10,000 K.
For giant stars (luminosity classes: II/III), direct $T_{\rm eff}$ measurements, also accurate to better than 2.5 per cent, of 61 stars are compiled from 9 publications.
The giants span spectral types from M5 to G5, with $T_{\rm eff}$ ranging from $\sim 3,100$ to  5,700 K.
 
\onecolumn
\begin{landscape}
\begin{center}
\scriptsize
\begin{longtable}{ccccccccccccccccccc}
\caption{Compiled data of dwarf and giant sample stars employed in the current work}\\
\cline{1-19}\\
Name&Spectral type&$U$&$B$&$V$&$R_{\rm J}$&$I_{\rm J}$&$R_{\rm C}$&$I_{\rm C}$&$J$&$H$&$K$&$E(B-V)$&$T_{\rm eff}$&$\sigma_{T_{\rm eff}}$&[Fe/H]&Flag1&Flag2&Flag3\\
&&(mag)&(mag)&(mag)&(mag)&(mag)&(mag)&(mag)&(mag)&(mag)&(mag)&(mag)&(K)&(K)&&&&\\
\cline{1-19}\\
\endfirsthead
\caption{Continued.}\\
\cline{1-19}\\
Name&Spectral type&$U$&$B$&$V$&$R_{\rm J}$&$I_{\rm J}$&$R_{\rm C}$&$I_{\rm C}$&$J$&$H$&$K$&$E(B-V)$&$T_{\rm eff}$&$\sigma_{T_{\rm eff}}$&[Fe/H]&Flag1&Flag2&Flag3\\
&&(mag)&(mag)&(mag)&(mag)&(mag)&(mag)&(mag)&(mag)&(mag)&(mag)&(mag)&(K)&(K)&&&&\\
\cline{1-19}\\
\endhead
\cline{1-19}\\
\endfoot
\multicolumn{19}{c}{Dwarf stars} \\
 HD173740&       M5V &        -- & $  11.28$ & $   9.69$ &        -- &        -- &        -- &        -- &      $   5.72$ &      $   5.24$ &      $   4.98$& 0.001&  3104&  28& $-0.54$&   B12b& PASTEL& 2 \\
HIP087937&      M4Ve &        -- & $  11.26$ & $   9.53$ &        -- &        -- & $   8.32$ & $   6.76$ &      $   5.30$ &      $   4.77$ &      $   4.50$& 0.002&  3222&  10& $-0.15$&   B12b& PASTEL& 4 \\
 HD173739&       M3V &        -- & $  10.44$ & $   8.90$ &        -- &        -- &        -- &        -- &      $   5.19$ &      $   4.68$ &      $   4.46$& 0.001&  3407&  15& $-0.49$&   B12b&    B12& 4 \\
HIP086162&       M3V &        -- & $  10.65$ & $   9.15$ &        -- &        -- &        -- &        -- &      $   5.37$ &      $   4.75$ &      $   4.54$& 0.000&  3413&  28& $ 0.00$&   B12b& PASTEL& 1 \\
HIP057087&       M3V &        -- & $  12.17$ & $  10.65$ &        -- &        -- &        -- &        -- &      $   6.90$ &      $   6.32$ &      $   6.07$& 0.000&  3416&  53& $ 0.04$&   B12b&    B12& 3 \\
HIP074995&     M2.5V &        -- & $  12.19$ & $  10.58$ & $   8.89$ & $   7.46$ & $   9.45$ & $   8.06$ &      $   6.68$ &      $   6.09$ &      $   5.83$& 0.000&  3442&  54& $-0.10$&   B12b&    B12& 3 \\
 HD095735&       M2V &        -- & $   8.91$ & $   7.42$ & $   5.94$ & $   4.77$ & $   6.40$ & $   5.28$ &      $   4.11$ &      $   3.55$ &      $   3.34$& 0.023&  3464&  15& $-0.31$&   B12b& PASTEL& 1 \\
HIP054211&       M2V &        -- & $  10.27$ & $   8.73$ & $   7.26$ & $   6.20$ &        -- &        -- &      $   5.55$ &      $   4.94$ &      $   4.76$& 0.012&  3497&  39& $-0.43$&   B12b& PASTEL& 1 \\
 HD001326&     M1.5V &        -- & $   9.63$ & $   8.07$ & $   6.69$ & $   5.53$ &        -- &        -- &      $   4.86$ &      $   4.25$ &      $   4.02$& 0.000&  3567&  11& $-0.36$&   B12b&    B12& 0 \\
 HD119850&     M1.5V &        -- & $   9.93$ & $   8.50$ & $   7.06$ & $   5.92$ & $   7.49$ & $   6.39$ &      $   5.26$ &      $   4.64$ &      $   4.46$& 0.000&  3618&  31& $-0.10$&   B12b& PASTEL& 1 \\
 HD217987&     M0.5V &        -- & $   8.82$ & $   7.34$ &        -- &        -- & $   6.35$ & $   5.31$ &      $   4.20$ &      $   3.60$ &      $   3.36$& 0.004&  3676&  35& $-0.22$&   B12b& PASTEL& 1 \\
 HD199305&     M0.5V &        -- & $  10.03$ & $   8.57$ & $   7.19$ & $   6.13$ &        -- &        -- &      $   5.52$ &      $   4.81$ &      $   4.64$& 0.004&  3692&  22& $-0.13$&   B12b& PASTEL& 1 \\
 HD216899&     M1.5V &        -- & $  10.19$ & $   8.68$ &        -- &        -- & $   7.69$ & $   6.57$ &      $   5.41$ &      $   4.78$ &      $   4.58$& 0.000&  3713&  11& $ 0.06$&   B12b&    B12& 0 \\
 HD036395&     M1.5V &        -- & $   9.44$ & $   7.97$ & $   6.53$ & $   5.39$ & $   6.98$ & $   5.89$ &      $   4.77$ &      $   4.06$ &      $   3.86$& 0.000&  3801&   9& $ 0.21$&   B12b& PASTEL& 1 \\
 HD079211&       K7V &        -- & $   9.04$ & $   7.70$ &        -- &        -- &        -- &        -- &      $   4.78$ &      $   4.30$ &      $   4.15$& 0.001&  3867&  37& $-0.40$&   B12b&  MILES& 2 \\
 HD079210&       M0V &        -- & $   9.05$ & $   7.64$ &        -- &        -- &        -- &        -- &      $   4.89$ &      $   4.25$ &      $   4.09$& 0.000&  3907&  35& $-0.18$&   B12b&    B12& 3 \\
 HD201092&       K7V & $   8.60$ & $   7.38$ & $   6.02$ & $   4.87$ & $   4.07$ &        -- &        -- &      $   3.58$ &      $   2.93$ &      $   2.73$& 0.000&  3932&  25& $-0.39$&   B12b& PASTEL& 1 \\
 HD088230&       K7V &        -- & $   7.92$ & $   6.57$ & $   5.35$ & $   4.55$ &        -- &        -- &      $   3.98$ &      $   3.32$ &      $   3.19$& 0.005&  4085&  14& $-0.03$&   B12b& PASTEL& 1 \\
 HD201091&       K5V & $   7.51$ & $   6.39$ & $   5.23$ & $   4.20$ & $   3.56$ &        -- &        -- &      $   3.16$ &      $   2.61$ &      $   2.40$& 0.000&  4361&  17& $-0.25$&   B12b& PASTEL& 2 \\
 HD131977&       K4V & $   7.95$ & $   6.88$ & $   5.78$ & $   4.85$ & $   4.28$ & $   4.96$ & $   4.32$ &      $   3.82$ &      $   3.27$ &      $   3.15$& 0.000&  4507&  58& $ 0.07$&   B12b& PASTEL& 1 \\
 HD209100&       K5V &        -- & $   5.75$ & $   4.69$ & $   3.81$ & $   3.25$ & $   4.05$ & $   3.53$ &      $   2.83$ &      $   2.30$ &      $   2.18$& 0.000&  4555&  24& $-0.08$&   B12b& PASTEL& 1 \\
 HD122563&      F8IV & $   7.26$ & $   6.92$ & $   6.06$ & $   5.27$ & $   4.72$ & $   5.55$ & $   5.04$ &      $   4.29$ &      $   3.74$ &      $   3.67$& 0.044&  4598&  42& $-2.65$&    R13& PASTEL& 1 \\
 HD016160&       K3V &        -- & $   6.79$ & $   5.82$ & $   4.99$ & $   4.46$ & $   5.24$ & $   4.75$ &      $   4.07$ &      $   3.52$ &      $   3.45$& 0.000&  4662&  17& $-0.11$&   B12b& PASTEL& 1 \\
 HD219134&       K3V & $   7.34$ & $   6.47$ & $   5.50$ & $   4.69$ & $   4.18$ &        -- &        -- &      $   3.84$ &      $   3.39$ &      $   3.22$& 0.023&  4699&  16& $ 0.08$&   B12b& PASTEL& 1 \\
 HD222404&       K1V & $   5.19$ & $   4.24$ & $   3.21$ & $   2.46$ &        -- &        -- &        -- &             -- &             -- &             --& 0.000&  4744&  21& $ 0.12$&   Ba09& PASTEL& 1 \\
 HD103095&       K1V & $   7.36$ & $   7.18$ & $   6.44$ & $   5.78$ & $   5.33$ &        -- &        -- &      $   4.95$ &      $   4.44$ &      $   4.40$& 0.004&  4791&  28& $-1.33$&    B13& PASTEL& 1 \\
 HD198149&      K0IV & $   4.91$ & $   4.31$ & $   3.40$ & $   2.74$ & $   2.25$ &        -- &        -- &      $   1.89$ &      $   1.49$ &      $   1.28$& 0.011&  4835&  37& $-0.12$&    B13& PASTEL& 1 \\
 HD188512&    G8IV-V & $   5.06$ & $   4.58$ & $   3.72$ & $   3.06$ & $   2.57$ & $   3.26$ & $   2.83$ &      $   2.26$ &      $   1.71$ &      $   1.71$& 0.000&  4920& 102& $-0.18$&    B13& PASTEL& 1 \\
 HD004628&       K2V &        -- & $   6.62$ & $   5.75$ & $   4.98$ & $   4.51$ & $   5.21$ & $   4.76$ &      $   4.24$ &      $   3.72$ &      $   3.61$& 0.004&  4950&  14& $-0.27$&   B12b& PASTEL& 1 \\
 HD023249&      K1IV &        -- & $   4.46$ & $   3.54$ & $   2.82$ & $   2.32$ & $   3.02$ & $   2.59$ &      $   1.96$ &      $   1.52$ &      $   1.40$& 0.000&  4955&  30& $ 0.15$&    B13& PASTEL& 1 \\
 HD011964&       G9V &        -- & $   7.23$ & $   6.41$ &        -- &        -- & $   5.97$ & $   5.56$ &      $   5.02$ &      $   4.64$ &      $   4.49$& 0.002&  5013&  62& $ 0.12$&    B13& PASTEL& 1 \\
 HD003651&       K0V &        -- & $   6.71$ & $   5.86$ & $   5.21$ & $   4.82$ &        -- &        -- &      $   4.48$ &      $   4.03$ &      $   3.97$& 0.000&  5046&  86& $ 0.15$&    B13& PASTEL& 1 \\
 HD022049&       K2V &        -- & $   4.62$ & $   3.74$ & $   3.01$ & $   2.55$ & $   3.23$ & $   2.79$ &      $   2.20$ &      $   1.75$ &      $   1.65$& 0.000&  5077&  35& $-0.06$&   B12b& PASTEL& 1 \\
 HD026965&      K1Ve &        -- & $   5.25$ & $   4.43$ & $   3.74$ & $   3.29$ & $   3.96$ & $   3.54$ &      $   2.95$ &      $   2.48$ &      $   2.41$& 0.000&  5147&  14& $-0.28$&   B12b& PASTEL& 1 \\
 HD075732&    K0IV-V &        -- & $   6.80$ & $   5.94$ &        -- &        -- &        -- &        -- &      $   4.59$ &      $   4.14$ &      $   4.07$& 0.000&  5172&  18& $ 0.37$&    B13& PASTEL& 1 \\
 HD158633&       K0V &        -- & $   7.10$ & $   6.36$ &        -- &        -- &        -- &        -- &             -- &             -- &             --& 0.023&  5203&  46& $-0.47$&    B13& PASTEL& 1 \\
 HD128621&      K2IV & $   2.84$ & $   2.19$ & $   1.30$ &        -- &        -- &        -- &        -- &      $  -0.01$ &      $  -0.41$ &      $  -0.61$& 0.015&  5232&   8& $ 0.24$&    B13& PASTEL& 1 \\
 HD182736&      G0IV &        -- & $   7.63$ & $   6.86$ &        -- &        -- &        -- &        -- &  $$   5.53$^a$ &             -- &  $$   5.04$^a$& 0.047&  5239&  37& $-0.11$&    B13& PASTEL& 1 \\
 HD010476&       K0V &        -- & $   6.08$ & $   5.24$ & $   4.55$ & $   4.12$ &        -- &        -- &      $   3.85$ &      $   3.44$ &      $   3.21$& 0.000&  5242&  12& $-0.04$&    B13& PASTEL& 1 \\
 HD185144&       K0V & $   5.86$ & $   5.49$ & $   4.69$ & $   4.04$ & $   3.63$ &        -- &        -- &      $   3.32$ &      $   3.04$ &      $   2.78$& 0.000&  5246&  26& $-0.23$&   B12b& PASTEL& 1 \\
 HD021019&       G2V &        -- & $   6.81$ & $   6.13$ &        -- &        -- &        -- &        -- &             -- &             -- &             --& 0.023&  5261&  65& $-0.44$&    B13& PASTEL& 1 \\
 HD010700&       G8V &        -- & $   4.22$ & $   3.50$ & $   2.88$ & $   2.41$ & $   3.06$ & $   2.68$ &      $   2.16$ &      $   1.72$ &      $   1.68$& 0.000&  5290&  39& $-0.52$&    B13& PASTEL& 1 \\
 HD173701&       K0V &        -- & $   8.26$ & $   7.45$ &        -- &        -- &        -- &        -- &  $$   6.11$^a$ &  $$   5.73$^a$ &  $$   5.69$^a$& 0.029&  5297&  53& $ 0.33$&    B13& PASTEL& 1 \\
 HD101501&       G8V & $   6.31$ & $   6.08$ & $   5.34$ & $   4.73$ & $   4.37$ &        -- &        -- &      $   4.02$ &      $   3.61$ &      $   3.60$& 0.000&  5309&  27& $-0.03$&    B13& PASTEL& 1 \\
 HD000166&       G8V &        -- & $   6.89$ & $   6.14$ &        -- &        -- &        -- &        -- &             -- &             -- &             --& 0.000&  5327&  39& $ 0.11$&    B13& PASTEL& 1 \\
 HD149661&       K0V & $   7.00$ & $   6.55$ & $   5.74$ & $   5.13$ & $   4.74$ & $   5.30$ & $   4.89$ &      $   4.32$ &      $   3.86$ &      $   3.83$& 0.000&  5337&  41& $ 0.05$&   B12b& PASTEL& 1 \\
 HD006582&       G5V &        -- & $   5.79$ & $   5.12$ & $   4.51$ & $   4.11$ &        -- &        -- &      $   3.84$ &      $   3.38$ &      $   3.35$& 0.019&  5348&  26& $-0.88$&   B12b& PASTEL& 1 \\
 HD082885&       G8V &        -- & $   6.18$ & $   5.41$ & $   4.79$ & $   4.42$ &        -- &        -- &      $   4.14$ &      $   3.77$ &      $   3.70$& 0.000&  5376&  43& $ 0.36$&    B13& PASTEL& 1 \\
 HD217107&    G8IV-V &        -- & $   6.89$ & $   6.15$ &        -- &        -- &        -- &        -- &             -- &             -- &             --& 0.003&  5391&  40& $ 0.36$&    B13& PASTEL& 1 \\
 HD010780&       K0V &        -- & $   6.44$ & $   5.63$ & $   4.99$ & $   4.60$ &        -- &        -- &      $   4.31$ &      $   3.88$ &      $   3.84$& 0.000&  5398&  75& $ 0.02$&   B12b& PASTEL& 1 \\
 HD117176&       G5V & $   5.90$ & $   5.65$ & $   4.95$ & $   4.35$ & $   3.96$ &        -- &        -- &      $   3.64$ &      $   3.25$ &      $   3.24$& 0.010&  5406&  64& $-0.06$&    B13& PASTEL& 1 \\
 HD165341&      K0Ve &        -- & $   4.95$ & $   4.12$ &        -- &        -- & $   3.45$ & $   3.01$ &             -- &             -- &      $   2.29$& 0.026&  5407&  52& $ 0.04$&   B12b& PASTEL& 1 \\
 HD195564&       G2V &        -- & $   6.32$ & $   5.64$ &        -- &        -- & $   5.27$ & $   4.91$ &             -- &             -- &             --& 0.003&  5421& 118& $ 0.05$&    B13& PASTEL& 1 \\
 HD010697&      G3Va &        -- & $   6.95$ & $   6.23$ &        -- &        -- &        -- &        -- &      $   4.97$ &      $   4.66$ &      $   4.58$& 0.009&  5442&  65& $ 0.15$&    B13& PASTEL& 1 \\
 HD190360&    G7IV-V &        -- & $   6.41$ & $   5.70$ &        -- &        -- &        -- &        -- &      $   4.45$ &      $   4.11$ &      $   4.05$& 0.002&  5461&  75& $ 0.23$&    B13& PASTEL& 1 \\
 HD131156&       G7V & $   5.60$ & $   5.31$ & $   4.54$ & $   3.91$ & $   3.48$ &        -- &        -- &      $   3.01$ &      $   2.59$ &      $   2.57$& 0.001&  5483&  32& $-0.15$&    B13& PASTEL& 1 \\
 HD161797&      G5IV & $   4.56$ & $   4.17$ & $   3.42$ & $   2.89$ & $   2.51$ &        -- &        -- &      $   2.18$ &      $   1.81$ &      $   1.77$& 0.000&  5502&  55& $ 0.25$&    B13& PASTEL& 1 \\
 HD086728&       G4V &        -- & $   6.01$ & $   5.35$ &        -- &        -- &        -- &        -- &             -- &             -- &             --& 0.000&  5619&  44& $ 0.21$&    B13& PASTEL& 1 \\
 HD182572&      G8IV & $   6.25$ & $   5.86$ & $   5.10$ &        -- &        -- &        -- &        -- &      $   3.82$ &      $   3.54$ &      $   3.48$& 0.020&  5643&  84& $ 0.39$&    B13& PASTEL& 1 \\
 HD038858&       G2V &        -- & $   6.56$ & $   5.93$ &        -- &        -- &        -- &        -- &             -- &             -- &             --& 0.012&  5646&  45& $-0.23$&    B13& PASTEL& 1 \\
 HD150680&      G2IV & $   3.50$ & $   3.31$ & $   2.70$ & $   2.22$ & $   1.93$ &        -- &        -- &      $   1.67$ &      $   1.32$ &      $   1.29$& 0.036&  5656&  63& $ 0.01$&    B13& PASTEL& 1 \\
 HD136202&      F8IV &        -- & $   5.57$ & $   5.04$ & $   4.63$ & $   4.37$ & $   4.73$ & $   4.43$ &             -- &             -- &             --& 0.008&  5661&  87& $-0.05$&    B13& PASTEL& 1 \\
 HD186427&       G3V & $   7.06$ & $   6.85$ & $   6.19$ & $   5.75$ & $   5.42$ &        -- &        -- &      $   5.04$ &      $   4.70$ &      $   4.65$& 0.002&  5678&  66& $ 0.07$&    B13& PASTEL& 1 \\
 HD140538&       G5V &        -- & $   6.55$ & $   5.87$ &        -- &        -- & $   5.46$ & $   5.11$ &             -- &             -- &             --& 0.005&  5692&  74& $ 0.05$&    B13& PASTEL& 1 \\
 HD109358&       G0V & $   4.90$ & $   4.85$ & $   4.26$ & $   3.72$ & $   3.42$ &        -- &        -- &      $   3.23$ &      $   2.85$ &      $   2.84$& 0.002&  5700&  95& $-0.21$&    B13& PASTEL& 1 \\
 HD217014&       G3V & $   6.39$ & $   6.17$ & $   5.50$ & $   4.96$ & $   4.62$ &        -- &        -- &      $   4.36$ &      $   4.03$ &      $   3.99$& 0.000&  5706&  95& $ 0.21$&    B13& PASTEL& 1 \\
 HD140283$^{b}$&      F3VI & $   7.52$ & $   7.71$ & $   7.22$ & $   6.63$ & $   6.32$ & $   6.87$ & $   6.51$ &      $   6.04$ &      $   5.69$ &      $   5.66$& 0.000&  5720&  29& $-2.46$&    R13& PASTEL& 1 \\
 HD020630&       G5V &        -- & $   5.51$ & $   4.83$ & $   4.26$ & $   3.91$ & $   4.44$ & $   4.10$ &      $   3.71$ &      $   3.35$ &      $   3.34$& 0.003&  5723&  76& $ 0.04$&    B13& PASTEL& 1 \\
 HD157214&       G0V & $   6.07$ & $   6.00$ & $   5.38$ & $   4.87$ & $   4.53$ &        -- &        -- &      $   4.22$ &      $   3.86$ &      $   3.84$& 0.000&  5738&  48& $-0.39$&    B13&  MILES& 2 \\
 HD186408&     G1.5V & $   6.79$ & $   6.59$ & $   5.95$ & $   5.50$ & $   5.17$ &        -- &        -- &      $   4.91$ &      $   4.44$ &      $   4.52$& 0.000&  5760&  57& $ 0.07$&    B13& PASTEL& 1 \\
 HD190406&       G0V &        -- & $   6.41$ & $   5.80$ &        -- &        -- &        -- &        -- &             -- &             -- &             --& 0.000&  5763&  49& $ 0.05$&    B13& PASTEL& 1 \\
 HD034411&       G1V &        -- & $   5.30$ & $   4.69$ & $   4.16$ & $   3.85$ &        -- &        -- &      $   3.61$ &      $   3.33$ &      $   3.28$& 0.007&  5774&  44& $ 0.04$&    B13& PASTEL& 1 \\
 HD022879$^{b}$&       F9V & $   7.02$ & $   7.11$ & $   6.60$ & $   6.14$ & $   5.82$ & $   6.28$ & $   5.97$ &      $   5.57$ &      $   5.25$ &      $   5.22$& 0.026&  5786&  16& $-0.84$&    R13& PASTEL& 1 \\
 HD130948&    F9IV-V &        -- & $   6.41$ & $   5.85$ &        -- &        -- &        -- &        -- &      $   4.79$ &      $   4.53$ &      $   4.48$& 0.000&  5787&  57& $-0.05$&    B13& PASTEL& 1 \\
 HD128620&       G2V & $   0.72$ & $   0.50$ & $  -0.15$ &        -- &        -- & $  -0.47$ & $  -0.77$ &      $  -1.14$ &      $  -1.32$ &      $  -1.42$& 0.047&  5793&   7& $ 0.20$&    B13& PASTEL& 1 \\
 HD019373&    G0IV-V &        -- & $   4.65$ & $   4.05$ & $   3.52$ & $   3.23$ &        -- &        -- &      $   3.06$ &      $   2.73$ &      $   2.69$& 0.000&  5832&  33& $ 0.08$&    B13& PASTEL& 1 \\
 HD206860&    G0IV-V &        -- & $   6.53$ & $   5.94$ &        -- &        -- &        -- &        -- &             -- &             -- &             --& 0.000&  5860&  83& $-0.08$&    B13& PASTEL& 1 \\
 HD002151&      G2IV & $   3.47$ & $   3.37$ & $   2.76$ & $   2.26$ & $   1.92$ & $   2.42$ & $   2.10$ &      $   1.71$ &      $   1.39$ &      $   1.34$& 0.011&  5872&  44& $-0.11$&    N07& PASTEL& 1 \\
 HD039587&    G0IV-V &        -- & $   5.00$ & $   4.41$ & $   3.90$ & $   3.59$ &        -- &        -- &      $   3.34$ &      $   3.04$ &      $   2.97$& 0.000&  5898&  25& $-0.04$&    B13& PASTEL& 1 \\
 HD177153&       G0V &        -- & $   7.76$ & $   7.20$ &        -- &        -- &        -- &        -- &  $$   6.20$^a$ &  $$   5.92$^a$ &  $$   5.86$^a$& 0.001&  5909&  69& $-0.07$&    B13& PASTEL& 1 \\
 HD019994&     F8.5V &        -- & $   5.55$ & $   5.00$ &        -- &        -- & $   4.66$ & $   4.37$ &             -- &             -- &             --& 0.020&  5916&  98& $ 0.20$&    B13& PASTEL& 1 \\
 HD162003&    F5IV-V &        -- & $   5.01$ & $   4.58$ & $   4.20$ & $   3.97$ &        -- &        -- &      $   3.70$ &      $   3.47$ &      $   3.43$& 0.000&  5928&  81& $-0.07$&    B13& PASTEL& 1 \\
 HD114710&       G0V & $   4.92$ & $   4.84$ & $   4.26$ & $   3.77$ & $   3.47$ &        -- &        -- &      $   3.22$ &      $   2.95$ &      $   2.89$& 0.000&  5957&  29& $ 0.03$&    B13& PASTEL& 1 \\
 HD005015&       F8V &        -- & $   5.35$ & $   4.82$ & $   4.34$ & $   4.04$ &        -- &        -- &      $   3.85$ &      $   3.56$ &      $   3.54$& 0.001&  5965&  35& $ 0.01$&    B13& PASTEL& 1 \\
 HD004614&       F9V &        -- & $   3.95$ & $   3.38$ & $   2.90$ & $   2.55$ &        -- &        -- &      $   2.33$ &      $   2.01$ &      $   1.95$& 0.018&  5973&   8& $-0.30$&    B13& PASTEL& 1 \\
 HD022484&    F9IV-V &        -- & $   4.79$ & $   4.24$ & $   3.76$ & $   3.45$ & $   3.94$ & $   3.63$ &      $   3.28$ &      $   3.00$ &      $   2.91$& 0.014&  5998&  39& $-0.10$&    B13& PASTEL& 1 \\
 HD121370&      G0IV & $   3.46$ & $   3.26$ & $   2.68$ & $   2.24$ & $   1.95$ &        -- &        -- &      $   1.70$ &      $   1.38$ &      $   1.37$& 0.000&  6012&  45& $ 0.24$&    B13& PASTEL& 1 \\
 HD102870&  F8.5IV-V & $   4.26$ & $   4.15$ & $   3.60$ & $   3.12$ & $   2.84$ & $   3.28$ & $   2.99$ &      $   2.63$ &      $   2.35$ &      $   2.33$& 0.000&  6054&  13& $ 0.13$&    B13& PASTEL& 1 \\
 HD175726&       G5V &        -- & $   7.18$ & $   6.63$ &        -- &        -- &        -- &        -- &  $$   5.73$^a$ &  $$   5.40$^a$ &  $$   5.37$^a$& 0.027&  6067&  67& $-0.04$&    B13& PASTEL& 1 \\
 HD006210&       F6V &        -- & $   6.22$ & $   5.72$ &        -- &        -- &        -- &        -- &             -- &             -- &             --& 0.038&  6089&  35& $-0.17$&    B13& PASTEL& 1 \\
 HD215648&       F6V & $   4.62$ & $   4.65$ & $   4.16$ & $   3.74$ & $   3.44$ &        -- &        -- &      $   3.21$ &      $   3.06$ &      $   2.92$& 0.009&  6090&  22& $-0.29$&    B13& PASTEL& 1 \\
 HD009826&       F8V &        -- & $   4.64$ & $   4.10$ & $   3.64$ & $   3.35$ &        -- &        -- &      $   3.17$ &      $   2.99$ &      $   2.85$& 0.000&  6104&  75& $ 0.09$&    B13& PASTEL& 1 \\
 HD069897&       F6V &        -- & $   5.53$ & $   5.08$ &        -- &        -- &        -- &        -- &      $   4.15$ &      $   3.93$ &      $   3.90$& 0.020&  6130&  58& $-0.28$&    B13& PASTEL& 1 \\
 HD016895&       F7V &        -- & $   4.54$ & $   4.07$ & $   3.63$ & $   3.34$ &        -- &        -- &      $   3.32$ &      $   3.06$ &      $   2.97$& 0.019&  6153&  25& $ 0.00$&    B13& PASTEL& 1 \\
 HD187637&       F5V &        -- & $   8.03$ & $   7.52$ &        -- &        -- &        -- &        -- &  $$   6.60$^a$ &  $$   6.34$^a$ &  $$   6.32$^a$& 0.002&  6155&  85& $-0.14$&    B13& PASTEL& 1 \\
 HD222368&       F7V & $   4.61$ & $   4.62$ & $   4.11$ & $   3.68$ & $   3.37$ & $   3.82$ & $   3.54$ &             -- &             -- &             --& 0.005&  6192&  26& $-0.14$&    B13& PASTEL& 1 \\
 HD090839&       F8V &        -- & $   5.29$ & $   4.79$ & $   4.32$ & $   4.06$ &        -- &        -- &      $   3.83$ &      $   3.57$ &      $   3.53$& 0.016&  6203&  56& $-0.13$&    B13& PASTEL& 1 \\
 HD126660&       F7V & $   4.53$ & $   4.52$ & $   4.03$ & $   3.62$ & $   3.38$ &        -- &        -- &      $   3.09$ &      $   2.86$ &      $   2.82$& 0.009&  6211&  19& $-0.04$&    B13& PASTEL& 1 \\
 HD168151&       F5V &        -- & $   5.39$ & $   5.02$ &        -- &        -- &        -- &        -- &      $   4.09$ &      $   3.87$ &      $   3.84$& 0.023&  6221&  39& $-0.32$&    B13& PASTEL& 1 \\
 HD142860&       F6V & $   4.31$ & $   4.34$ & $   3.86$ & $   3.37$ & $   3.13$ &        -- &        -- &      $   2.93$ &      $   2.64$ &      $   2.65$& 0.000&  6222&  13& $-0.20$&    B13& PASTEL& 1 \\
 HD082328&  F5.5IV-V &        -- & $   3.57$ & $   3.13$ & $   2.70$ & $   2.44$ &        -- &        -- &      $   2.27$ &      $   2.02$ &      $   2.01$& 0.017&  6238&  10& $-0.18$&    B13& PASTEL& 1 \\
HD084937$^{b}$&      F5VI & $   8.47$ & $   8.67$ & $   8.31$ & $   7.96$ & $   7.69$ &        -- &        -- &      $   7.38$ &      $   7.13$ &      $   7.09$& 0.004&  6275&  17& $-2.12$&    R13& PASTEL& 2 \\
 HD181420&       F2V &        -- & $   6.97$ & $   6.54$ &        -- &        -- &        -- &        -- &  $$   5.79$^a$ &  $$   5.55$^a$ &  $$   5.55$^a$& 0.010&  6283& 106& $ 0.00$&    B13& PASTEL& 1 \\
 HD219623&       F8V &        -- & $   6.10$ & $   5.58$ &        -- &        -- &        -- &        -- &             -- &             -- &             --& 0.000&  6285&  94& $ 0.02$&    B13& PASTEL& 1 \\
 HD173667&  F5.5IV-V & $   4.58$ & $   4.58$ & $   4.14$ & $   3.76$ & $   3.52$ &        -- &        -- &      $   3.29$ &      $   3.07$ &      $   3.03$& 0.016&  6296&  19& $-0.05$&    B13& PASTEL& 1 \\
 HD210027&       F5V & $   4.17$ & $   4.20$ & $   3.76$ & $   3.36$ & $   3.11$ &        -- &        -- &      $   2.98$ &      $   2.71$ &      $   2.66$& 0.000&  6324& 139& $-0.18$&    B13& PASTEL& 1 \\
 HD016765&       F7V &        -- & $   6.23$ & $   5.71$ &        -- &        -- &        -- &        -- &             -- &             -- &             --& 0.000&  6356&  46& $-0.15$&    B13&    B13& 3 \\
 HD128167&       F4V & $   4.76$ & $   4.84$ & $   4.47$ & $   4.13$ & $   3.94$ &        -- &        -- &      $   3.65$ &      $   3.50$ &      $   3.49$& 0.000&  6435&  50& $-0.40$&    B13& PASTEL& 1 \\
 HD030652&    F6IV-V &        -- & $   3.60$ & $   3.15$ & $   2.74$ & $   2.49$ & $   2.89$ & $   2.64$ &      $   2.34$ &      $   2.14$ &      $   2.07$& 0.013&  6441&  19& $ 0.01$&    B13& PASTEL& 1 \\
 HD089449&    F6VI/V & $   5.26$ & $   5.25$ & $   4.80$ & $   4.35$ & $   4.12$ &        -- &        -- &      $   3.84$ &             -- &      $   3.65$& 0.000&  6450& 140& $ 0.09$&    M13& PASTEL& 1 \\
 HD164259&       F2V &        -- & $   4.98$ & $   4.60$ & $   4.27$ & $   4.09$ & $   4.38$ & $   4.16$ &      $   3.86$ &      $   3.70$ &      $   3.67$& 0.007&  6454& 113& $-0.12$&    B13& PASTEL& 1 \\
 HD048737&    F5IV-V &        -- & $   3.78$ & $   3.35$ & $   2.96$ & $   2.74$ &        -- &        -- &      $   2.57$ &      $   1.87$ &      $   2.30$& 0.003&  6478&  21& $ 0.14$&    B13&    B13& 4 \\
 HD061421&    F5IV-V &        -- & $   0.73$ & $   0.33$ & $  -0.08$ & $  -0.30$ & $   0.09$ & $  -0.14$ &      $  -0.41$ &      $  -0.51$ &      $  -0.60$& 0.014&  6582&  16& $-0.01$&    B13& PASTEL& 1 \\
 HD120136&    F7IV-V & $   5.02$ & $   4.98$ & $   4.50$ & $   4.09$ & $   3.85$ &        -- &        -- &      $   3.61$ &      $   3.40$ &      $   3.35$& 0.000&  6620&  67& $ 0.29$&    B13& PASTEL& 1 \\
 HD049933&       F2V &        -- & $   6.14$ & $   5.75$ &        -- &        -- &        -- &        -- &      $   4.90$ &      $   4.71$ &      $   4.67$& 0.006&  6635&  90& $-0.39$&    B13& PASTEL& 1 \\
 HD081937&      F0IV &        -- & $   4.00$ & $   3.67$ & $   3.33$ & $   3.15$ &        -- &        -- &      $   3.01$ &      $   3.00$ &      $   2.82$& 0.000&  6651&  27& $ 0.17$&    B13&    B13& 3 \\
 HD058946&       F0V &        -- & $   4.50$ & $   4.18$ & $   3.86$ & $   3.67$ &        -- &        -- &      $   3.58$ &      $   3.34$ &      $   3.36$& 0.000&  6738&  55& $-0.25$&    B13&    B13& 0 \\
 HD218396&       F0V &        -- & $   6.22$ & $   5.96$ &        -- &        -- &        -- &        -- &      $   5.46$ &      $   5.30$ &      $   5.28$& 0.006&  7163&  84& $-9.00$&    B13&     --& 4 \\
 HD187642&       A7V & $   0.96$ & $   0.89$ & $   0.69$ & $   0.57$ & $   0.46$ & $   0.58$ & $   0.46$ &      $   0.33$ &      $   0.23$ &      $   0.23$& 0.023&  7361&  91& $-0.24$&    M03& PASTEL& 1 \\
 HD219080&       F1V & $   4.55$ & $   4.56$ & $   4.33$ & $   4.09$ & $   3.98$ &        -- &        -- &      $   3.90$ &             -- &      $   3.75$& 0.062&  7380&  90& $-0.02$&    M13& PASTEL& 1 \\
 HD128898& A7VpSrCrE & $   3.49$ & $   3.38$ & $   3.15$ & $   2.93$ & $   2.84$ & $   3.03$ & $   2.78$ &      $   2.83$ &      $   2.72$ &      $   2.74$& 0.012&  7420& 170& $ 0.13$&   Ba08&    N94& 4 \\
 HD222603&       A7V & $   4.79$ & $   4.72$ & $   4.51$ & $   4.33$ & $   4.23$ & $   4.39$ & $   4.17$ &      $   4.10$ &      $   4.20$ &      $   4.00$& 0.000&  7734&  80& $-9.00$&    B13&     --& 3 \\
 HD210418&       A2V & $   3.67$ & $   3.58$ & $   3.52$ & $   3.47$ & $   3.44$ & $   3.46$ & $   3.42$ &      $   3.37$ &      $   3.37$ &      $   3.33$& 0.011&  7872&  82& $-0.38$&    B13& PASTEL& 1 \\
 HD097603&      A5IV &        -- & $   2.68$ & $   2.56$ & $   2.43$ & $   2.40$ &        -- &        -- &      $   2.33$ &      $   2.27$ &      $   2.27$& 0.000&  7889&  60& $-0.18$&    B13&    B13& 3 \\
 HD141795&      A2mV &        -- & $   3.81$ & $   3.66$ & $   3.59$ & $   3.55$ & $   3.61$ & $   3.56$ &             -- &             -- &             --& 0.012&  7928&  88& $ 0.32$&    B13& PASTEL& 1 \\
 HD118098&     A2Van & $   3.58$ & $   3.50$ & $   3.38$ & $   3.31$ & $   3.25$ & $   3.32$ & $   3.26$ &      $   3.18$ &      $   3.05$ &      $   3.06$& 0.000&  8097&  43& $-0.05$&    B13& PASTEL& 1 \\
 HD005448&       A6V & $   4.12$ & $   3.97$ & $   3.86$ & $   3.71$ & $   3.63$ &        -- &        -- &      $   3.57$ &      $   3.50$ &      $   3.49$& 0.004&  8320& 150& $ 0.03$&    M13&    M12& 2 \\
 HD216956&       A4V & $   1.19$ & $   1.15$ & $   1.08$ & $   1.04$ & $   1.04$ & $   1.04$ & $   1.03$ &      $   1.00$ &      $   1.04$ &      $   0.96$& 0.024&  8459&  44& $ 0.34$&    B13& PASTEL& 1 \\
 HD213558&       A1V & $   3.69$ & $   3.70$ & $   3.71$ & $   3.73$ & $   3.77$ &        -- &        -- &             -- &             -- &             --& 0.019&  9050& 157& $-9.00$&    B13&     --& 4 \\
 HD177724&    A0IV-V & $   2.92$ & $   2.94$ & $   2.95$ & $   2.95$ & $   2.96$ &        -- &        -- &      $   2.92$ &      $   3.02$ &      $   2.91$& 0.014&  9078&  86& $-0.52$&    B13&    B13& 4 \\
 HD095418&      A1IV &        -- & $   2.24$ & $   2.29$ & $   2.25$ & $   2.31$ &        -- &        -- &      $   2.33$ &      $   2.34$ &      $   2.34$& 0.026&  9193&  56& $ 0.16$&    B13& PASTEL& 1 \\
 HD097633&       A2V & $   3.40$ & $   3.33$ & $   3.35$ & $   3.32$ & $   3.34$ &        -- &        -- &      $   3.34$ &             -- &      $   3.30$& 0.000&  9480& 120& $-0.03$&    M13& PASTEL& 1 \\
 HD172167&       A0V & $   0.03$ & $   0.03$ & $   0.03$ & $   0.07$ & $   0.10$ & $   0.04$ & $   0.03$ &      $   0.00$ &      $   0.00$ &      $   0.00$& 0.000&  9657& 119& $-0.62$&    M03& PASTEL& 1 \\
 HD048915&      A0Va &        -- & $  -1.46$ & $  -1.46$ & $  -1.46$ & $  -1.43$ & $  -1.41$ & $  -1.40$ &      $  -1.34$ &      $  -1.33$ &      $  -1.31$& 0.000&  9711&  23& $ 0.34$&    B13& PASTEL& 1 \\
 \multicolumn{19}{c}{Giant stars} \\
  HD175865&     M5III & $   7.00$ & $   5.59$ & $   4.00$ & $   1.95$ & $   0.04$ &        -- &        -- &      $  -0.90$ &      $  -1.80$ &      $  -2.08$& 0.000&  3174&  41& $ 0.14$&          M03&        MILES& 2 \\
 HD132813&   M4.5III & $   7.86$ & $   6.19$ & $   4.54$ & $   2.69$ & $   0.97$ &        -- &        -- &      $   0.19$ &      $  -0.78$ &      $  -1.00$& 0.000&  3281&  45& $-9.00$&          M03&           --& 3 \\
 HD042995&     M3III & $   6.42$ & $   4.78$ & $   3.20$ & $   1.73$ & $   0.44$ &        -- &        -- &      $  -0.35$ &      $  -1.23$ &      $  -1.44$& 0.027&  3462&  43& $ 0.04$&          M03&       PASTEL& 4 \\
 HD011695&     M4III & $   7.59$ & $   5.89$ & $   4.33$ & $   2.62$ & $   1.13$ & $   2.94$ & $   0.46$ &      $   0.50$ &      $  -0.44$ &      $  -0.65$& 0.026&  3550&  50& $-9.00$&          W04&           --& 2 \\
 HD018884&  M1.5IIIa & $   5.55$ & $   3.70$ & $   2.17$ & $   0.91$ & $  -0.15$ & $   1.30$ & $   0.35$ &      $  -0.72$ &      $  -1.47$ &      $  -1.71$& 0.115&  3578&  53& $-0.22$&          M03&       PASTEL& 3 \\
 HD112300&     M3III & $   6.61$ & $   4.84$ & $   3.28$ & $   1.79$ & $   0.48$ & $   2.31$ & $   1.08$ &      $  -0.24$ &      $  -1.07$ &      $  -1.25$& 0.031&  3602&  44& $-0.16$&          M03&       PASTEL& 3 \\
 HD102212&   M1IIIab & $   7.32$ & $   5.54$ & $   4.04$ & $   2.79$ & $   1.78$ & $   3.10$ & $   1.31$ &      $   1.09$ &      $   0.25$ &      $   0.07$& 0.000&  3610&  53& $-0.41$&          M03&          K12& 3 \\
 HD216386& M2.5IIIaF & $   7.20$ & $   5.44$ & $   3.79$ & $   2.37$ & $   1.18$ & $   2.60$ & $   0.53$ &      $   0.44$ &      $  -0.38$ &      $  -0.64$& 0.000&  3639&  47& $-9.00$&          M03&           --& 3 \\
 HD100029&     M0III & $   7.44$ & $   5.47$ & $   3.85$ & $   2.54$ & $   1.55$ &        -- &        -- &      $   0.87$ &      $   0.18$ &      $  -0.12$& 0.000&  3675&  46& $-9.00$&          M03&           --& 3 \\
 HD025025& M1IIIbCA- & $   6.52$ & $   4.54$ & $   2.94$ & $   1.68$ & $   0.68$ & $   2.06$ & $   0.28$ &      $   0.07$ &      $  -0.74$ &      $  -0.93$& 0.000&  3703&  54& $-9.00$&          M03&           --& 3 \\
 HD146051&   M0.5III & $   6.30$ & $   4.34$ & $   2.75$ & $   1.46$ & $   0.43$ & $   1.83$ & $   0.91$ &      $  -0.24$ &      $  -0.98$ &      $  -1.26$& 0.000&  3721&  47& $ 0.32$&          M03&       PASTEL& 2 \\
 HD183439&     M0III & $   7.77$ & $   5.95$ & $   4.45$ & $   3.24$ & $   2.27$ &        -- &        -- &      $   1.53$ &             -- &      $   0.52$& 0.000&  3769&  46& $-0.38$&          M03&       PASTEL& 1 \\
 HD089758&     M0III & $   6.48$ & $   4.59$ & $   3.02$ & $   1.73$ & $   0.78$ &        -- &        -- &      $   0.09$ &      $  -0.69$ &      $  -0.86$& 0.011&  3793&  47& $ 0.00$&          M03&       PASTEL& 3 \\
 HD017709&     K5III & $   7.83$ & $   5.93$ & $   4.40$ & $   3.23$ & $   2.31$ &        -- &        -- &      $   1.73$ &             -- &      $   0.71$& 0.040&  3799&  47& $-0.36$&          M03&       PASTEL& 4 \\
 HD080493&   K7IIIab & $   6.13$ & $   4.26$ & $   2.82$ & $   1.67$ & $   0.85$ &        -- &        -- &      $   0.23$ &      $  -0.53$ &      $  -0.68$& 0.105&  3836&  47& $-0.26$&          M03&       PASTEL& 1 \\
 HD131873&     K4III & $   5.32$ & $   3.55$ & $   2.08$ & $   0.97$ & $   0.21$ &        -- &        -- &      $  -0.45$ &             -- &      $  -1.39$& 0.000&  3849&  47& $-0.29$&          M03&       PASTEL& 1 \\
 HD189319&     M0III & $   6.88$ & $   4.96$ & $   3.41$ & $   2.23$ & $   1.32$ &        -- &        -- &      $   0.68$ &             -- &      $  -0.26$& 0.019&  3836&  36& $-0.31$&     M03\&W06&       PASTEL& 4 \\
 HD029139&     K5III & $   4.20$ & $   2.30$ & $   0.78$ & $  -0.43$ & $  -1.35$ &        -- &        -- &      $  -1.89$ &      $  -2.62$ &      $  -2.86$& 0.025&  3871&  48& $-0.22$&          M03&       PASTEL& 1 \\
 HD069267&     K4III & $   6.32$ & $   4.62$ & $   3.23$ & $   2.19$ & $   1.47$ & $   2.51$ & $   1.87$ &      $   0.98$ &      $   0.23$ &      $   0.10$& 0.098&  4012&  52& $-0.16$&          M03&       PASTEL& 1 \\
 HD164058&     K5III & $   5.26$ & $   3.43$ & $   1.99$ & $   0.90$ & $   0.12$ &        -- &        -- &      $  -0.51$ &             -- &      $  -1.38$& 0.076&  4013&  52& $-0.11$&          M03&       PASTEL& 1 \\
 HD098262&     K3III & $   6.44$ & $   4.89$ & $   3.49$ & $   2.43$ & $   1.73$ &        -- &        -- &      $   1.18$ &             -- &      $   0.31$& 0.000&  4091&  50& $-0.11$&          M03&       PASTEL& 1 \\
 HD124897& K1.5IIIFe & $   2.46$ & $   1.18$ & $  -0.05$ & $  -1.03$ & $  -1.68$ & $  -0.76$ & $  -1.34$ &      $  -2.22$ &      $  -2.91$ &      $  -3.00$& 0.000&  4226&  53& $-0.55$&          M03&       PASTEL& 1 \\
 HD023319&   K2.5III & $   7.03$ & $   5.73$ & $   4.55$ &        -- &        -- & $   3.95$ & $   3.44$ &             -- &             -- &             --& 0.012&  4294&  58& $ 0.30$&          C12&       PASTEL& 1 \\
 HD073108&     K1III & $   6.94$ & $   5.78$ & $   4.61$ & $   3.72$ & $   3.09$ &        -- &        -- &      $   2.48$ &      $   1.94$ &      $   1.88$& 0.000&  4336&  99& $-0.23$&        Ba10b&       PASTEL& 1 \\
 HD102328&     K3III & $   8.05$ & $   6.56$ & $   5.28$ &        -- &        -- &        -- &        -- &             -- &             -- &             --& 0.000&  4358&  97& $ 0.09$&        Ba10b&       PASTEL& 1 \\
 HD003627&     K3III & $   5.90$ & $   4.44$ & $   3.19$ & $   2.29$ & $   1.66$ &        -- &        -- &      $   1.15$ &             -- &      $   0.42$& 0.030&  4392&  54& $ 0.26$&          M03&       PASTEL& 1 \\
 HD085503&     K3III & $   6.42$ & $   5.03$ & $   3.83$ & $   2.93$ & $   2.37$ &        -- &        -- &      $   1.92$ &             -- &      $   1.21$& 0.017&  4433&  51& $ 0.34$&          R13&       PASTEL& 1 \\
 HD160290&     K1III & $   7.57$ & $   6.51$ & $   5.36$ &        -- &        -- &        -- &        -- &             -- &             -- &             --& 0.000&  4493&  98& $-0.21$&        Ba10b&       PASTEL& 1 \\
 HD012929&     K2III & $   4.15$ & $   3.05$ & $   1.92$ & $   1.08$ & $   0.50$ &        -- &        -- &      $   0.04$ &      $  -0.54$ &      $  -0.64$& 0.029&  4493&  55& $-0.23$&          M03&       PASTEL& 1 \\
 HD137759&     K2III & $   5.56$ & $   4.35$ & $   3.21$ & $   2.45$ & $   1.87$ &        -- &        -- &      $   1.36$ &             -- &      $   0.67$& 0.025&  4545& 110& $ 0.09$&         Ba11&       PASTEL& 1 \\
 HD096833&     K1III & $   5.07$ & $   3.98$ & $   2.88$ & $   2.08$ & $   1.55$ &        -- &        -- &      $   1.12$ &             -- &      $   0.42$& 0.042&  4563&  42& $-0.07$&   M03\&Ba10a&       PASTEL& 1 \\
 HD140573& K2IIIbCN1 & $   5.08$ & $   3.82$ & $   2.65$ & $   1.82$ & $   1.25$ & $   2.05$ & $   1.55$ &      $   0.74$ &      $   0.18$ &      $   0.06$& 0.000&  4558&  56& $ 0.05$&          M03&       PASTEL& 1 \\
 HD003712&    K0IIIa & $   4.13$ & $   3.07$ & $   1.98$ & $   1.26$ & $   0.73$ &        -- &        -- &      $   0.35$ &             -- &      $  -0.28$& 0.082&  4602&  57& $-0.15$&          M03&       PASTEL& 1 \\
 HD095689&    K0IIIa & $   3.77$ & $   2.85$ & $   1.78$ & $   0.98$ & $   0.40$ &        -- &        -- &      $  -0.01$ &      $  -0.61$ &      $  -0.64$& 0.002&  4637&  62& $-0.20$&          M03&       PASTEL& 1 \\
 HD175955&     K0III &        -- & $   7.95$ & $   6.84$ &        -- &        -- &        -- &        -- &  $$   5.00$^a$ &             -- &  $$   4.32$^a$& 0.057&  4688&  66& $ 0.12$&          H12&       PASTEL& 1 \\
 HD215665&   G8IIIa* & $   5.69$ & $   4.79$ & $   3.77$ & $   3.04$ & $   2.58$ &        -- &        -- &      $   2.21$ &             -- &      $   1.63$& 0.050&  4699&  71& $-0.08$&          M03&       PASTEL& 1 \\
 HD188310&     G9III & $   6.64$ & $   5.75$ & $   4.70$ &        -- &        -- & $   4.17$ & $   3.69$ &             -- &             -- &             --& 0.000&  4742&  26& $-0.32$&         Ba09&       PASTEL& 1 \\
 HD197989&     K0III & $   4.30$ & $   3.44$ & $   2.42$ & $   1.70$ & $   1.17$ &        -- &        -- &      $   0.72$ &      $   0.19$ &      $   0.09$& 0.012&  4756&  59& $-0.13$&          M03&       PASTEL& 1 \\
 HD177151&     K0III &        -- & $   7.97$ & $   6.99$ &        -- &        -- &        -- &        -- &  $$   5.30$^a$ &  $$   4.81$^a$ &  $$   4.71$^a$& 0.016&  4761&  70& $-0.09$&          H12&       PASTEL& 1 \\
 HD028307&    K0IIIb & $   5.52$ & $   4.79$ & $   3.84$ &        -- &        -- &        -- &        -- &      $   2.29$ &             -- &      $   1.73$& 0.000&  4811&  50& $ 0.11$&          B09&       PASTEL& 1 \\
 HD186815&     K2III &        -- & $   7.13$ & $   6.26$ &        -- &        -- &        -- &        -- &             -- &             -- &             --& 0.006&  4823&  81& $-0.32$&        Ba10b&          B10& 2 \\
 HD221345&     G8III & $   6.93$ & $   6.09$ & $   5.10$ &        -- &        -- &        -- &        -- &             -- &             -- &             --& 0.039&  4826&  40& $-0.37$&         Ba09&       PASTEL& 1 \\
 HD148387&   G8IIIab & $   4.35$ & $   3.65$ & $   2.74$ & $   2.12$ & $   1.66$ &        -- &        -- &      $   1.17$ &             -- &      $   0.61$& 0.001&  4826&  71& $-0.11$&          M03&       PASTEL& 1 \\
 HD027697&     K0III & $   5.56$ & $   4.74$ & $   3.76$ &        -- &        -- &        -- &        -- &      $   2.15$ &             -- &      $   1.59$& 0.000&  4826&  51& $ 0.09$&          B09&       PASTEL& 1 \\
 HD028305&   G9.5III & $   5.42$ & $   4.54$ & $   3.53$ &        -- &        -- &        -- &        -- &      $   1.90$ &             -- &      $   1.32$& 0.000&  4827&  44& $ 0.11$&          B09&       PASTEL& 1 \\
 HD150997& G7.5IIIbF & $   5.03$ & $   4.42$ & $   3.50$ & $   2.83$ & $   2.35$ &        -- &        -- &      $   1.98$ &             -- &      $   1.35$& 0.000&  4841&  63& $-0.17$&          M03&       PASTEL& 1 \\
 HD028305&   G9.5III & $   5.42$ & $   4.54$ & $   3.53$ &        -- &        -- &        -- &        -- &      $   1.90$ &             -- &      $   1.32$& 0.000&  4843&  62& $ 0.11$&          M03&        MILES& 2 \\
 HD027371&     K0III & $   5.46$ & $   4.64$ & $   3.65$ &        -- &        -- &        -- &        -- &      $   2.02$ &             -- &      $   1.49$& 0.000&  4844&  47& $ 0.13$&          B09&       PASTEL& 1 \\
 HD135722&     G8III & $   5.07$ & $   4.40$ & $   3.46$ & $   2.74$ & $   2.24$ &        -- &        -- &      $   1.79$ &      $   1.26$ &      $   1.19$& 0.009&  4850&  60& $-0.36$&          M03&       PASTEL& 1 \\
 HD180711&     G9III & $   4.84$ & $   4.06$ & $   3.06$ & $   2.37$ & $   1.86$ &        -- &        -- &      $   1.43$ &             -- &      $   0.80$& 0.002&  4851&  67& $-0.20$&          M03&       PASTEL& 1 \\
 HD062509&    K0IIIb & $   3.00$ & $   2.14$ & $   1.14$ & $   0.39$ & $  -0.11$ &        -- &        -- &      $  -0.52$ &      $  -1.00$ &      $  -1.11$& 0.000&  4858&  60& $ 0.06$&          M03&       PASTEL& 1 \\
 HD027697&     K0III & $   5.52$ & $   4.71$ & $   3.74$ &        -- &        -- &        -- &        -- &      $   2.14$ &             -- &      $   1.59$& 0.007&  4897&  65& $ 0.09$&          M03&       PASTEL& 4 \\
 HD133208& G8IIIaBa0 & $   5.10$ & $   4.38$ & $   3.45$ & $   2.80$ & $   2.34$ &        -- &        -- &      $   1.88$ &      $   1.42$ &      $   1.33$& 0.022&  5017&  53& $-0.05$&   M03\&Ba10a&       PASTEL& 1 \\
 HD148856&    G7IIIa & $   4.22$ & $   3.56$ & $   2.66$ & $   2.05$ & $   1.60$ &        -- &        -- &      $   1.18$ &      $   0.67$ &      $   0.61$& 0.034&  4979&  61& $-0.16$&          M03&       PASTEL& 1 \\
 HD113226&   G8IIIab & $   4.45$ & $   3.71$ & $   2.79$ & $   2.16$ & $   1.71$ &        -- &        -- &      $   1.29$ &      $   0.77$ &      $   0.80$& 0.000&  4981&  61& $ 0.13$&          M03&       PASTEL& 1 \\
 HD202109& G8III/III & $   4.82$ & $   4.08$ & $   3.12$ & $   2.45$ & $   2.00$ &        -- &        -- &      $   1.58$ &      $   1.13$ &      $   1.07$& 0.029&  5002&  62& $-0.06$&          M03&       PASTEL& 1 \\
 HD216131&     G8III & $   5.08$ & $   4.41$ & $   3.47$ & $   2.79$ & $   2.32$ &        -- &        -- &      $   1.93$ &             -- &      $   1.37$& 0.004&  5084& 51& $-0.05$&   M03\&Ba10a&       PASTEL& 1 \\
 HD181827&     K0III &        -- & $   8.20$ & $   7.19$ &        -- &        -- &        -- &        -- &  $$   5.50$^a$ &             -- &  $$   4.90$^a$& 0.000&  5039&  66& $ 0.14$&          H12&       PASTEL& 1 \\
 HD100407&     G7III & $   5.17$ & $   4.47$ & $   3.54$ & $   2.84$ & $   2.36$ & $   3.06$ & $   2.63$ &      $   2.04$ &      $   1.57$ &      $   1.45$& 0.000&  5044&  33& $ 0.08$&     R13&       PASTEL& 1 \\
 HD189349&     G5III &        -- & $   7.90$ & $   7.09$ &        -- &        -- &        -- &        -- &  $$   5.63$^a$ &  $$   5.13$^a$ &  $$   5.13$^a$& 0.070&  5282&  72& $-0.56$&          H12&          H12& 2 \\
 HD205435&     G5III & $   5.47$ & $   4.91$ & $   4.02$ & $   3.31$ & $   2.81$ &        -- &        -- &      $   2.47$ &      $   2.06$ &      $   1.96$& 0.000&  5663&  74& $-0.13$&        Ba10a&       PASTEL& 1 \\
\end{longtable}
\end{center}
\scriptsize
\textbf{Notes.} Spectral types are taken from the SIMBAD database (\url{http://simbad.u-strasbg.fr/simbad/}) or from Ducati (2002).\\
Photometric magnitudes of the Johnson system ($UBVRIJHK$) are taken from B12b, B13, Ducati (2002), Mermilliod et al. (1997) and from the Lausanne photometric data base (\url{http://obswww.unige.ch/gcpd/gcpd.html}).\\
Flag1 and Flag2 give respectively the references from which the value of $T_{\rm eff}$ and [Fe/H] is taken from.\\
The references are: N94--North et al. (1994); M03--Mozurkewich et al. (2003); W04--Wittkowski et al. (2004);  MILES--S\'anchez-Bl\'azquez et al. (2006); W06--Wittkowski et al. (2006); N07--North et al. (2007); Ba08--Baines et al. (2008); Ba09--Baines et al. (2009); B09--Boyajian et al. (2009); Ba10a--Baines et al. (2010a); Ba10b--Baines et al. (2010b); PASTEL--Soubiran et al. (2010); Ba11--Baines et al. (2011); B12b--Boyajian et al. (2012b); C12--Cusano et al. (2012); H12--Huber et al. (2012); K12--Koleva \& Vazdekis (2012); R13--Ruchti et al. (2013); B13--Boyajian et al. (2013); M13--Maestro et al. (2013).\\
Flag3 indicates the method used for the extinction determination. See Table 2 for details.\\
$^{a}$ The Johnson $JHK$ magnitudes are converted from the 2MASS photometry. See Section 2 for details.\\
$^{b}$Stars with angular diameters determined using the surface--brightness relations of Kervella et al. (2004).\\
$^{*}$  Stars with multiple direct effective temperature measurements. The adopted value is the mean weighted by the uncertainties of individual measurements.

\end{landscape}
\twocolumn
\normalsize

\begin{figure}
\centering
\includegraphics[scale=0.5,angle=0]{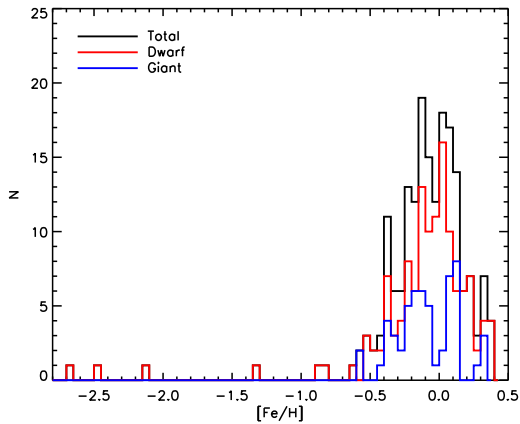}
\caption{Metallicity distributions of our sample stars compiled  from the literature. The black, red and blue histograms represent the distributions of the whole, the dwarf and the giant samples, respectively.}
\end{figure}

 The metallicity [Fe/H] of these stars are mainly taken from the PASTEL catalog (Soubiran et al. 2010), a bibliographical compilation of measurements of stellar atmospheric parameters ($T_{\rm eff}$, log $g$, [Fe/H]), mostly obtained from the analysis of high resolution ($R \ge$ 30,\,000) and high signal-to-noise ratio (S/N $\ge 100$) spectra.
 As of the version of April, 2013, over 8000 stars catalogued by the PASTEL have at least one set of determinations for all the three parameters $T_{\rm eff}$, log $g$ and [Fe/H]. 
 Amongst them, more than 3000 (36 per cent) stars have at least two independent  determinations available for [Fe/H].
 There are166  stars (85 per cent) stars in our sample that have all three atmospheric parameters available from the PASTEL.
 For those stars with multiple [Fe/H] determinations, we simply take the mean values with 3$\sigma$ clipping. 
The [Fe/H] values of the remaining 21 (11 per cent) stars are taken from the literature not included in PASTEL. 
There are still 8 (4 per cent) stars without a [Fe/H] determination.
The distributions of [Fe/H] values of dwarfs and giants in our sample range from $\sim$ $-0.8$ to $0.4$ (Population I) except for a few metal-poor dwarfs as shown in Fig.\,1.

 Four photometric systems are used in the current work to derive the empirical metallicity-dependent $T_{\rm eff}$--colour relations: the standard Johnson ($UBVR_{\rm J}I_{\rm J}JHK$),  the Cousins ($R_{\rm C}I_{\rm C}$), the SDSS ($gr$) and the 2MASS ($JHK_{\rm s}$, hereafter $J_{2}$, $H_{2}$ and $K_{2}$  in order to distinguish from the Johnson $J$, $H$ and $K$).
Most stars in our sample have high quality multiple measurements in the Johnson system, especially in $B$ and $V$ bands (the magnitudes of the two bands are available for all stars in our sample).
Cousins ($R_{\rm C}I_{\rm C}$) magnitudes available for the current sample of stars are quite limited, especially for giants (about 10 stars). 
The $J_{2}$, $H_{2}$ and $K_{2}$ magnitudes for most stars are saturated because they are too bright for 2MASS.  
For the same reason, there are no SDSS photometry for all sample stars.
On the other hand, the later two photometric systems are of particular interest for the potential applications of the empirical relations discussed here, considering that  2MASS is one of the major near-infrared all sky survey that yields high precision photometry for nearly 471 million point-like sources (Skrutskie et al. 2006) and that the SDSS-like filters are now the most widely used in modern large-scale digital sky surveys  including the Xuyi Schmidt Telescope Photometric Survey of the Galactic Anti-centre (XSTPS-GAC; Zhang et al. 2013, 2014; Liu et al. 2014),  the Panoramic Survey Telescope \& Rapid Response System survey (PanSTARRS; Kaiser et al. 2002), the SkyMapper (Keller et al. 2007) and the future Large Synoptic Survey Telescope survey (LSST; LSST collaboration 2009).
It is thus essential to derive the empirical metallicity-dependent $T_{\rm eff}$-colour relations in term of the SDSS and 2MASS photometric colours.
In order to obtain magnitudes in the SDSS and 2MASS photometric systems, we have applied transformation equations that convert the Johnson magnitudes to those two  systems for our sample stars.
For the SDSS photometry system, $g$ and $r$ magnitudes are deduced from the Johnson $B$ and $V$ magnitudes using the transformations derived by Jester et al. (2005).
For the 2MASS $J_{2}$, $H_{2}$ and $K_{2}$, we first convert the Johnson magnitudes to the system of Bessell \& Brett using the transformations given by Bessell \& Brett (1988) and then to the 2MASS system using the transformations of Carpenter (2001)\footnote{We adopt the latest transformations available from \url{http://www.astro.caltech.edu/~jmc/2mass/v3/transformations/}.}.
Only a few stars in our sample have Johnson $J$, $H$ and $K$ photometry unavailable, but instead have good 2MASS photometry\footnote{Only stars with {\it ph\_qual} flag flagged `A' and photometric error less than 0.05 mag in all three 2MASS bands are considered as having good photometry.}.
For those stars, we obtain their Johnson $J$, $H$ and $K$ magnitudes from the 2MASS ones by reversing the aforementioned  transformations.

\begin{table}
\begin{center}
\caption{Summary of $E(B-V)$ determinations}
\begin{threeparttable}
\begin{tabular}{ccc}
\hline
Flag&Method\tnote{*}&$N_{\rm star}$\\
\hline
0&$d\,\le\,20$ pc, $A_{V}\,=\,0$&3\\
1&`star pairs'&154\\
2&literature&12\\
3&SFD98&15\\
4&extinction-distance relation&11\\
  \hline
 \end{tabular}
 \begin{tablenotes}
\item[*]When the extinction of a star has been determined with multiple methods, we adopt the one with the highest priority. 
               The priority, from high to low, is given by Flag value sequence 1, 2, 3, 4 and 0. 
 \end{tablenotes}
\end{threeparttable}
\end{center}
\end{table}

\begin{table*}
 \begin{center}
\begin{scriptsize}
  \caption{Fit coefficients, applicability ranges in metallicity and colour of the metallicity-dependent  $T_{\rm eff}$-colour relations for dwarfs}
  \begin{tabular}{cccccccccccc}
  \hline
Colour &[Fe/H] range\tnote{*} &Colour range &$a_0$ &$a_1$ &$a_2$ &$a_3$ &$a_4$ &$a_5$ &$N$ & s.d.$(\%)$ &Type \\
\hline
$    U-V $ &  [$-0.8$, $  0.4 $]   &  [$+0.11$, $  2.58 $]  &$  0.64612 $&$  0.32629 $&$ -0.05032 $&$ -0.02288 $&$ -0.14090 $&$ -0.04187  $&  42 & 2.18 &  AFGKM \\
$    U-V $ &  [$-0.8$, $  0.4 $]   &  [$+0.16$, $  2.58 $]  &$  0.68567 $&$  0.25708 $&$ -0.02644 $&$ -0.03637 $&$ -0.09993 $&$ -0.03003  $&  40 & 1.73 &   FGKM \\
$    B-V $ &  [$-0.9$, $  0.4 $]   &  [$-0.05$, $  1.73 $]  &$  0.59225 $&$  0.39926 $&$  0.07848 $&$ -0.04374 $&$ -0.04289 $&$ -0.01406  $& 120 & 2.02 &  AFGKM \\
$    B-V $ &  [$-0.9$, $  0.4 $]   &  [$+0.32$, $  1.73 $]  &$  0.63421 $&$  0.30538 $&$  0.12308 $&$ -0.06216 $&$ -0.01987 $&$ -0.00951  $& 111 & 1.96 &   FGKM \\
$    V-R_{\rm J} $ &  [$-0.9$, $  0.4 $]   &  [$-0.04$, $  1.69 $]  &$  0.53767 $&$  0.66928 $&$ -0.05870 $&$ -0.04605 $&$ -0.02025 $&$ -0.01402  $&  86 & 3.21 &  AFGKM \\
$    V-R_{\rm J} $ &  [$-0.9$, $  0.4 $]   &  [$+0.32$, $  1.69 $]  &$  0.50741 $&$  0.74628 $&$ -0.10012 $&$ -0.05966 $&$ -0.00411 $&$ -0.00297  $&  71 & 2.64 &   FGKM \\
$    R_{\rm J}-I_{\rm J} $ &  [$-0.9$, $  0.4 $]   &  [$-0.03$, $  1.43 $]  &$  0.55253 $&$  1.09402 $&$ -0.32146 $&$ -0.12854 $&$  0.03871 $&$ -0.00030  $&  84 & 2.89 &  AFGKM \\
$    R_{\rm J}-I_{\rm J} $ &  [$-0.9$, $  0.4 $]   &  [$+0.18$, $  1.43 $]  &$  0.52917 $&$  1.18647 $&$ -0.38691 $&$ -0.14505 $&$  0.05246 $&$  0.00312  $&  73 & 2.73 &   FGKM \\
$    V-I_{\rm J} $ &  [$-0.9$, $  0.4 $]   &  [$-0.07$, $  3.12 $]  &$  0.54700 $&$  0.41277 $&$ -0.03462 $&$ -0.04607 $&$  0.00706 $&$ -0.01032  $&  86 & 2.79 &  AFGKM \\
$    V-I_{\rm J} $ &  [$-0.9$, $  0.4 $]   &  [$+0.51$, $  3.12 $]  &$  0.50673 $&$  0.47658 $&$ -0.05391 $&$ -0.04491 $&$  0.01236 $&$ -0.00734  $&  73 & 2.43 &   FGKM \\
$   V-R_{\rm C} $ &  [$-0.8$, $  0.3 $]   &  [$-0.05$, $  1.21 $]  &$  0.56180 $&$  0.93029 $&$ -0.11466 $&$ -0.15858 $&$  0.02426 $&$ -0.00998  $&  41 & 2.51 &  AFGKM \\
$   V-R_{\rm C} $ &  [$-0.8$, $  0.2 $]   &  [$+0.22$, $  1.21 $]  &$  0.57684 $&$  0.87545 $&$ -0.07718 $&$ -0.22472 $&$  0.06639 $&$ -0.00347  $&  33 & 2.35 &   FGKM \\
$  R_{\rm C}-I_{\rm C} $ &  [$-0.8$, $  0.3 $]   &  [$-0.01$, $  1.56 $]  &$  0.53569 $&$  1.17516 $&$ -0.36436 $&$ -0.24823 $&$  0.05024 $&$ -0.02187  $&  41 & 2.93 &  AFGKM \\
$  R_{\rm C}-I_{\rm C} $ &  [$-0.8$, $  0.2 $]   &  [$+0.22$, $  1.56 $]  &$  0.52327 $&$  1.22375 $&$ -0.39607 $&$ -0.30792 $&$  0.08613 $&$ -0.01724  $&  35 & 2.69 &   FGKM \\
$   V-I_{\rm C} $ &  [$-0.8$, $  0.3 $]   &  [$-0.06$, $  2.77 $]  &$  0.54447 $&$  0.54119 $&$ -0.06920 $&$ -0.10748 $&$  0.04723 $&$ -0.01367  $&  40 & 2.46 &  AFGKM \\
$   V-I_{\rm C} $ &  [$-0.8$, $  0.2 $]   &  [$+0.44$, $  2.77 $]  &$  0.53066 $&$  0.56556 $&$ -0.07742 $&$ -0.13244 $&$  0.07752 $&$ -0.00969  $&  34 & 2.38 &   FGKM \\
$    V-J $ &  [$-0.9$, $  0.4 $]   &  [$-0.12$, $  4.24 $]  &$  0.54007 $&$  0.33983 $&$ -0.02512 $&$ -0.05359 $&$  0.06601 $&$ -0.00133  $& 107 & 2.28 &  AFGKM \\
$    V-J $ &  [$-0.9$, $  0.4 $]   &  [$+0.60$, $  4.24 $]  &$  0.51683 $&$  0.36567 $&$ -0.03065 $&$ -0.05507 $&$  0.06928 $&$ -0.00047  $&  95 & 2.08 &   FGKM \\
$    V-H $ &  [$-0.9$, $  0.4 $]   &  [$-0.13$, $  4.76 $]  &$  0.55015 $&$  0.24801 $&$ -0.00887 $&$ -0.04204 $&$  0.06782 $&$ -0.00168  $& 103 & 2.31 &  AFGKM \\
$    V-H $ &  [$-0.9$, $  0.4 $]   &  [$+0.67$, $  4.76 $]  &$  0.57826 $&$  0.22238 $&$ -0.00473 $&$ -0.07342 $&$  0.13820 $&$  0.01264  $&  92 & 2.00 &   FGKM \\
$    V-K $ &  [$-0.9$, $  0.4 $]   &  [$-0.15$, $  5.03 $]  &$  0.55261 $&$  0.23293 $&$ -0.00757 $&$ -0.03760 $&$  0.05030 $&$ -0.00516  $& 107 & 1.96 &  AFGKM \\
$    V-K $ &  [$-0.9$, $  0.4 $]   &  [$+0.82$, $  5.03 $]  &$  0.56784 $&$  0.22038 $&$ -0.00585 $&$ -0.06218 $&$  0.10765 $&$  0.00651  $&  97 & 1.87 &   FGKM \\
$    g-r $ &  [$-0.9$, $  0.4 $]   &  [$-0.28$, $  1.53 $]  &$  0.68692 $&$  0.42617 $&$  0.07504 $&$ -0.04372 $&$ -0.05220 $&$ -0.01390  $& 120 & 2.02 &  AFGKM \\
$    g-r $ &  [$-0.9$, $  0.4 $]   &  [$+0.10$, $  1.53 $]  &$  0.70726 $&$  0.33962 $&$  0.13127 $&$ -0.04405 $&$ -0.03616 $&$ -0.00935  $& 109 & 1.86 &   FGKM \\
$    g-J $ &  [$-0.9$, $  0.4 $]   &  [$-0.24$, $  5.15 $]  &$  0.57513 $&$  0.24031 $&$ -0.01020 $&$ -0.03323 $&$  0.03011 $&$ -0.00494  $& 108 & 2.06 &  AFGKM \\
$    g-J $ &  [$-0.9$, $  0.4 $]   &  [$+0.67$, $  5.15 $]  &$  0.56499 $&$  0.24982 $&$ -0.01194 $&$ -0.03654 $&$  0.03880 $&$ -0.00298  $&  95 & 1.88 &   FGKM \\
$    g-H $ &  [$-0.9$, $  0.4 $]   &  [$-0.25$, $  5.68 $]  &$  0.57538 $&$  0.19009 $&$ -0.00388 $&$ -0.02659 $&$  0.03663 $&$ -0.00357  $& 102 & 2.11 &  AFGKM \\
$    g-H $ &  [$-0.9$, $  0.4 $]   &  [$+0.75$, $  5.68 $]  &$  0.60421 $&$  0.16755 $&$ -0.00064 $&$ -0.05271 $&$  0.10246 $&$  0.00965  $&  93 & 1.93 &   FGKM \\
$    g-K $ &  [$-0.9$, $  0.4 $]   &  [$-0.27$, $  5.95 $]  &$  0.57596 $&$  0.18108 $&$ -0.00342 $&$ -0.02443 $&$  0.02469 $&$ -0.00617  $& 107 & 1.90 &  AFGKM \\
$    g-K $ &  [$-0.9$, $  0.4 $]   &  [$+0.89$, $  5.95 $]  &$  0.59712 $&$  0.16510 $&$ -0.00123 $&$ -0.04565 $&$  0.07963 $&$  0.00481  $&  96 & 1.76 &   FGKM \\
$  V-J_2 $ &  [$-0.9$, $  0.4 $]   &  [$-0.06$, $  4.28 $]  &$  0.52082 $&$  0.34241 $&$ -0.02495 $&$ -0.05376 $&$  0.06846 $&$ -0.00170  $& 107 & 2.26 &  AFGKM \\
$  V-J_2 $ &  [$-0.9$, $  0.4 $]   &  [$+0.66$, $  4.28 $]  &$  0.49543 $&$  0.36971 $&$ -0.03068 $&$ -0.05506 $&$  0.07191 $&$ -0.00066  $&  95 & 2.07 &   FGKM \\
$  V-H_2 $ &  [$-0.9$, $  0.4 $]   &  [$-0.13$, $  4.77 $]  &$  0.54872 $&$  0.24869 $&$ -0.00896 $&$ -0.04212 $&$  0.06615 $&$ -0.00232  $& 103 & 2.25 &  AFGKM \\
$  V-H_2 $ &  [$-0.9$, $  0.4 $]   &  [$+0.69$, $  4.77 $]  &$  0.57565 $&$  0.22418 $&$ -0.00501 $&$ -0.07282 $&$  0.13574 $&$  0.01211  $&  92 & 1.95 &   FGKM \\
$  V-K_2 $ &  [$-0.9$, $  0.4 $]   &  [$-0.10$, $  5.05 $]  &$  0.54042 $&$  0.23676 $&$ -0.00796 $&$ -0.03798 $&$  0.05413 $&$ -0.00448  $& 113 & 1.98 &  AFGKM \\
$  V-K_2 $ &  [$-0.9$, $  0.4 $]   &  [$+0.86$, $  5.05 $]  &$  0.55470 $&$  0.22536 $&$ -0.00647 $&$ -0.06318 $&$  0.11273 $&$  0.00697  $& 103 & 1.88 &   FGKM \\
$  g-J_2 $ &  [$-0.9$, $  0.4 $]   &  [$-0.18$, $  5.19 $]  &$  0.56153 $&$  0.24125 $&$ -0.01009 $&$ -0.03327 $&$  0.03191 $&$ -0.00499  $& 108 & 2.05 &  AFGKM \\
$  g-J_2 $ &  [$-0.9$, $  0.4 $]   &  [$+0.73$, $  5.19 $]  &$  0.55158 $&$  0.25030 $&$ -0.01172 $&$ -0.03662 $&$  0.04096 $&$ -0.00296  $&  95 & 1.87 &   FGKM \\
$  g-H_2 $ &  [$-0.9$, $  0.4 $]   &  [$-0.25$, $  5.68 $]  &$  0.57407 $&$  0.19046 $&$ -0.00390 $&$ -0.02663 $&$  0.03644 $&$ -0.00342  $& 101 & 2.02 &  AFGKM \\
$  g-H_2 $ &  [$-0.9$, $  0.4 $]   &  [$+0.77$, $  5.68 $]  &$  0.60082 $&$  0.16907 $&$ -0.00078 $&$ -0.04698 $&$  0.07517 $&$  0.00001  $&  91 & 1.77 &   FGKM \\
$  g-K_2 $ &  [$-0.9$, $  0.4 $]   &  [$-0.22$, $  5.96 $]  &$  0.56653 $&$  0.18358 $&$ -0.00365 $&$ -0.02477 $&$  0.02794 $&$ -0.00552  $& 113 & 1.91 &  AFGKM \\
$  g-K_2 $ &  [$-0.9$, $  0.4 $]   &  [$+0.94$, $  5.96 $]  &$  0.58681 $&$  0.16856 $&$ -0.00163 $&$ -0.04666 $&$  0.08483 $&$  0.00564  $& 102 & 1.79 &   FGKM \\
  \hline
\end{tabular}
\end{scriptsize}
\end{center}
\end{table*}

\begin{table*}
 \begin{center}
 \begin{scriptsize}
  \caption{Fit coefficients, applicability ranges in metallicity and colour of the metallicity-dependent  $T_{\rm eff}$-colour relations for giants}
  \begin{tabular}{cccccccccccc}
  \hline
Colour &[Fe/H] range &Colour range &$a_0$ &$a_1$ &$a_2$ &$a_3$ &$a_4$ &$a_5$ &$N$ & s.d.$(\%)$ &Type \\
\hline
$    U-V $ &  [$-0.6$, $  0.3 $]   &  [$ 1.53$, $  3.55 $]  &$  0.85926 $&$  0.06128 $&$  0.02054 $&$  0.00341 $&$ -0.04250 $&$  0.08760  $&  45 & 1.55 &    GKM \\
$    B-V $ &  [$-0.6$, $  0.3 $]   &  [$ 0.88$, $  1.59 $]  &$  0.81784 $&$  0.03096 $&$  0.19350 $&$ -0.03292 $&$  0.00762 $&$  0.11140  $&  51 & 1.86 &    GKM \\
$    V-R_{\rm J} $ &  [$-0.6$, $  0.3 $]   &  [$ 0.61$, $  2.05 $]  &$  0.50982 $&$  0.86731 $&$ -0.16385 $&$  0.00099 $&$ -0.05206 $&$ -0.03531  $&  39 & 2.14 &    GKM \\
$    R-I_{\rm J} $ &  [$-0.6$, $  0.3 $]   &  [$ 0.45$, $  1.91 $]  &$  0.58534 $&$  1.07395 $&$ -0.30028 $&$  0.08547 $&$ -0.07481 $&$  0.10417  $&  39 & 2.34 &    GKM \\
$    V-I_{\rm J} $ &  [$-0.6$, $  0.3 $]   &  [$ 1.06$, $  3.96 $]  &$  0.51361 $&$  0.51406 $&$ -0.06282 $&$  0.03408 $&$ -0.09130 $&$  0.01118  $&  39 & 1.96 &    GKM \\
$    V-J $ &  [$-0.6$, $  0.3 $]   &  [$ 1.45$, $  4.90 $]  &$  0.46448 $&$  0.42236 $&$ -0.03986 $&$ -0.01037 $&$  0.00000 $&$ -0.06615  $&  48 & 1.92 &    GKM \\
$    V-H $ &  [$-0.6$, $  0.3 $]   &  [$ 1.95$, $  5.80 $]  &$  0.42078 $&$  0.33924 $&$ -0.02323 $&$ -0.04718 $&$  0.11736 $&$ -0.05597  $&  23 & 0.93 &    GKM \\
$    V-K $ &  [$-0.6$, $  0.3 $]   &  [$ 1.96$, $  6.08 $]  &$  0.47934 $&$  0.29738 $&$ -0.01888 $&$ -0.02526 $&$  0.06059 $&$ -0.04110  $&  48 & 1.68 &    GKM \\
$    g-r $ &  [$-0.6$, $  0.3 $]   &  [$ 0.66$, $  1.39 $]  &$  0.82134 $&$  0.14378 $&$  0.17360 $&$ -0.02994 $&$ -0.00477 $&$  0.10091  $&  51 & 1.88 &    GKM \\
$    g-J $ &  [$-0.6$, $  0.3 $]   &  [$ 1.81$, $  5.73 $]  &$  0.49851 $&$  0.30054 $&$ -0.01921 $&$ -0.01956 $&$  0.02561 $&$ -0.04232  $&  48 & 1.72 &    GKM \\
$    g-H $ &  [$-0.6$, $  0.3 $]   &  [$ 2.31$, $  6.63 $]  &$  0.46461 $&$  0.25124 $&$ -0.01189 $&$ -0.03594 $&$  0.09960 $&$ -0.03641  $&  23 & 0.93 &    GKM \\
$    g-K $ &  [$-0.6$, $  0.3 $]   &  [$ 2.32$, $  6.91 $]  &$  0.51612 $&$  0.22256 $&$ -0.00966 $&$ -0.01988 $&$  0.04938 $&$ -0.03274  $&  48 & 1.59 &    GKM \\
$  V-J_2 $ &  [$-0.6$, $  0.3 $]   &  [$ 1.51$, $  4.95 $]  &$  0.44863 $&$  0.41980 $&$ -0.03858 $&$ -0.02540 $&$  0.03704 $&$ -0.05304  $&  48 & 1.90 &    GKM \\
$  V-H_2 $ &  [$-0.6$, $  0.3 $]   &  [$ 1.94$, $  5.79 $]  &$  0.42367 $&$  0.33746 $&$ -0.02298 $&$ -0.04768 $&$  0.11713 $&$ -0.05342  $&  23 & 0.92 &    GKM \\
$  V-K_2 $ &  [$-0.6$, $  0.3 $]   &  [$ 1.99$, $  6.09 $]  &$  0.46447 $&$  0.30156 $&$ -0.01918 $&$ -0.02526 $&$  0.06132 $&$ -0.04036  $&  48 & 1.68 &    GKM \\
$  g-J_2 $ &  [$-0.6$, $  0.3 $]   &  [$ 1.87$, $  5.78 $]  &$  0.47230 $&$  0.30872 $&$ -0.02003 $&$ -0.01081 $&$ -0.00000 $&$ -0.05341  $&  48 & 1.72 &    GKM \\
$  g-H_2 $ &  [$-0.6$, $  0.3 $]   &  [$ 2.30$, $  6.63 $]  &$  0.46683 $&$  0.25008 $&$ -0.01175 $&$ -0.03545 $&$  0.09634 $&$ -0.03581  $&  23 & 0.93 &    GKM \\
$  g-K_2 $ &  [$-0.6$, $  0.3 $]   &  [$ 2.35$, $  6.92 $]  &$  0.50481 $&$  0.22511 $&$ -0.00980 $&$ -0.01982 $&$  0.04878 $&$ -0.03720  $&  48 & 1.59 &    GKM \\
  \hline
\end{tabular}
\end{scriptsize}
\end{center}
\end{table*}
\normalsize

Accurate reddening corrections are essential for constructing robust metallicity-dependent $T_{\rm eff}$--colour empirical relations, particularly for giant stars, which suffer from relatively high extinction given their large distances.
For most of the  sample stars, we have adopted extinction values deduced from the photometric colours using the `standard pair' technique (Stecher 1965; Massa et al. 1983; Yuan et al. 2013).
The technique assumes that stars of similar atmospheric parameters ($T_{\rm eff}$, log $g$, [Fe/H]) should have similar colours.
For this purpose, we first define a  `reference library'  by selecting stars with nil/low extinction and with all the three atmospheric parameters available from the PASTEL catalog. 
Then the extinction values of the current sample stars are estimated by comparing their observed colours with those in the `reference library'  that have similar atmospheric parameters.
A comparison with values given by the integrated extinction map of  Schlegel, Finkbeiner \& Davis (1998, hereafter SFD98) for high Galactic latitude stars shows the technique has achieved a precision of about 0.02 mag in $E(B-V)$.
The reddening values of $E(B-V)$ for  154 (79  per cent) sample stars are estimated with this method.
For the remaining stars, reddening values are adopted either from the literature, from the SFD98 map for stars of high Galactic latitudes, from extinction--distance relation that we derive for that particular line-of-sight\footnote{We first determine the extinction values for all stars in the PASTEL catalog that have all the three atmospheric parameters available.
Then we construct a library consisting of stars that have extinction determinations and geometric distances from the Hipparcos (van Leeuwen 2007).
An extinction--distance relation is then constructed using the stars in the library that fall within a solid angle (typically 4 square degree) of  line-of-sight to a sample star of concern here.}, or simply set to zero when they are close (within 20 pc).
Table\,2 presents a summary of the number of stars with their extinction determined by each method.
Finally, we correct the measured photometric magnitudes of each band for the interstellar reddening using the above estimated values of  $E(B-V)$ and the extinction law of Fitzpatrick (1999)  for $R=3.1$.

All relevant  information of our sample compiled here,  including $T_{\rm eff}$, photometric magnitudes\footnote{ Magnitudes converted from photometric measurements in other bands, i.e. those of SDSS $g,r$ and 2MASS $J_{2}H_{2}K_{2}$, are not listed in the Table.
One can obtain their values using the transformations described in Section\,2.} after reddening corrections, extinction, [Fe/H] and etc., is presented in Table\,1.
Unless specified otherwise, all magnitudes and colours presented in the paper refer to dereddened values.

\section{Empirical calibrations}

To obtain the metallicity-dependent $T_{\rm eff}$--colour relations,  data for dwarf and giant stars are fitted separately following the conventional approach (e.g. Alonso et al. 1996, 1999; Ram\'ires \& Mel\'endez 2005; Casagrande et al. 2006; Gonz\'alez Hern\'andes \& Bonifacio 2009; Casagrande et al. 2010):
\begin{equation}
\begin{split}
\theta_{\rm eff}\,=\,a_{0}\,+\,a_{1}X\,+\,a_{2}X^{2}\,+\,a_{3}X\rm{[Fe/H]}\, \\
+\,a_{4}{\rm[Fe/H]}\,+\,a_{5}\rm{[Fe/H]}^{2}\,,
\end{split}
\end{equation}
where $\theta_{\rm eff}\,=\,5040/T_{\rm eff}$, $X$ represents the colour of concern and $a_{i}$ ($i$ = 0,...,5) are the fit coefficients.
We iterate the fitting, discarding data points that deviate more than 2.5$\sigma$ from the fit.
In general, three to four iterations are sufficient. 

 \begin{figure*}
\centering
\includegraphics[scale=0.45,angle=0]{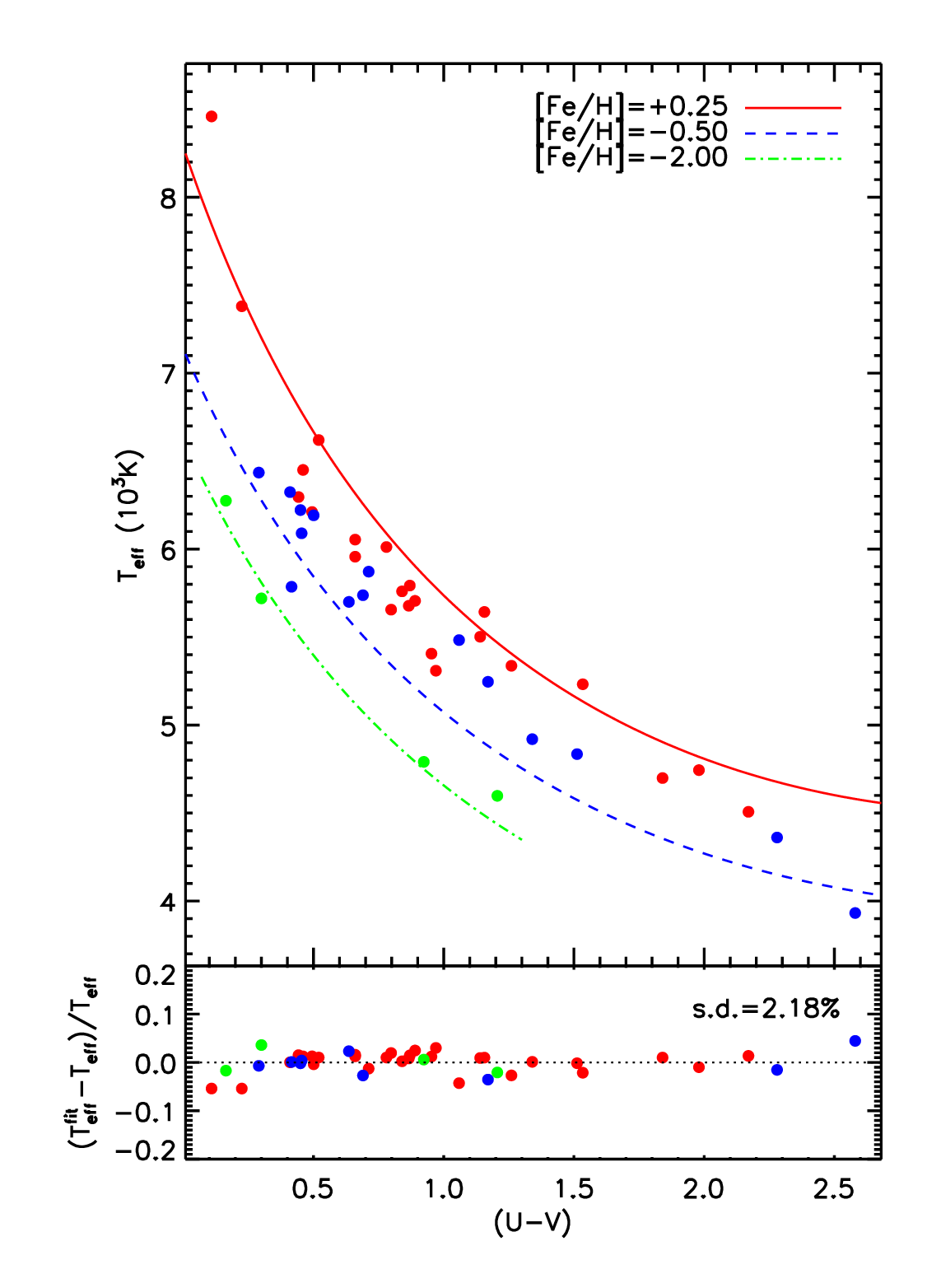}
\includegraphics[scale=0.45,angle=0]{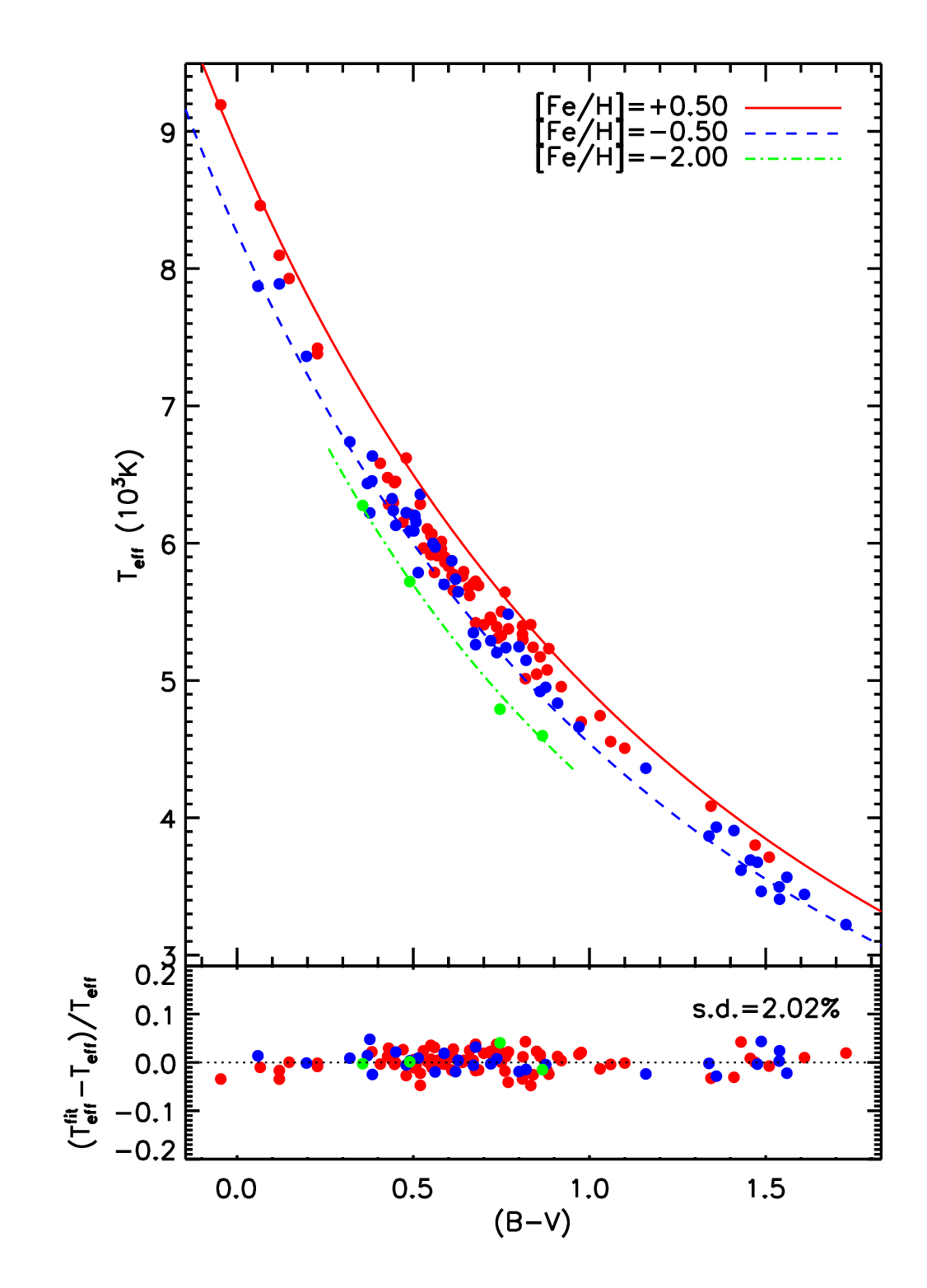}
\includegraphics[scale=0.45,angle=0]{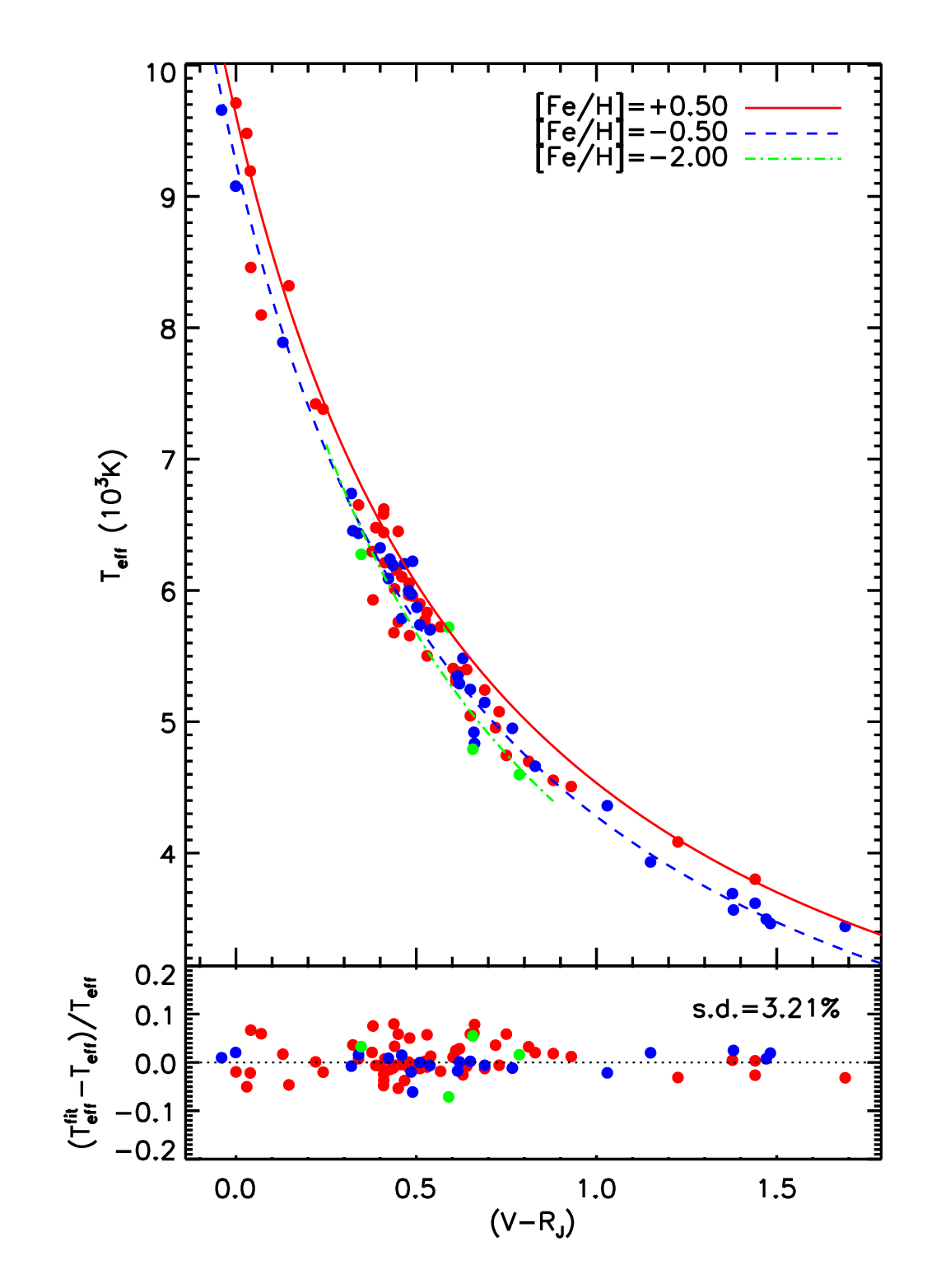}
\includegraphics[scale=0.45,angle=0]{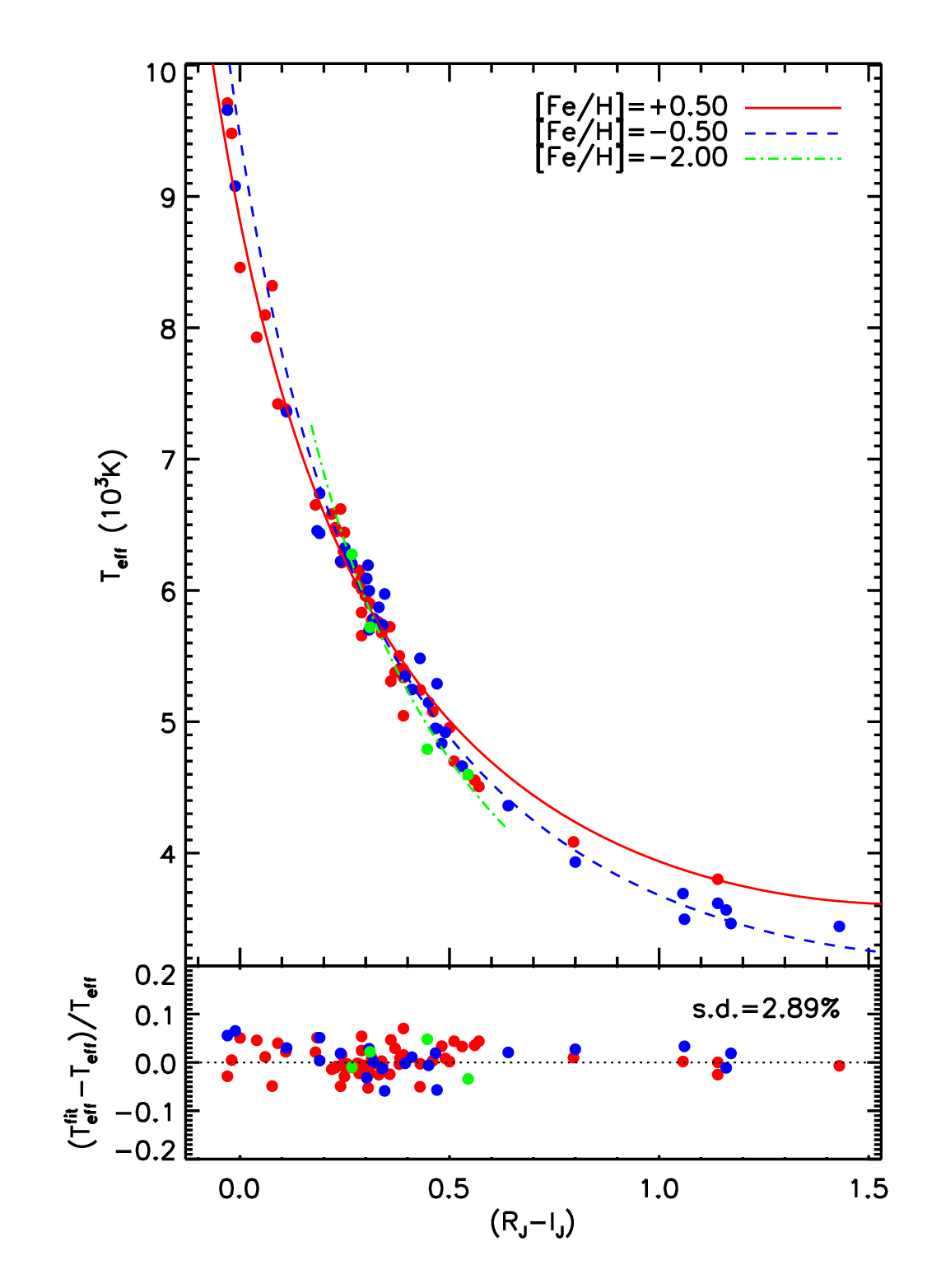}
\caption{$T_{\rm eff}$ plotted against colours $U-V$, $B-V$, $V-R_{\rm J}$ and $R_{\rm J}-I_{\rm J}$ for the dwarf sample in the metallicity bins $-0.1\,<\,{\rm [ Fe/H]}\,<\,0.5$ (red dots), $-1.0\,<\,{\rm [ Fe/H]}\,\leq\,-0.1$ (blue dots) and ${\rm [ Fe/H]} \leq -1.0$ (green dots). 
The lines represent our best fits for selected values of ${\rm [Fe/H]}$ as marked in each panel.
The lower part of each panel shows the relative residuals of the fit ($T_{\rm eff}^{\rm fit}-T_{\rm eff})/T_{\rm eff}$) as a function of colour.}
\end{figure*}

\begin{figure*}
\centering
\includegraphics[scale=0.45,angle=0]{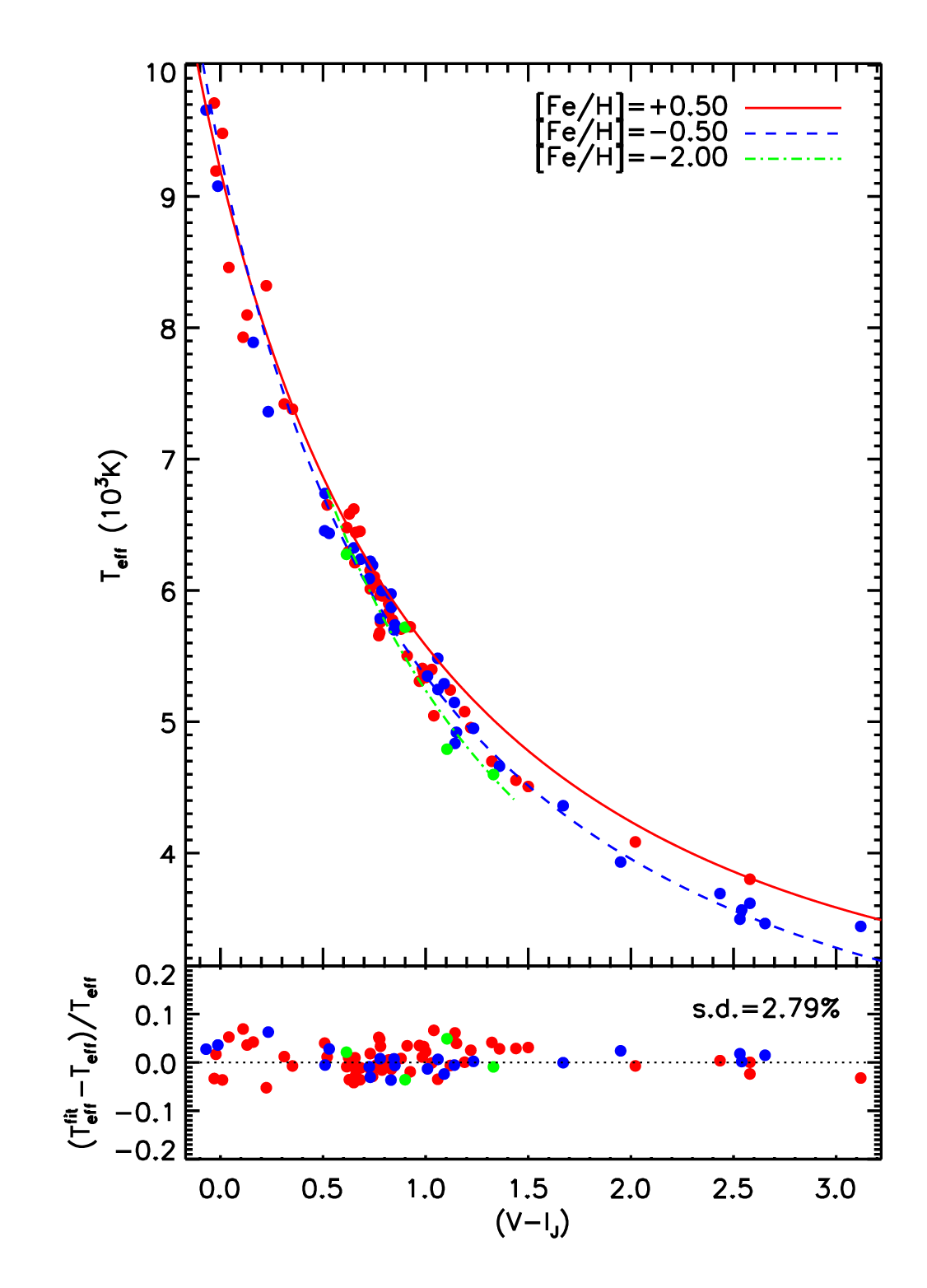}
\includegraphics[scale=0.45,angle=0]{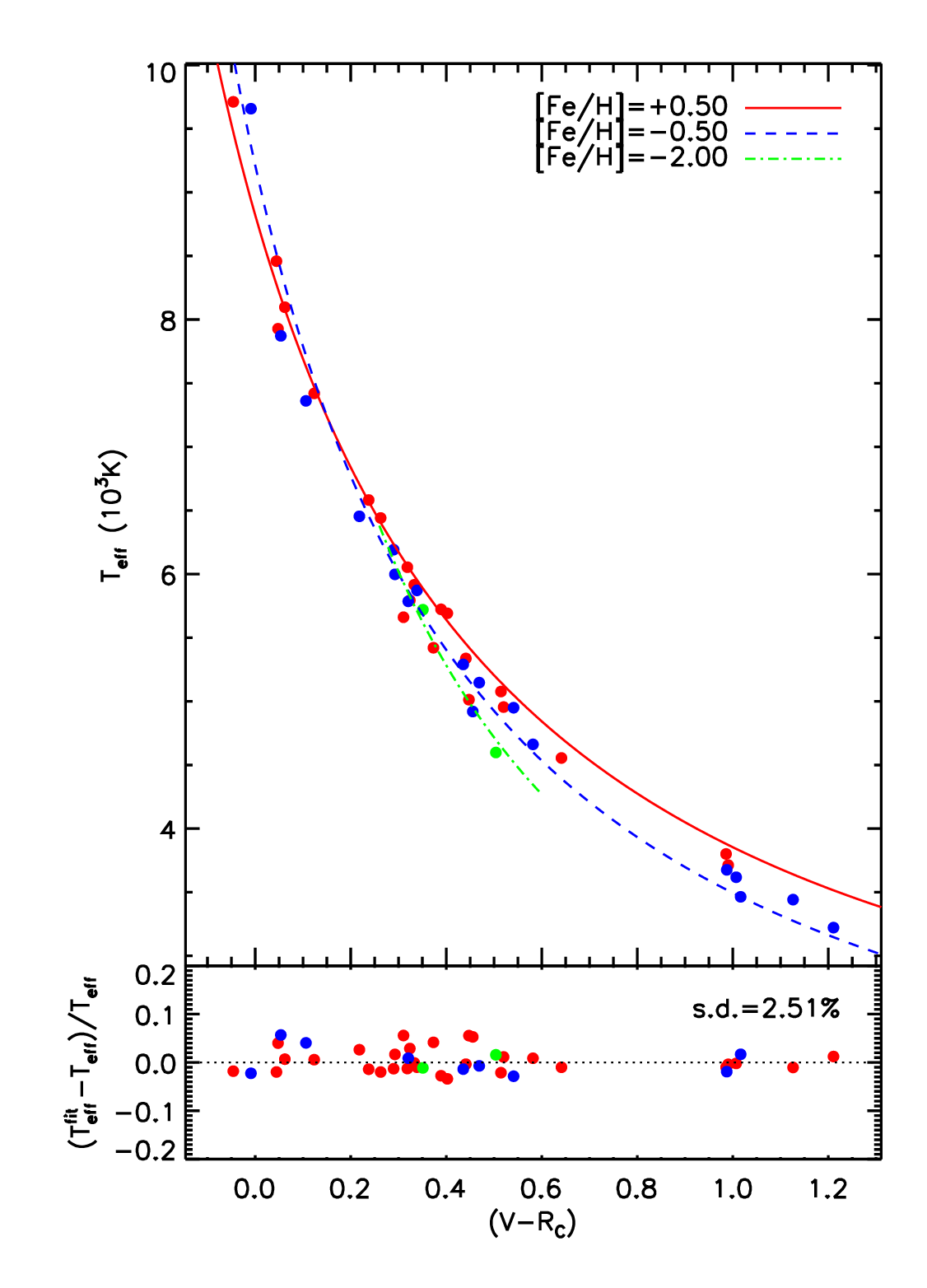}
\includegraphics[scale=0.45,angle=0]{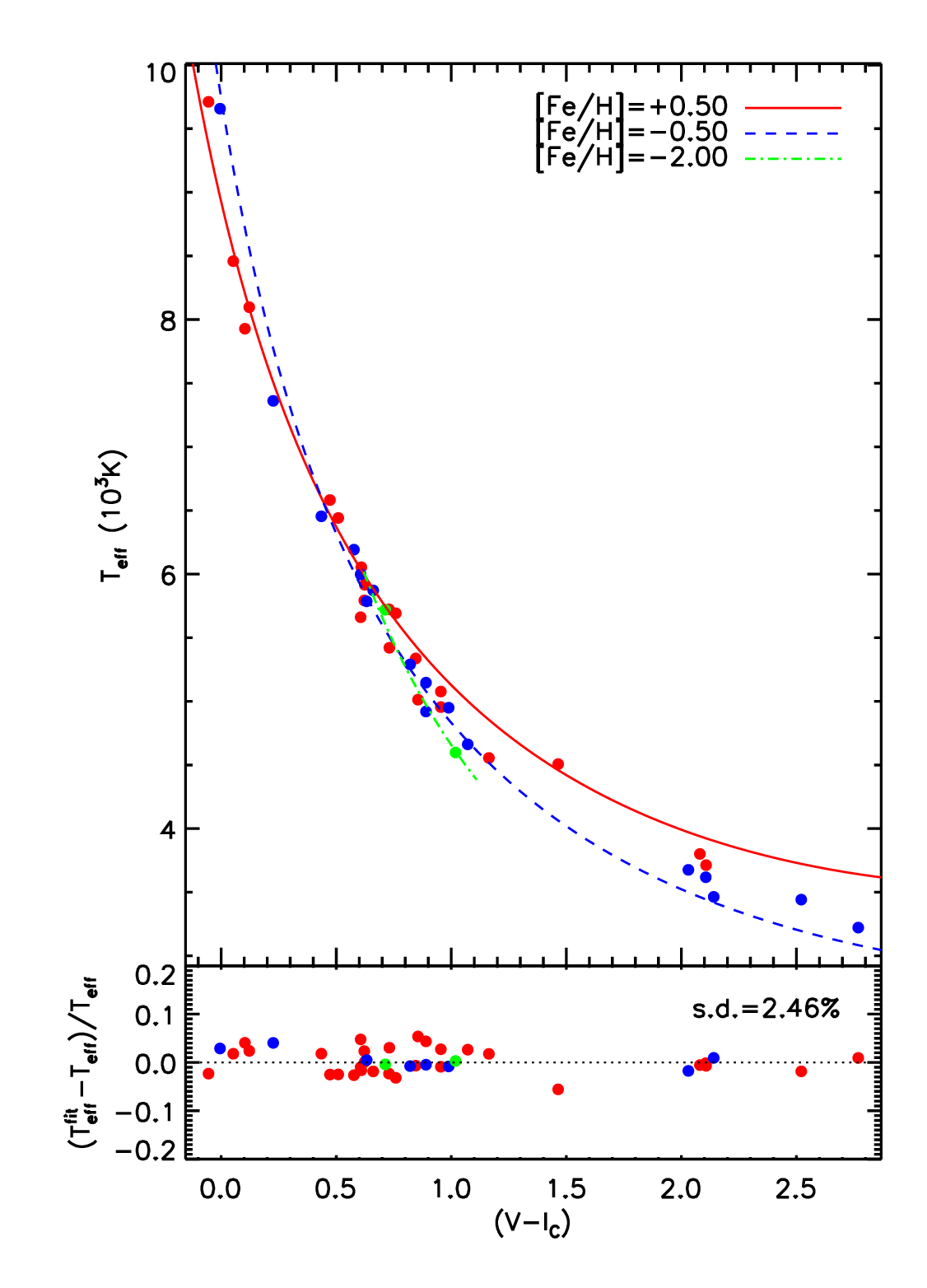}
\includegraphics[scale=0.45,angle=0]{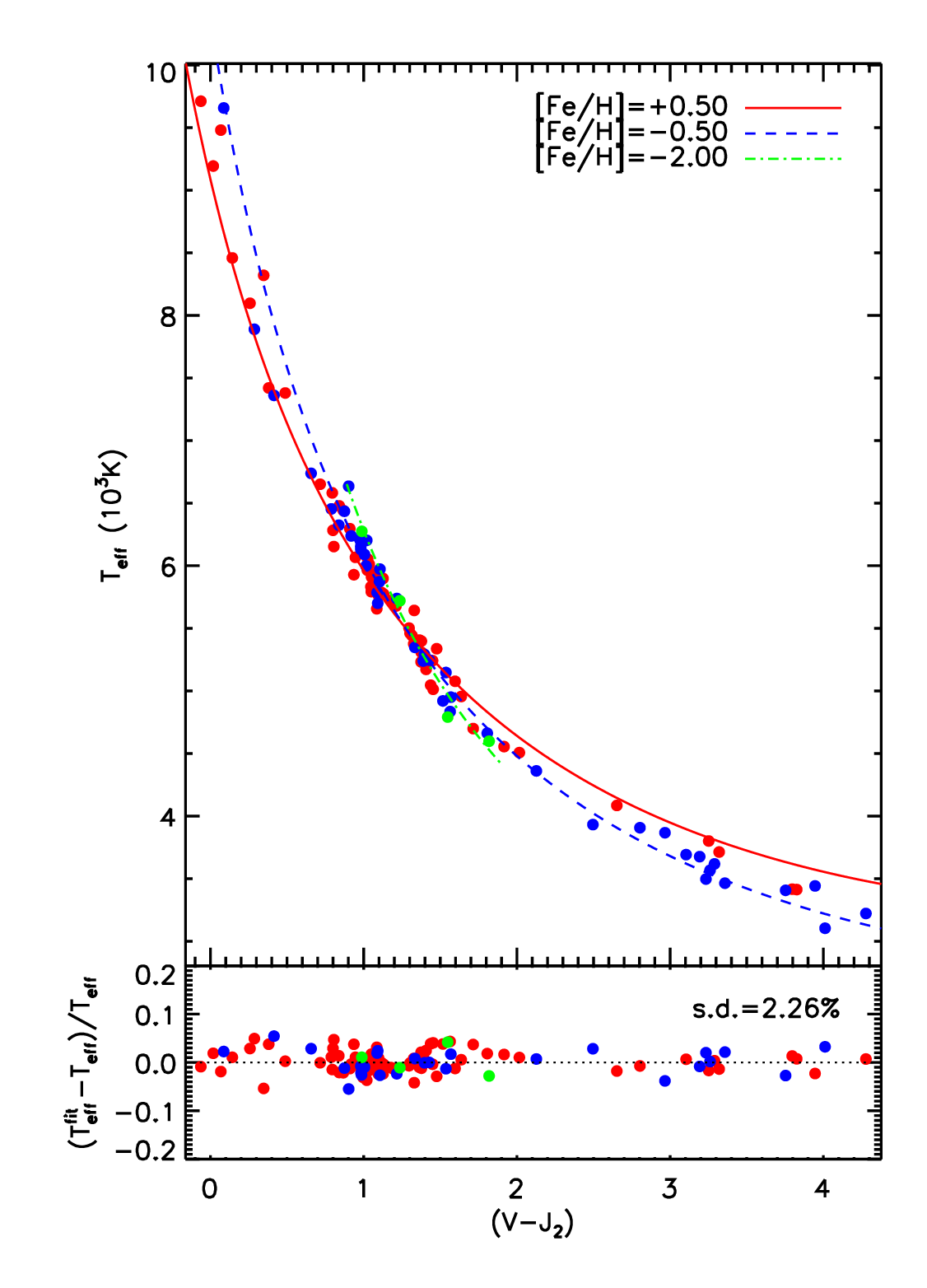}
\caption{Same as Fig. 2 colours but for $V-I_{\rm J}$, $V-R_{\rm C}$, $V-I_{\rm C}$ and $V-J_{2}$.}
\end{figure*}

\begin{figure*}
\centering
\includegraphics[scale=0.45,angle=0]{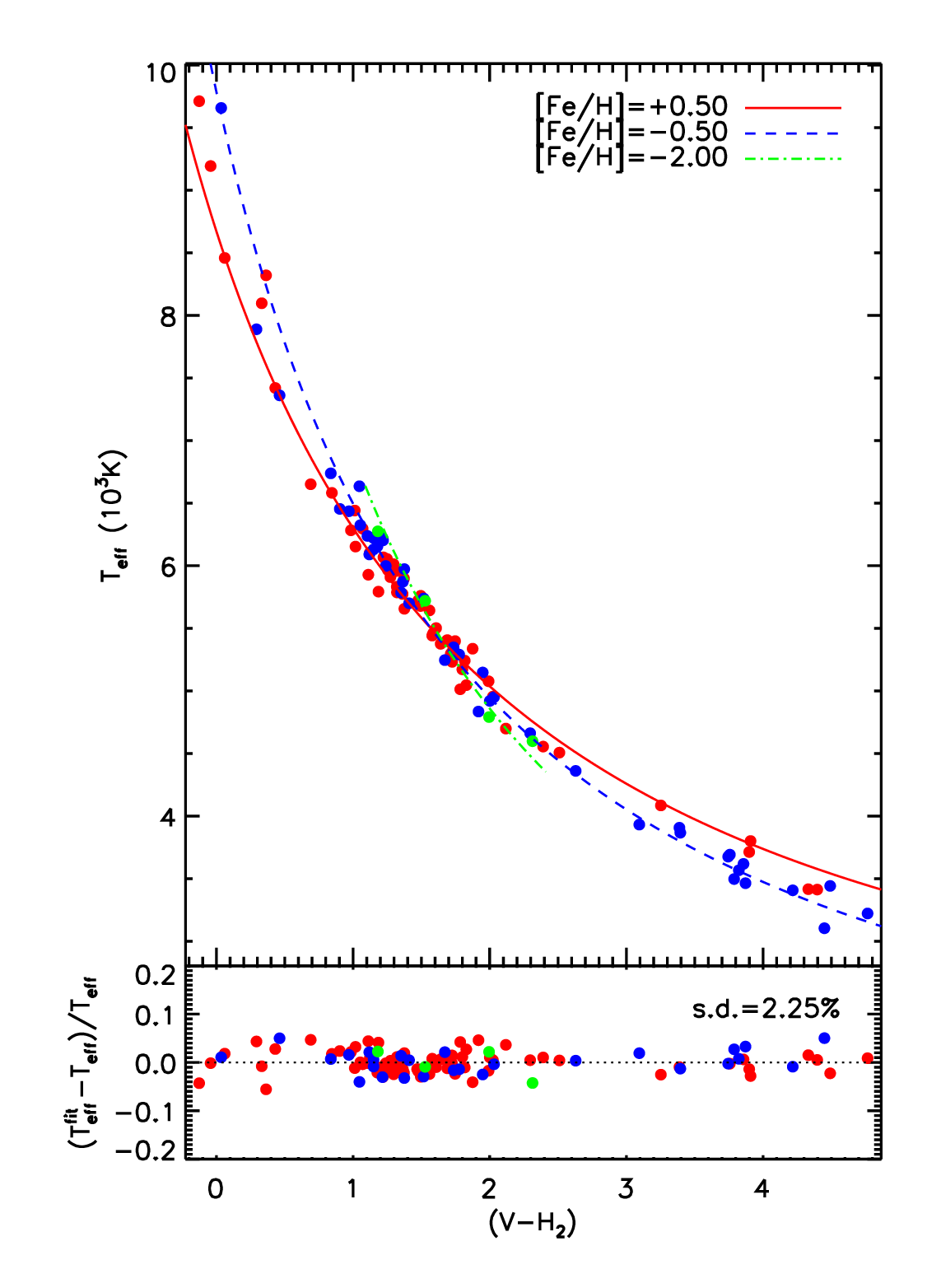}
\includegraphics[scale=0.45,angle=0]{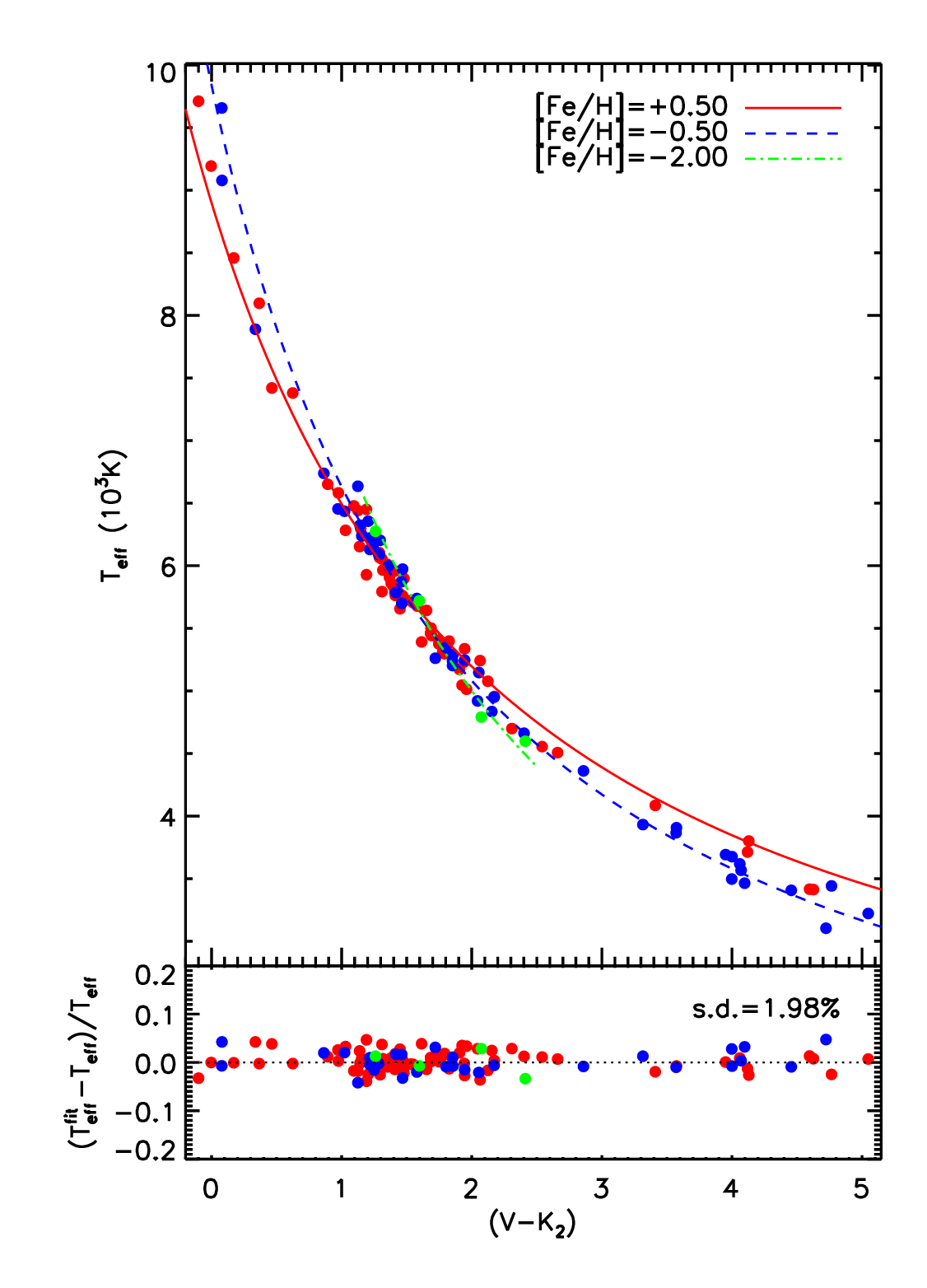}
\caption{Same as Fig. 2 colours but for  $V-H_{2}$ and $V-K_{2}$.}
\end{figure*}

\begin{figure*}
\centering
\includegraphics[scale=0.45,angle=0]{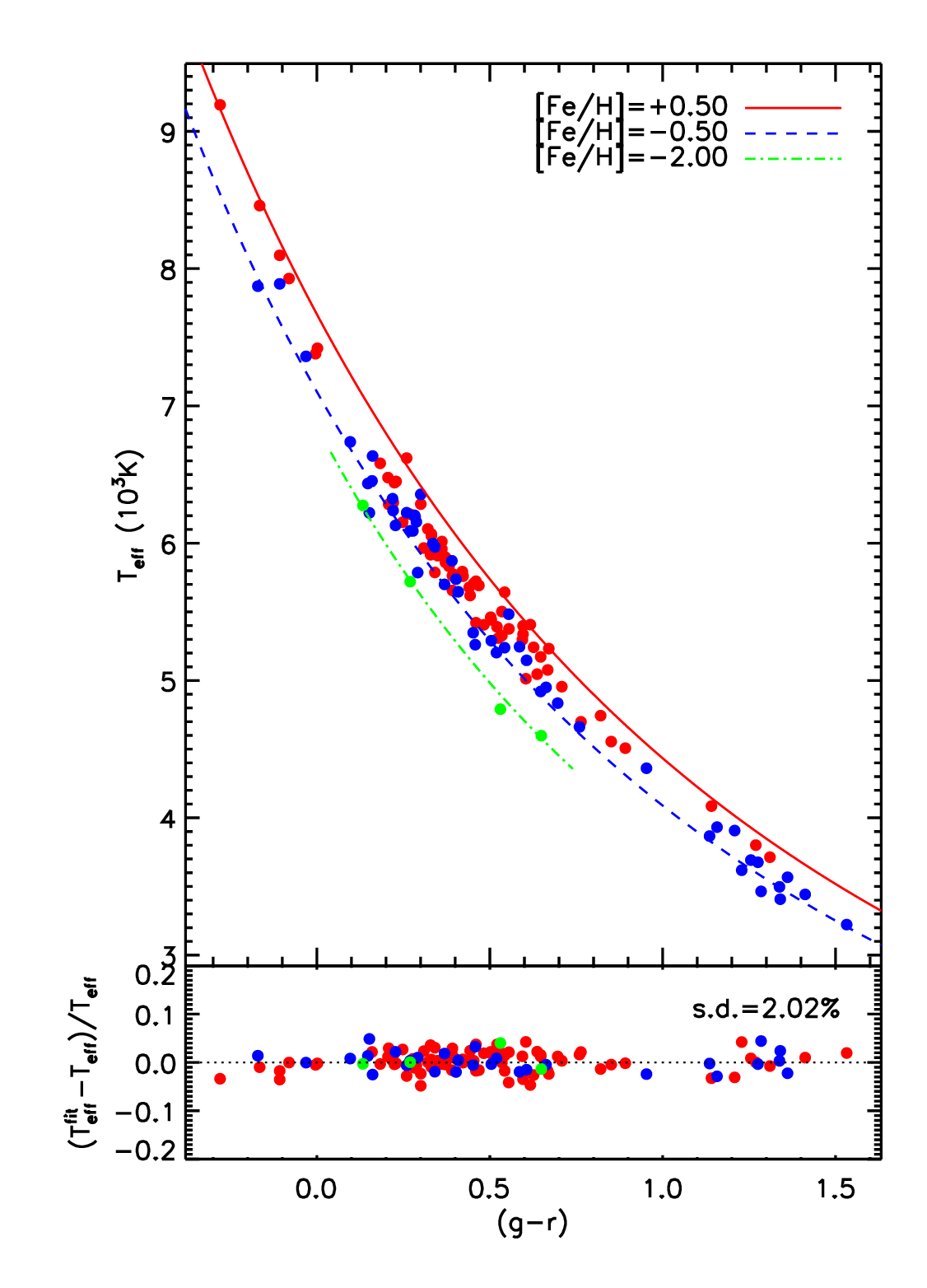}
\includegraphics[scale=0.45,angle=0]{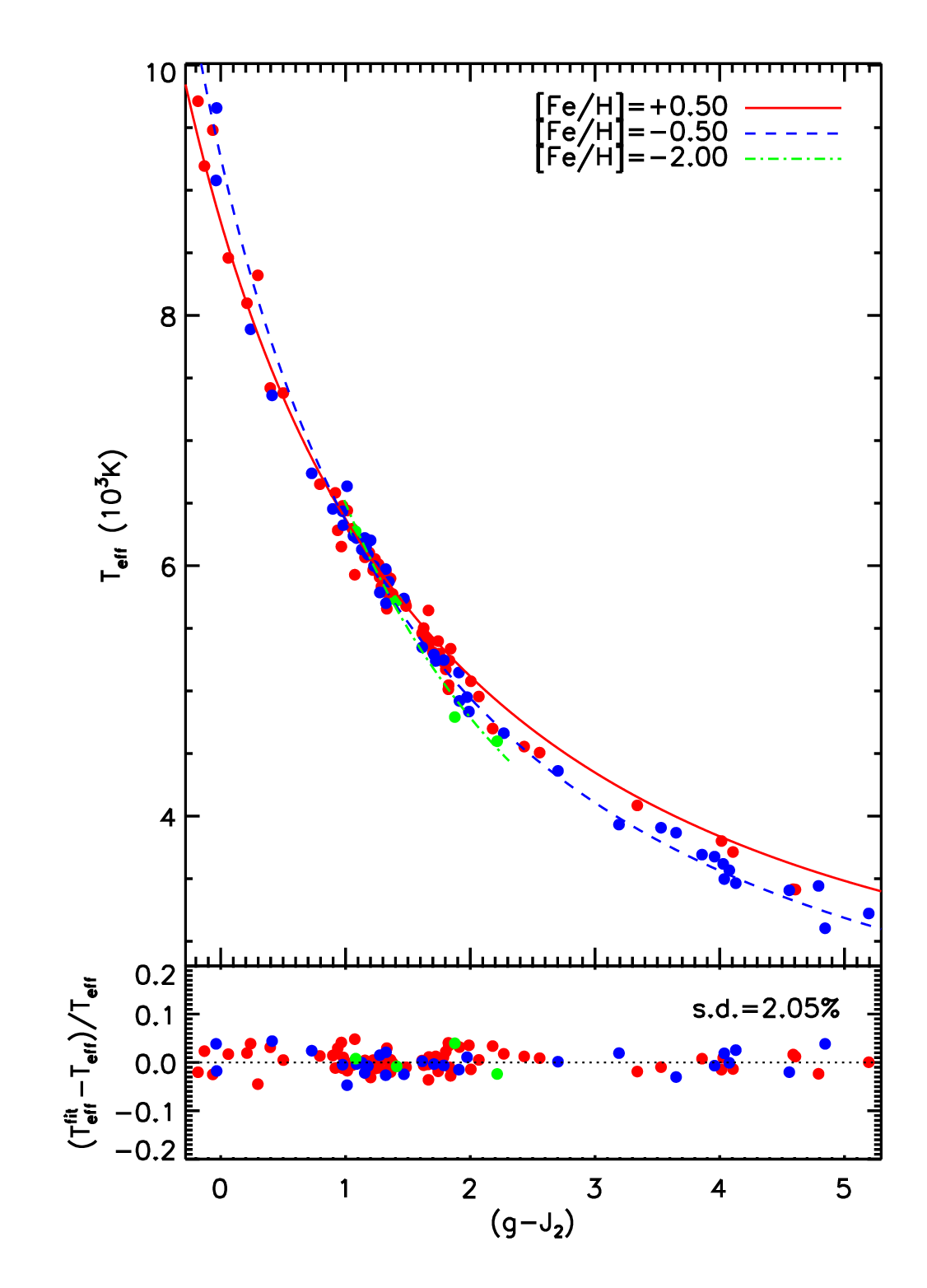}
\includegraphics[scale=0.45,angle=0]{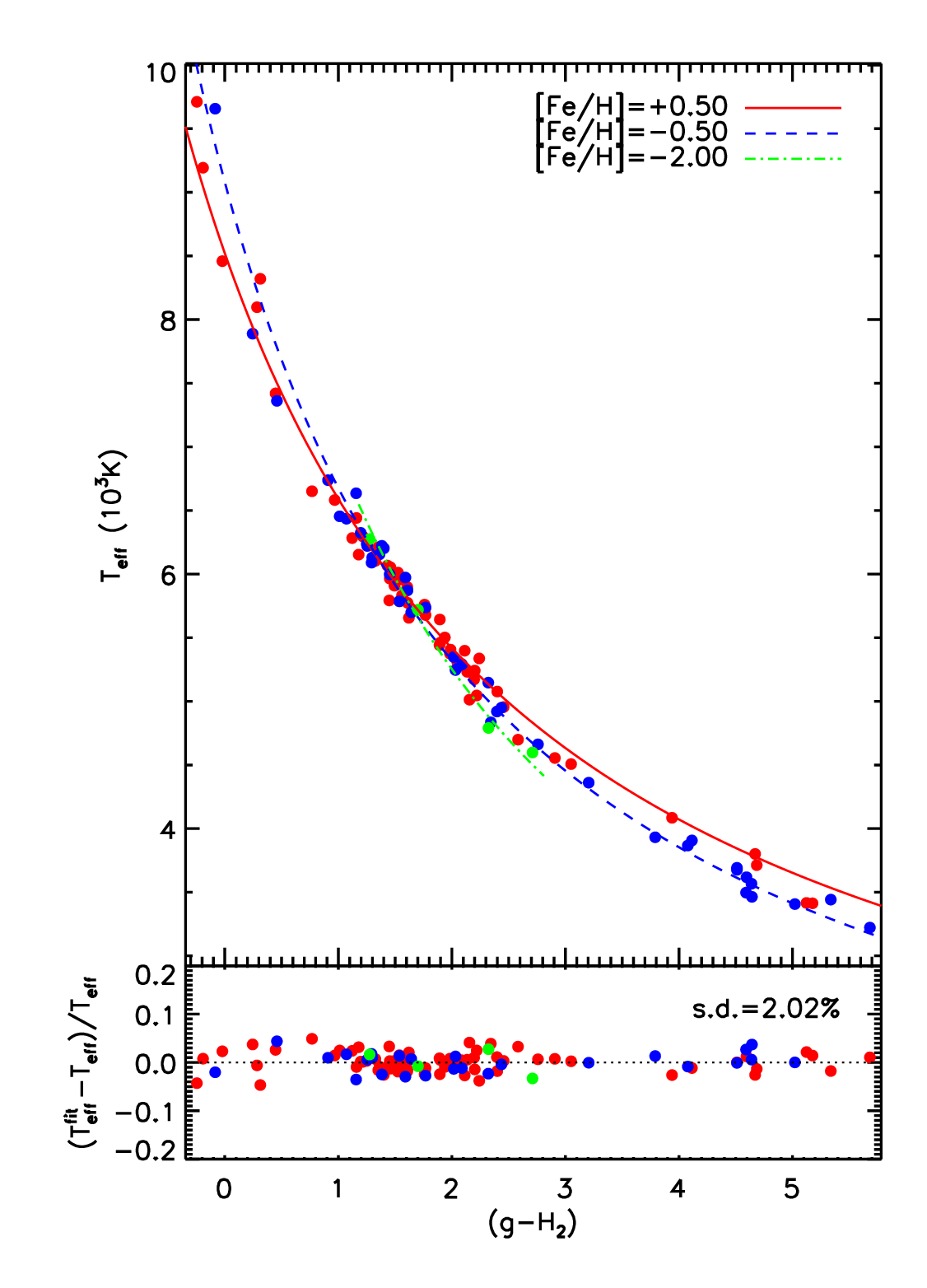}
\includegraphics[scale=0.45,angle=0]{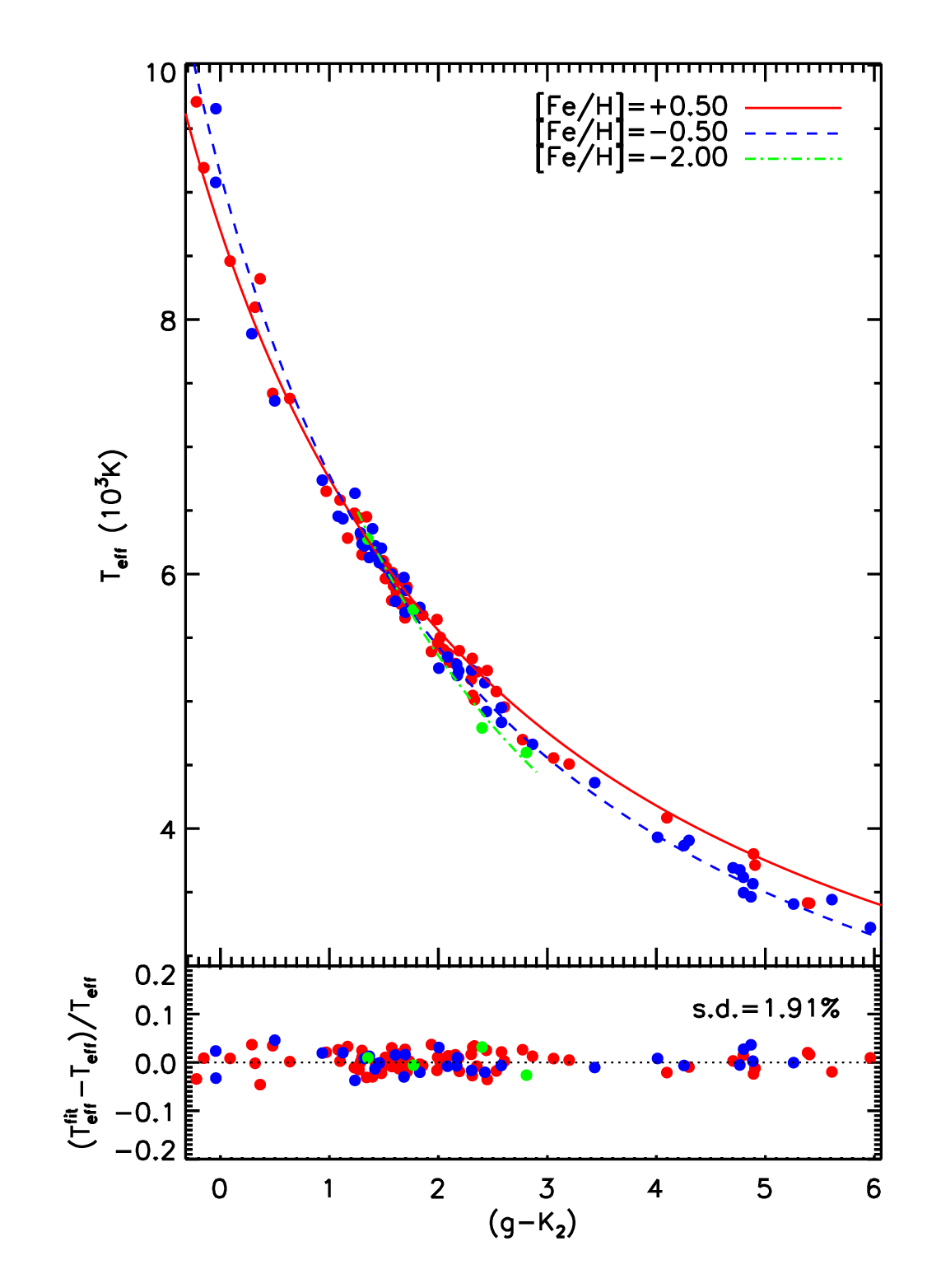}
\caption{Same as Fig. 2 colours but for $g-r$, $g-J_{2}$, $g-H_{2}$ and $g-K_{2}$.}
\end{figure*}

\begin{figure*}
\centering
\includegraphics[scale=0.45,angle=0]{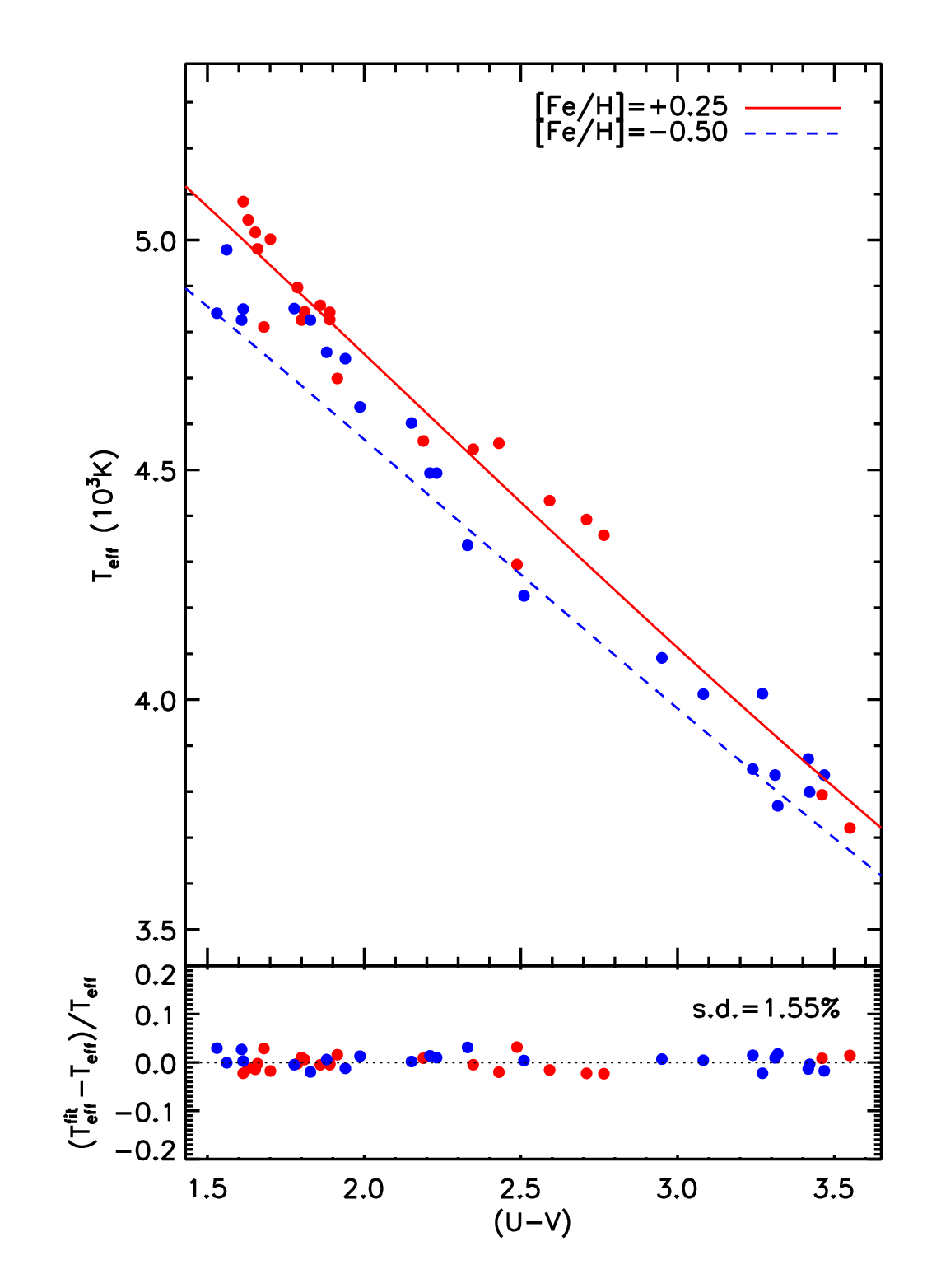}
\includegraphics[scale=0.45,angle=0]{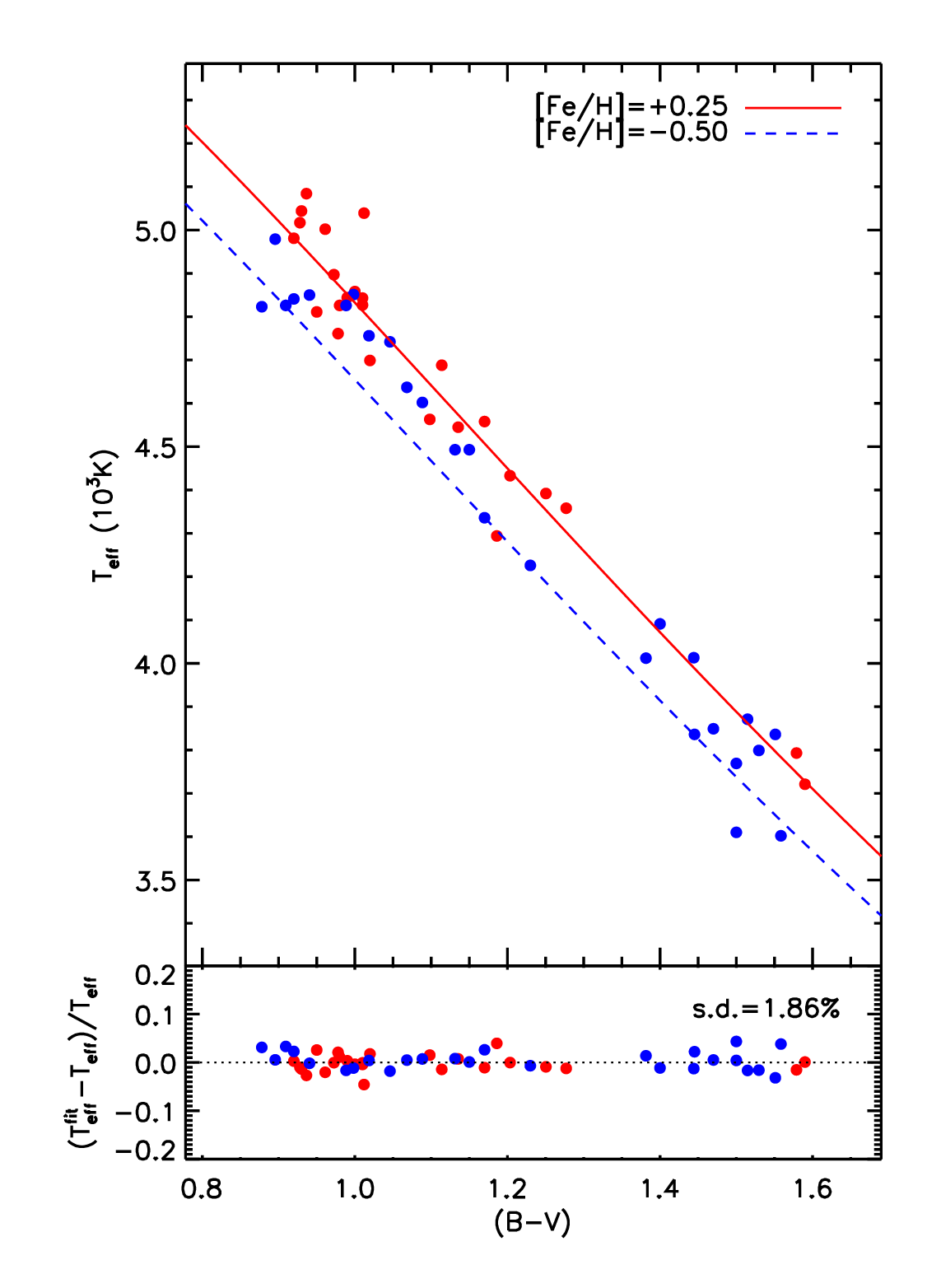}
\includegraphics[scale=0.45,angle=0]{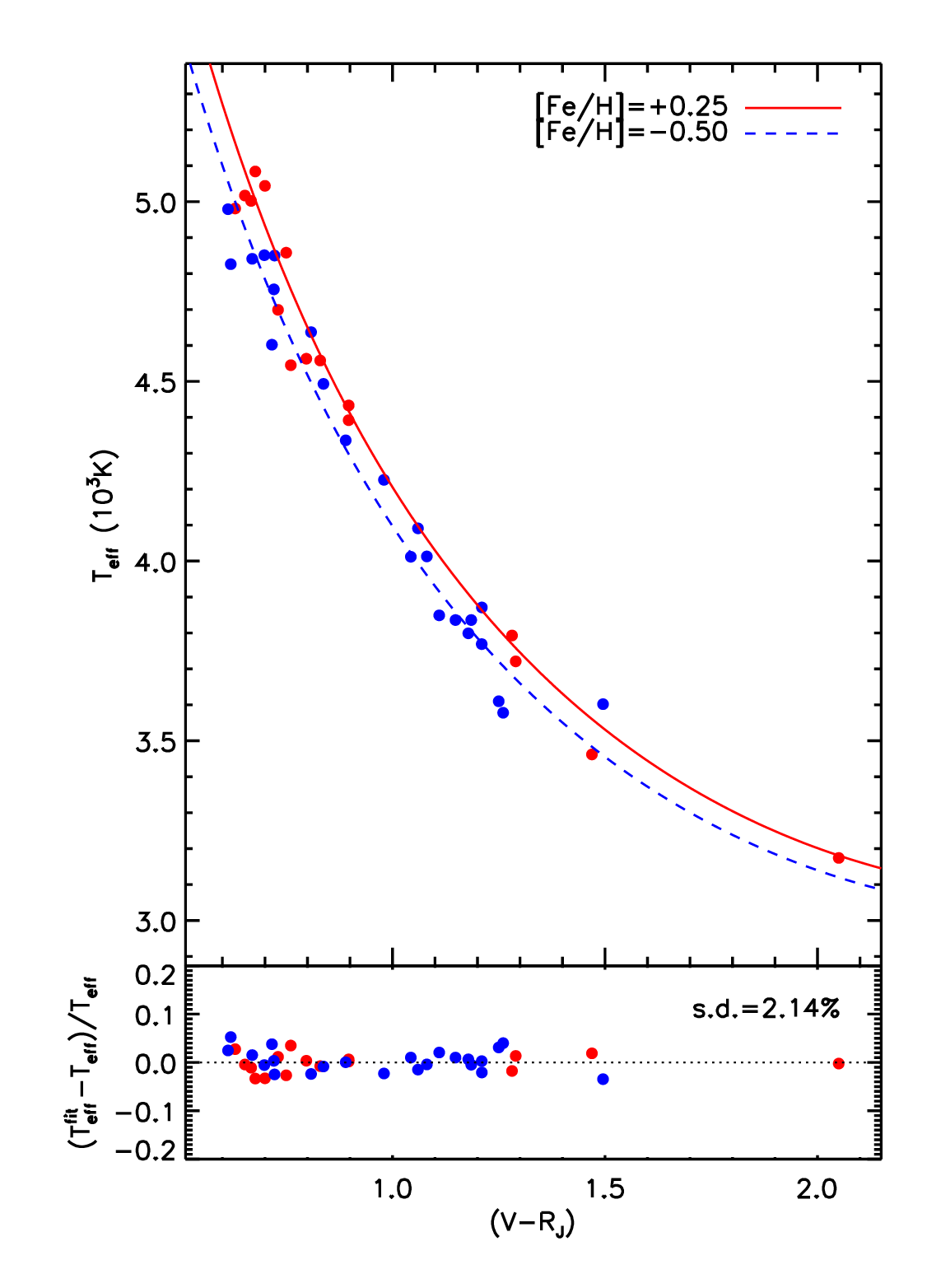}
\includegraphics[scale=0.45,angle=0]{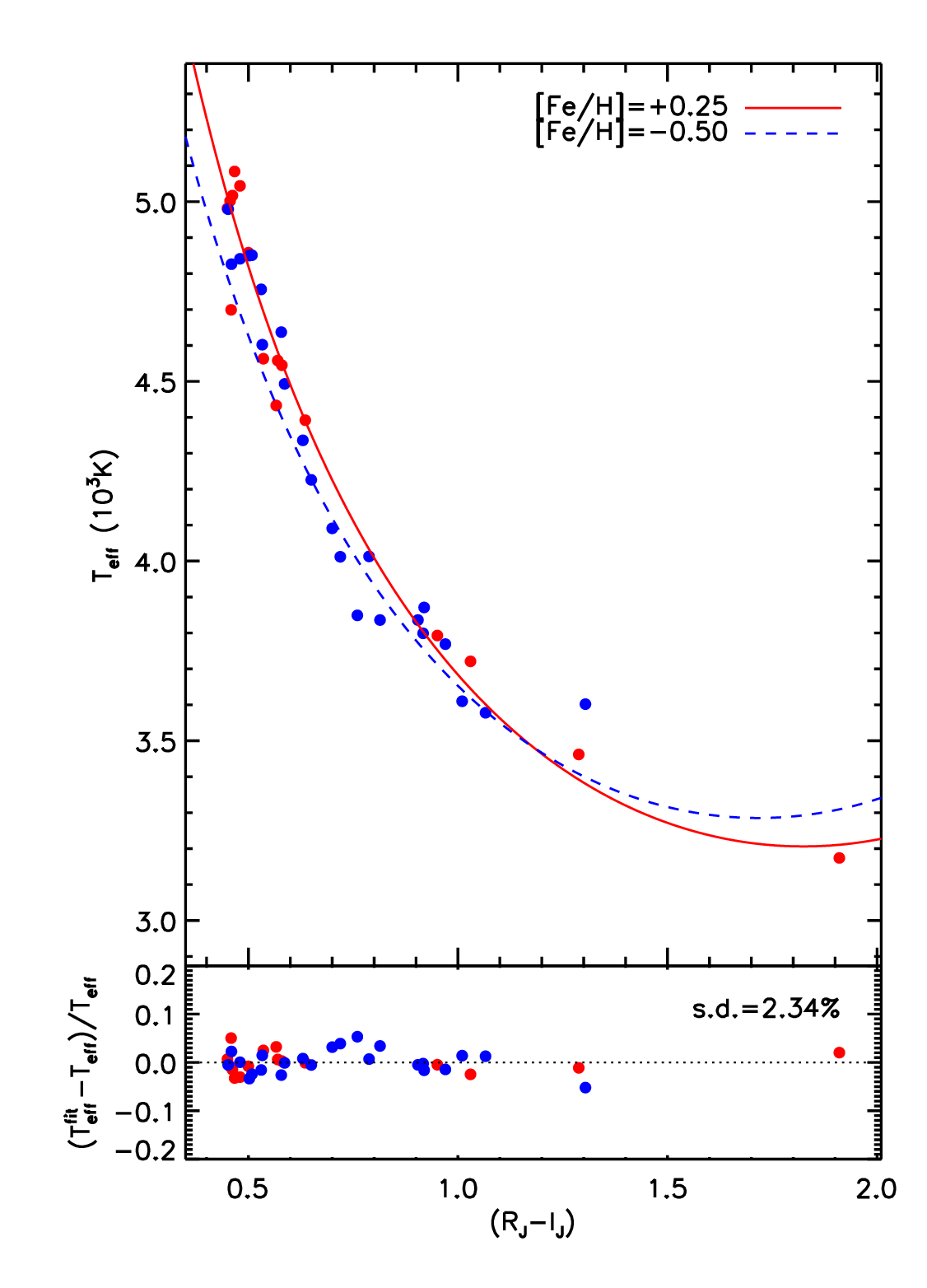}
\caption{Same as Fig.\,2 but for the sample of giants. Red and blue dots represent data points the metallicity bins $-0.1\,<\,{\rm [ Fe/H]}\,<\,0.5$ and ${\rm [ Fe/H]}\,\leq\,-0.1$, respectively.}
\end{figure*}

\begin{figure*}
\centering
\includegraphics[scale=0.45,angle=0]{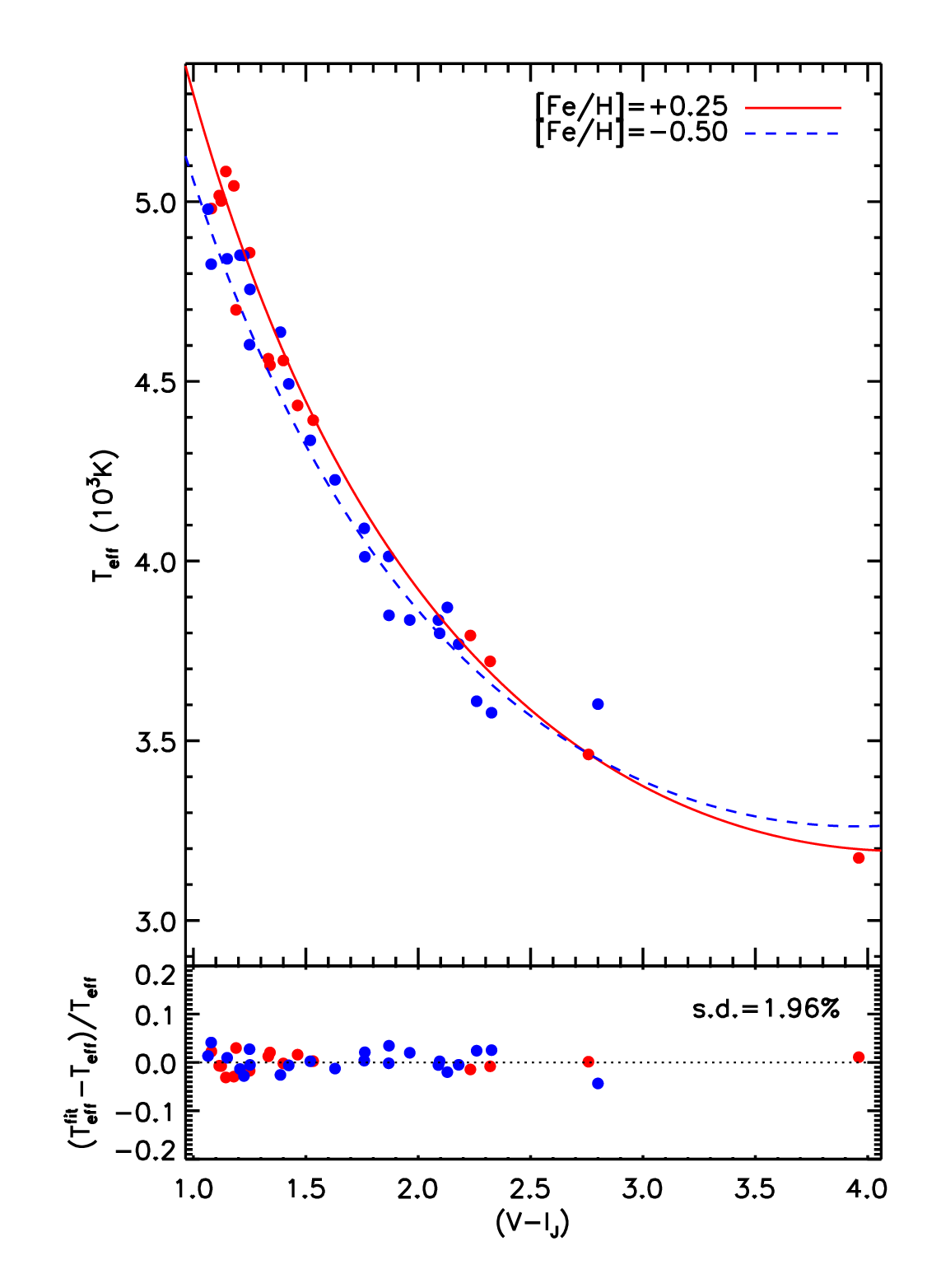}
\includegraphics[scale=0.45,angle=0]{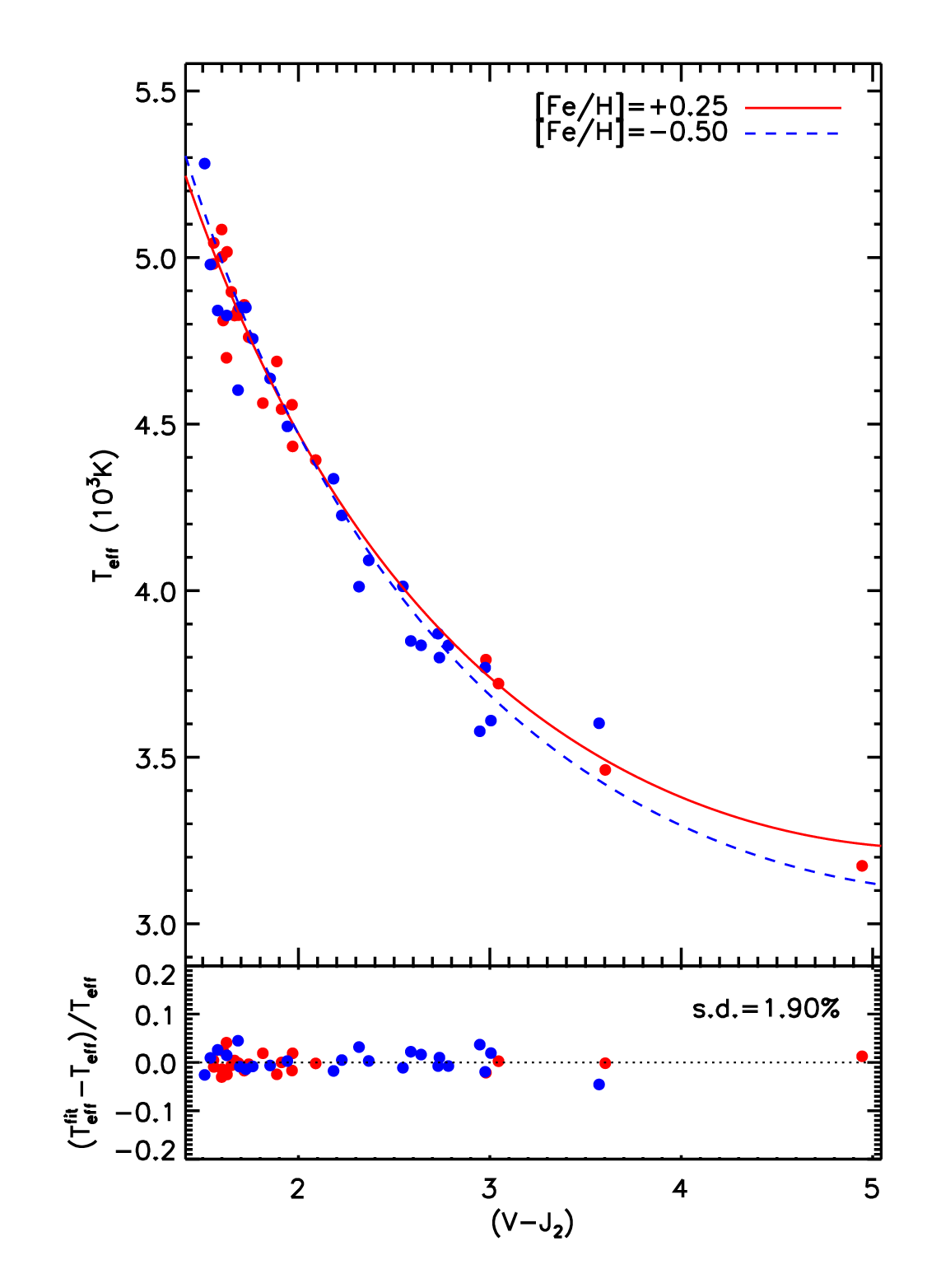}
\includegraphics[scale=0.45,angle=0]{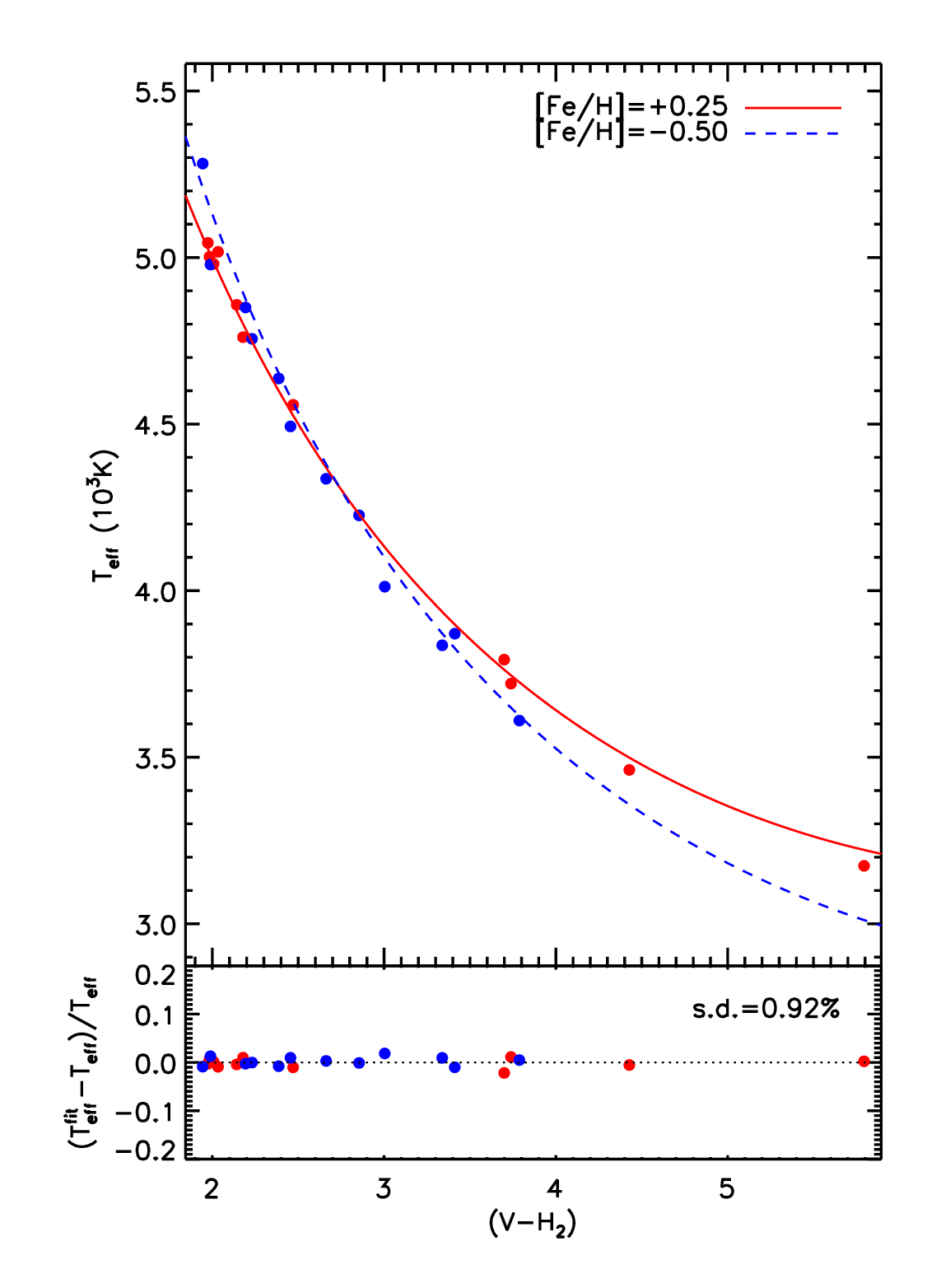}
\includegraphics[scale=0.45,angle=0]{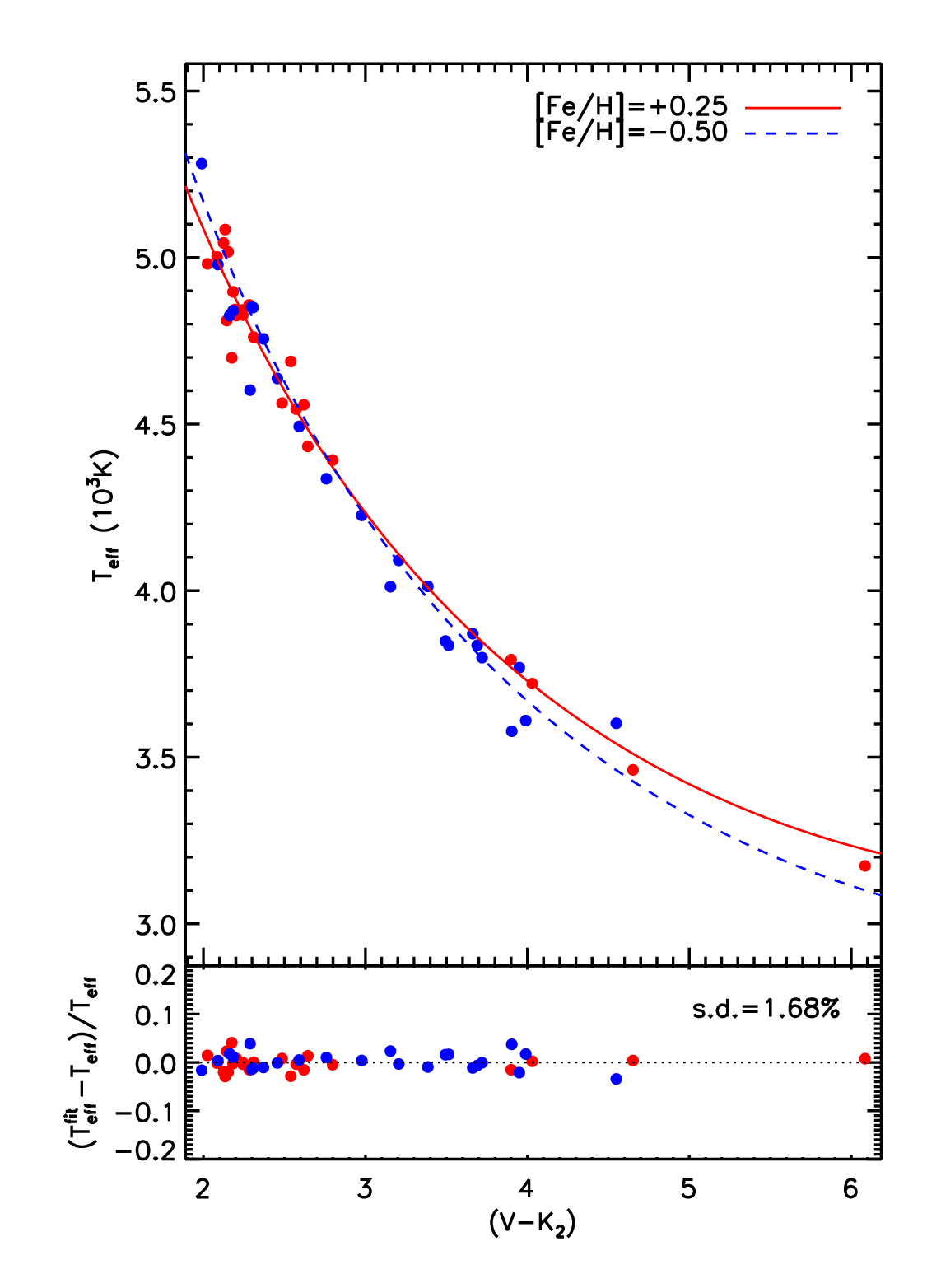}
\caption{Same as Fig.\,6, but for colours $V-I_{\rm J}$, $V-J_{2}$, $V-H_{2}$ and $V-K_{2}$.}
\end{figure*}

\begin{figure*}
\centering
\includegraphics[scale=0.45,angle=0]{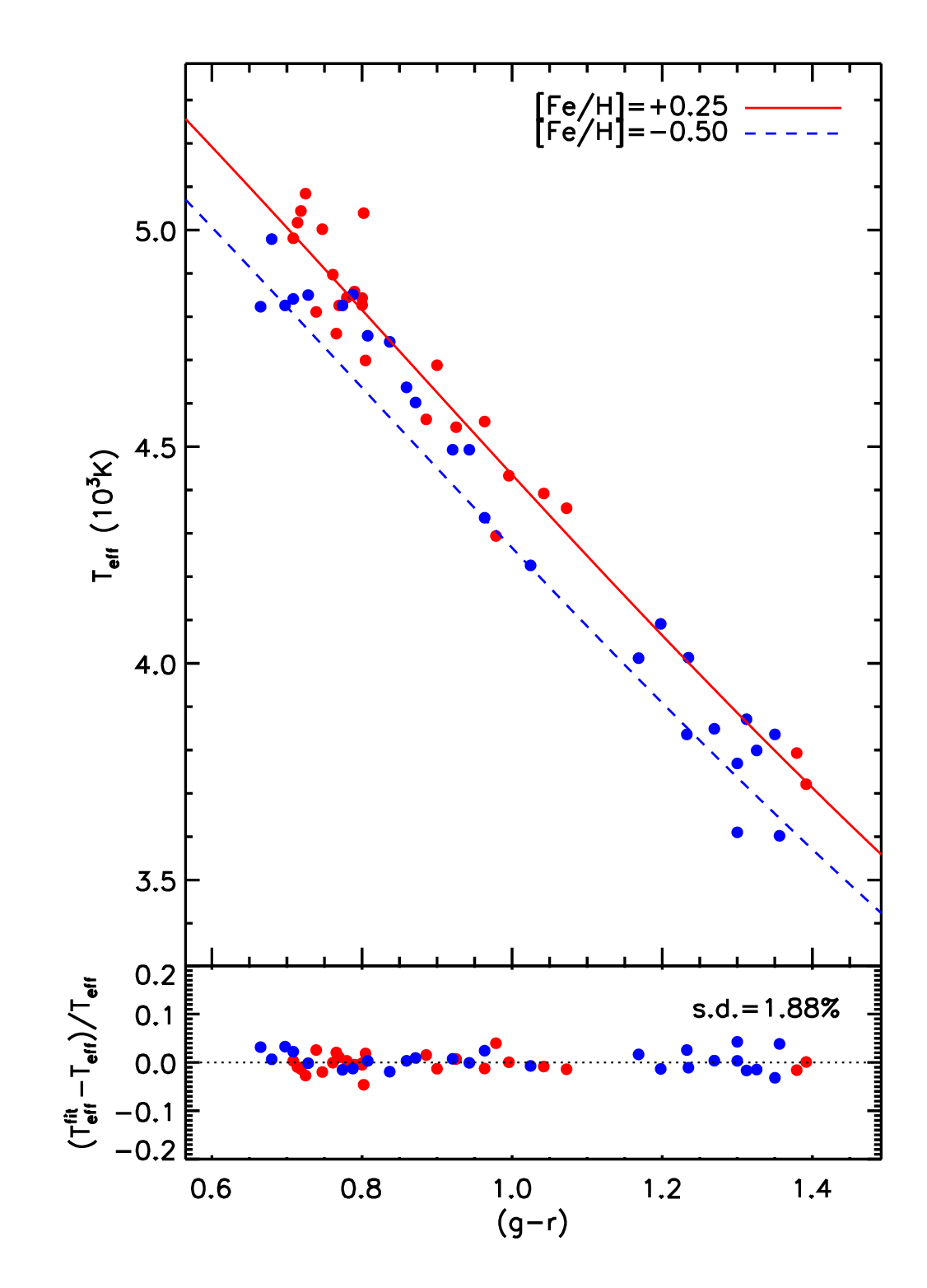}
\includegraphics[scale=0.45,angle=0]{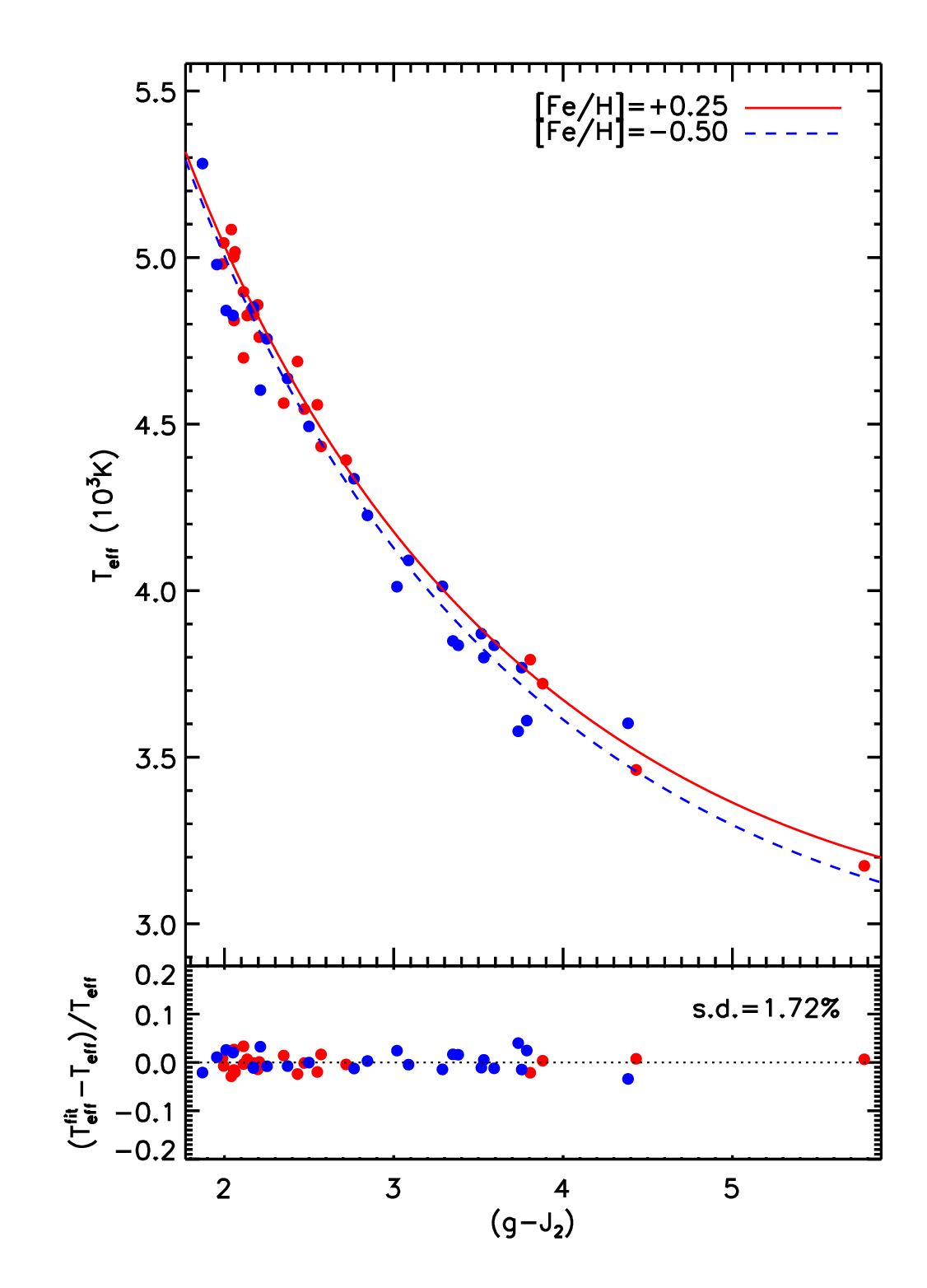}
\includegraphics[scale=0.45,angle=0]{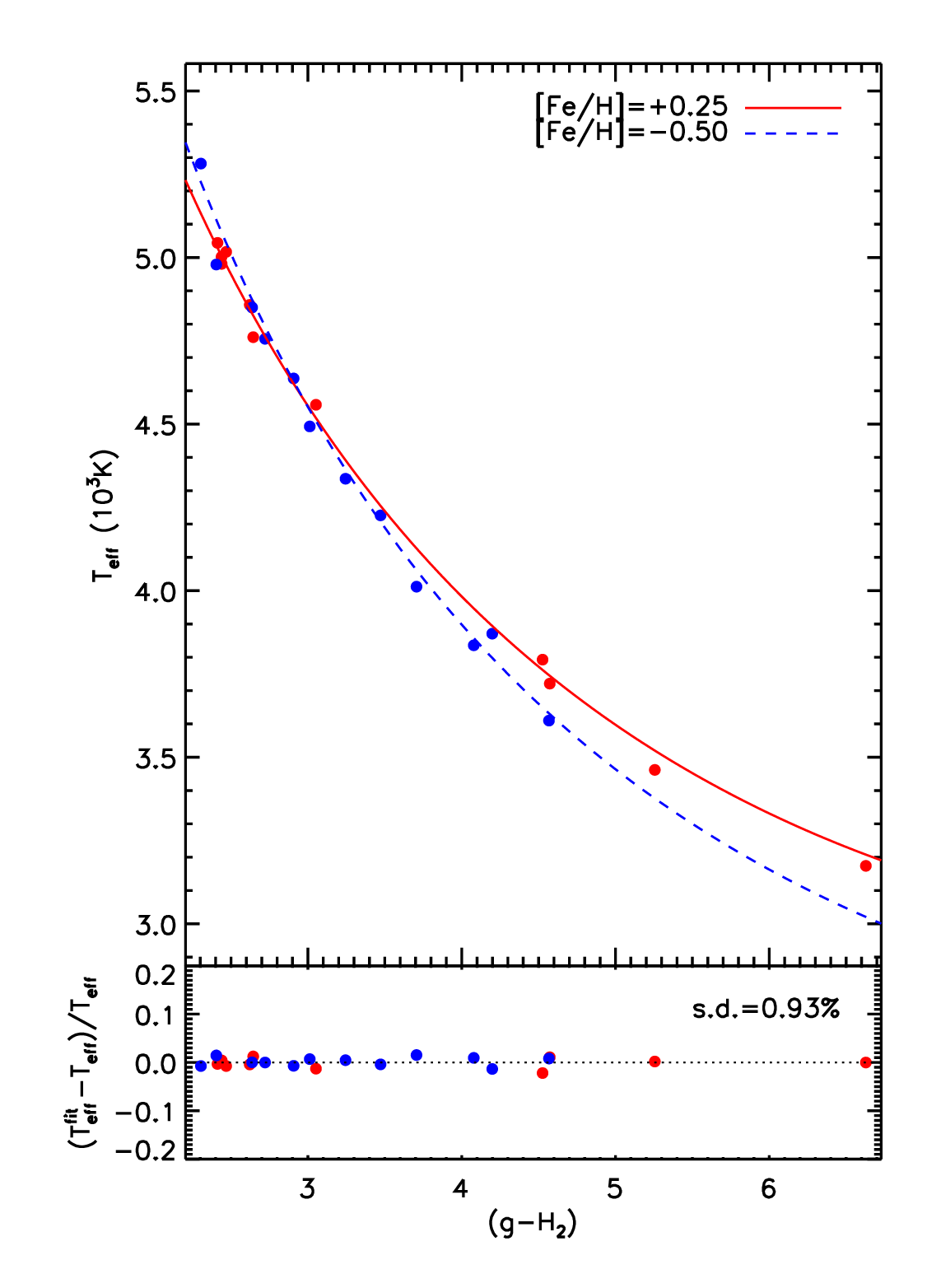}
\includegraphics[scale=0.45,angle=0]{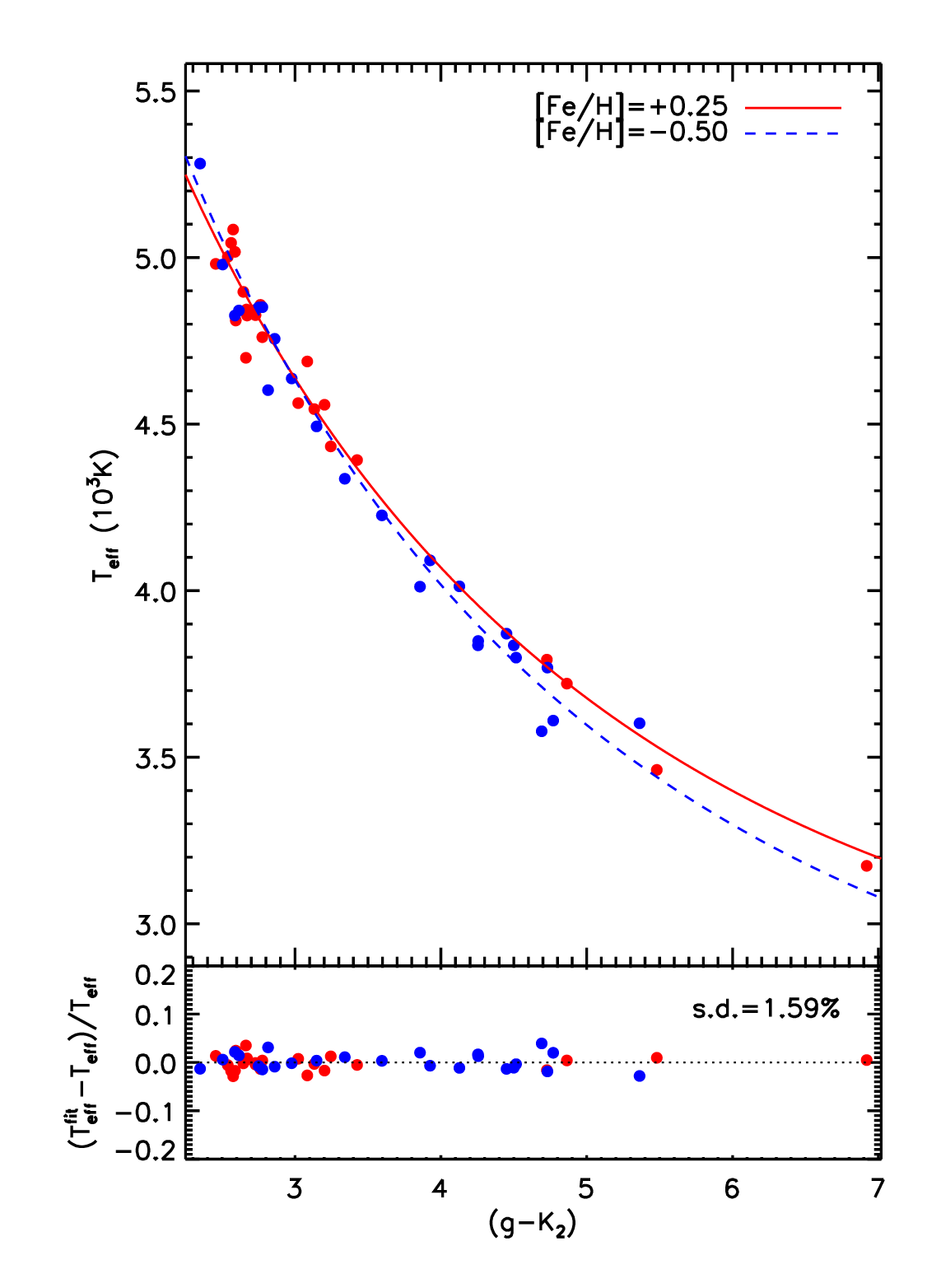}
\caption{Same as Fig.\,6, but for colours $g-r$, $g-J_{2}$, $g-H_{2}$ and $g-K_{2}$.}
\end{figure*}

The polynomial fits to the empirical relations are shown for dwarf stars in Figs. 2--5 and giants in Figs. 6--8.
The fit coefficients ($a_{i}$) for eighteen colours are given in Table 3 for dwarfs and Table 4 for giants, along with the standard deviation (s.d.), of the percentage errors of the fit, i.e. ($T_{\rm eff}^{\rm fit}-T_{\rm eff})/T_{\rm eff}\,\times\,100$, the number of stars used for the final fit (after the 2.5$\sigma$ clipping) $N$, the applicability ranges of the fit in colour and [Fe/H], and finally the spectral types corresponding to the colour range.
The fits for dwarf stars are performed for the full range of AFGKM stars.
However, as indicated by some recent studies (Monnier et al. 2007; van Belle 2012; B13), there may exist biases in the determinations of $T_{\rm eff}$ by interferometric imaging for stars earlier than mid-F due to their  rapid rotation.
Following B13, we have thus provided another set of fits, excluding stars hotter than 6750\,K.
The results are also given in Table 3 with the corresponding range of spectral type marked as FGKM.
For all the twenty-one colours, the maximum differences between the temperature given by the $T_{\rm eff}$--colour relations\footnote{For a metallicity [Fe/H] of $-0.1$, the typical value of the dwarf sample.} with and without early-type stars are all within a few percent (0.6--1.5 per cent). 
As mentioned earlier, the number of giant stars with the Cousins $R_{\rm C}I_{\rm C}$ magnitudes available is too small ($\sim$\,10) to fit the empirical  metallicity-dependent $T_{\rm eff}$--colour relations.
Thus we only consider the  Cousins $R_{\rm C}I_{\rm C}$ system in the calibrations for dwarfs.
The standard deviations of the percentage residuals of the fit are about 2.0 and 1.5 per cent  for dwarfs and giants, respectively.

Finally, we note that our sample includes four dwarfs (nicely sample FGK star covering a wide range of $T_{\rm eff}$ from $\sim$ 4,600 to 6,300 K) of metallicities [Fe/H] $\sim -2.0$, which provide some constraints on the relations at such low metallicities. 
Applying the relations to such low metallicities should be however treated with caution.

\subsection{Metallicity effects}

The intrinsic colours of a star are not only governed by $T_{\rm eff}$ but also by metallicity due to the line blanketing effects (e.g. Ram\'irez and Mel\'endez 2005, hereafter RM05).
As shown in Figs. 2--5, the effects of metallicity on the $T_{\rm eff}$ versus colour relations for dwarf stars show two distinct features.
Firstly, colours involving bands in the UV and optical (e.g.\,$U-V$,\,$B-V$,\,$V-R_{\rm J}/R_{\rm C}$,\,$g-r$) always get  redder (larger) with increasing metallicity at a given $T_{\rm eff}$.
On the other hand, for colours between a visual and a near-infrared band (e.g.\,$V-\,J/H/K$,\,$V-\,J_{2}/H_{2}/K_{2}$,\,$g-\,J/H/K$,\,$g-\,J_{2}/H_{2}/K_{2}$), while they get redder (larger) with increasing metallicity for cool stars (e.g.\,$T_{\rm eff}\,<\,6000$\,K) for a given effective temperature, they actually become bluer (smaller) for hot stars (e.g.\,$T_{\rm eff}\,>\,6000$\,K).
These behaviors are similar to those found by calibrations based on the IRFM $T_{\rm eff}$ scale and  have been explained in detail by RM05 using synthetic spectra.


Colour ($V-K$) (here $K$ can be either of the Johnson or the 2MASS system; also as shown in Figs.\,5 and 8, $g$ is very similar to $V$) is often considered the best $T_{\rm eff}$ indicator (e.g. Blackwell et al. 1990; Alonso et al. 1996, 1999; RM05), owning to its relatively weak dependence on metallicity and luminosity (see the next Section).
To investigate how well the colours between a visual (i.e. $V$ or $g$) and an infrared (i.e. $K$ or $K_{2}$) band  serve as a $T_{\rm eff}$ indicator without considering the metallicity effects, we explore in Fig.\,9 the relative errors that would be introduced in $T_{\rm eff}$ deduced from colour $(V - K_2)$ or $(g - K_2)$ for a dwarf of solar metallicity, if an incorrect metallicity has been assumed. 
The plot shows that if one can accept a  $T_{\rm eff}$ error of 3.0 per cent, then ($V-K_{2}$) or ($g-K_{2}$) is good enough even simply assuming solar metallicity without considering the metallicity effects, for disk stars ([Fe/H] $> -0.8$) of $T_{\rm eff}$ between  4,500 and 6,900 K in the case of $(V - K_2)$ and between 4,650 to 7,950 K in the case of $(g - K_2)$.
If one extrapolates and applies  the current calibrations for  metal-poor halo stars ([Fe/H] $\sim -2.0$), the corresponding usable range will be narrowed down to 5,000 -- 6,000 K and to 5,300 -- 6,500 K, for colours $(V - K_2)$ and $(g - K_2)$, respectively.
The effects of metallicity on colours are shown in the lower panel of Fig.\,9.
On the whole, the changes in colours due to the metallicity effects are similar to those found for effective temperature.
For the temperature range 5,000 $< T_{\rm eff} <$ 7,000\,K, the effects on colours ($V-K_{2}$) and ($g-K_{2}$) for disk-like dwarfs ([Fe/H] $> -0.8$) are smaller than $0.07$\,mag, comparable to the typical colour measurement uncertainties (assuming a  photometric error of 0.05 mag in each band).
For halo dwarfs ([Fe/H]\,$\sim\,-2.0$), the effects are bigger as one expects.
In summary, we conclude that colour  ($V-K_{2}$) or ($g-K_{2}$) can be a robust, metallicity-independent $T_{\rm eff}$ estimator for disk dwarfs ([Fe/H] $> -0.8$) of temperatures 5,000 $\lesssim T_{\rm eff} \lesssim$ 7,000\,K.
However, for halo dwarfs ([Fe/H]\,$\sim\,-2.0$), those two colours only work well for a restricted temperature range, 5,000 $\lesssim T_{\rm eff} \lesssim$ 6,000\,K,  if one ignores the metallicity effects.

For giants, the metallicity effects are insignificant either for ($V-K_{2}$) or ($g-K_{2}$), owning to the relative narrow range of metallicity and effective temperature covered by the current sample.
According to the results of RM05 based on the IRFM $T_{\rm eff}$ scale, the effects are similar to those of dwarf stars.
 
\subsection{Luminosity (surface gravity) effects}
The effects of luminosity (i.e. the surface gravity log $g$) on effective temperature and colours are illustrated in Fig.\,10. 
Colours ($V-K_{2}$) and ($g-K_{2}$) are insensitive to the luminosity effects in estimating $T_{\rm eff}$ for the whole parameter range of giants calibrated here if one accepts a 3.0 per cent uncertainty of determination. 
This is particularly true for colour ($g-K_{2}$), for which the luminosity effects are smaller or comparable to typical photometric uncertainties for stars in the temperature range $3,650 < T_{\rm eff} < 5,200$ K (the upper temperature limit for the calibration of giants).
 
We conclude that colour ($g-K_{2}$) can serve as an excellent effective temperature indicator for the currently on-going large scale stellar spectroscopic surveys, not only because of its weak dependence on metallicity and luminosity but also because it is easily available from high precision photometric surveys such as the SDSS, XSTPS-GAC, PanSTARRS and 2MASS.

\section{Comparisons with other temperature scales}

In this section, the direct $T_{\rm eff}$ scale from the interferometry  is compared to other $T_{\rm eff}$ scales from a variety of techniques,  including that based on the IRFM (Alonso 1996, 1999; RM05; Casagrande et al. 2011), on the line depth ratios (Kovtyukh et al. 2003, 2004, 2006, 2007), on the excitation equilibrium of iron lines (Santos et al. 2004) and that based on fitting with synthetic spectra (Valenti \& Fischer 2005).
We present the mean temperature difference, $\Delta T_{\rm eff}$ for objects analyzed by those alternative techniques that are in common with the current sample with direct temperature measurements from interferometry  along with the standard deviation of the mean, s.d., and the number of common objects $N$ in Table 5.
The actual data points used for the comparisons are also presented in Fig. 11.

\subsection{IRFM effective temperatures}
The IRFM effective temperatures of Alonso et al. (1996, 1999) and RM05 do not deviate significantly from the direct measurements, as Table 5 and  Fig. 11 show.
For dwarfs the values of Alonso et al. (1996) and RM05 are slightly hotter than direct measurements. 
The average difference and standard deviation  are $\Delta T_{\rm eff} = 41$\,K and s.d. = 129\,K for Alonso et al. (1996) ($N = 39$), and  $\Delta T_{\rm eff} = 57$\,K and s.d. = 135\,K for RM05 ($N = 54$).
For giants, the agreement is excellent for both Alonso et al. (1996) and RM05, with average difference  of just few tens Kelvin and standard deviations smaller than 80 K.
Note that in the above comparisons, stars of $T_{\rm eff} \leq 4,000$\,K have been excluded, because the IRFM does not  work well for stars cooler than $\sim$ 4000\,K.
Nevertheless, those cool stars are also plotted in Fig. 11 for completeness.
In general, for those stars the IRFM temperatures deviate somewhat from the direct measurements.

However, the effective temperatures of dwarfs of Casagrande et al. (2011)\footnote{The IRFM technique employed in Casagrande et al. (2011) is developed by Casagrande et al. (2010).}  based on the IRFM   or T$_{\rm eff}$--colour relations calibrated by the IRFM are significant hotter than direct measurements\footnote{For giants, there is no common sources between the two samples.}.
As Table 5 and Fig. 11 show, the average difference amounts to $\Delta T_{\rm eff} = 131$\,K, with a standard deviation s.d. = 127\,K ($N = 69$). 
The difference is significant that the IRFM temperature scale of Casagrande et al. (2011) may be too hot. 

\begin{figure}
\centering
\includegraphics[scale=0.4,angle=0]{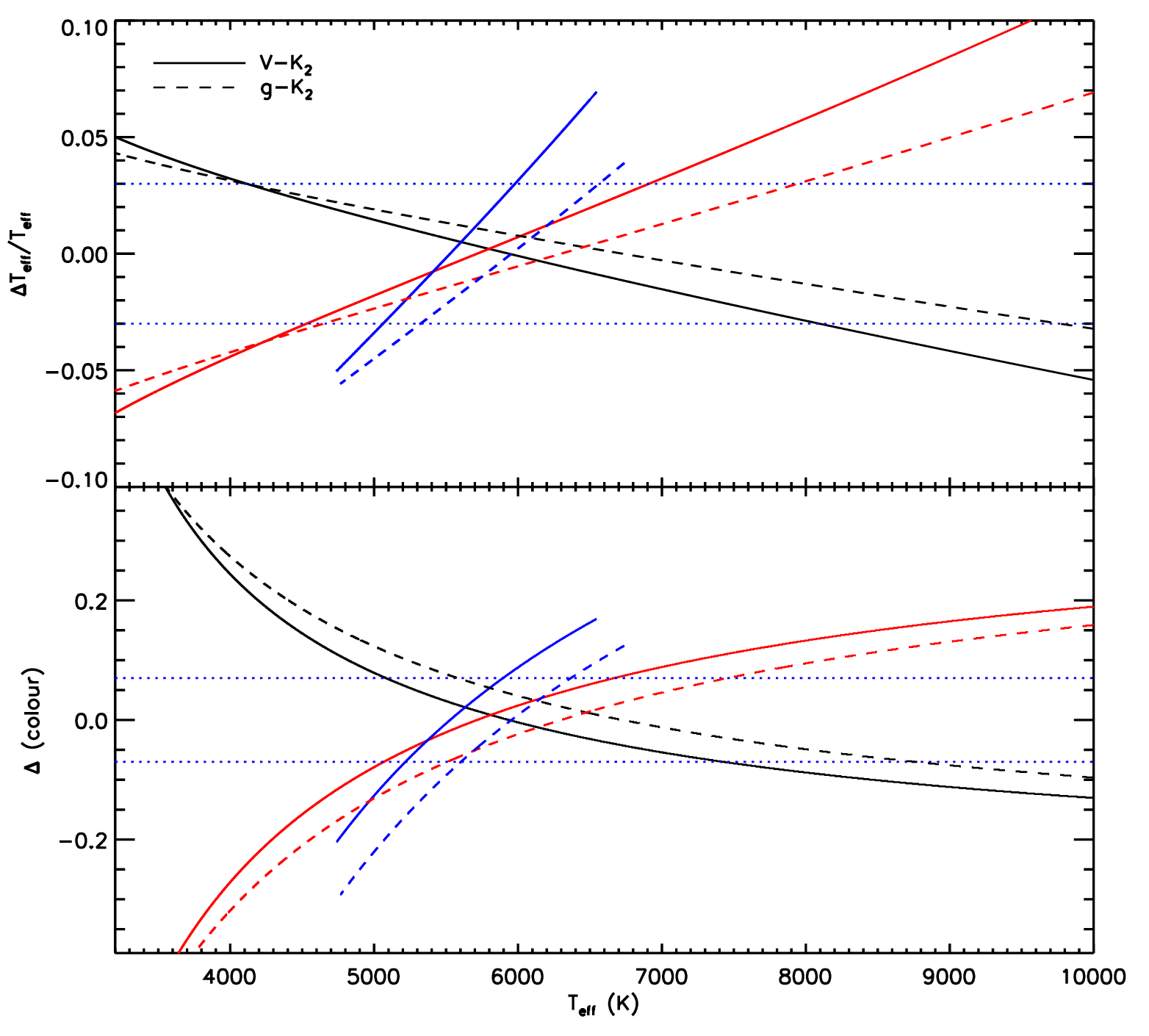}
\caption{{\em Upper panel:}\, The relative error in $T_{\rm eff}$ yielded by colour $(V - K_2)$ (solid lines) or $(g - K_2)$ (dash lines) for a dwarf of solar metallicity if one assumes an incorrect metallicity [Fe/H] of 0.5 (black), $-0.8$ (red), and $-2.0$ (blue). 
The blue dot lines represent a maximum acceptable relative uncertainty of 3 per cent.
{\em Bottom panel:}\, Effects of metallicity on colours $(V - K_2)$ and $(g - K_2)$. 
The meaning of the different line types and colours are the same as in the upper panel.  
The blue dot lines represent a typical photometric error of  0.07 mag.}
\end{figure}

\begin{figure}
\centering
\includegraphics[scale=0.4,angle=0]{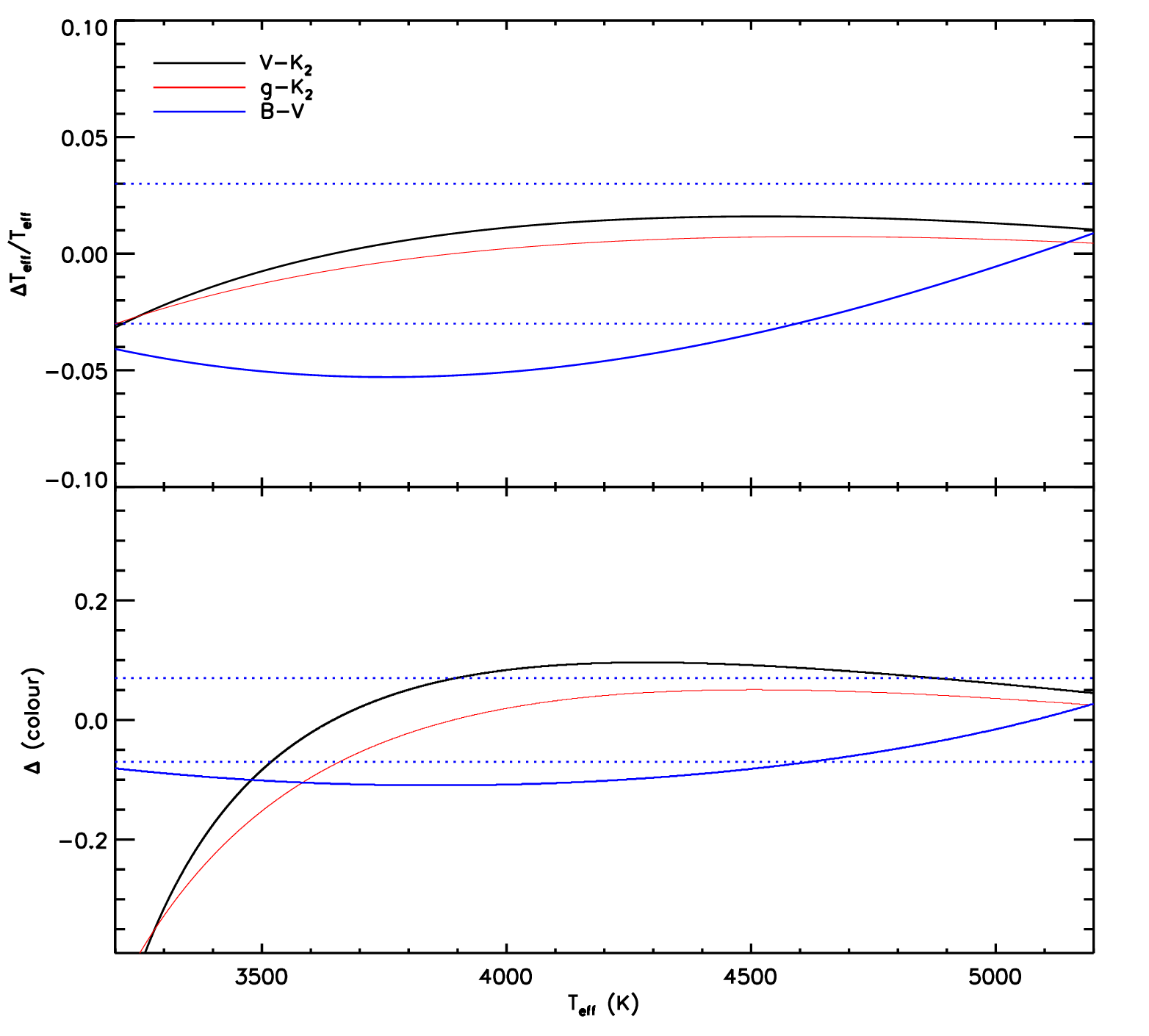}
\caption{{\em Upper panel:}\, The relative error in $T_{\rm eff}$ deduced from colour $(V-K_{2})$ (black lines), $(g-K_{2})$ (red lines) and $(B-V)$ (blue lines) for a dwarf of solar metallicity if the star is actually a giant, plotted as a function of $T_{\rm eff}$ of the dwarf. 
The blue dot lines represent a maximum acceptable relative uncertainty of 3.0 per cent.
{\em Bottom panel:} The differences in colours between a giant and a dwarf for different colours as described in the upper panel, plotted as a function of $T_{\rm eff}$.
The blue dot lines represent a typical photometric error of 0.07 mag.}
\end{figure}

As discussed in previous work (e.g. Casagrande et al. 2006, 2010), the systematic differences seen  in IRFM effective temperature scales given by different studies are mainly due to the different zero points adopted. 
Now with many stars available in the CALSPEC database\footnote{http://www.stsci.edu/hst/observatory/cdbs/calspec.html/} that have spectrophotometric fluxes between 0.3 -- 2.5 $\mu$m accurate to better than 1 -- 2 per cent (Bohlin 2010), the calibration of the zero-point of IRFM effective temperature scale can be much improved.
Alternatively, one can also renormalize the zero-point of IRFM effective temperature scale such that it gives consistent results with direct measurements from interferometry.

\begin{table*}
\caption{Comparisons with other effective temperature scales}
\begin{threeparttable}
\begin{tabular}{lcccc}
\hline
Source&$\Delta T_{\rm eff}$ (K)&s.d. (K)&$N$& Method\\
\hline
\multicolumn{5}{c}{Dwarf stars} \\
Alonso et al. (1996) & 41 & 139 & 39& IRFM\\
Santos et al. (2004) & 127 & 129 & 24 & Excitation equilibrium of Fe~{\sc i} lines\\
RM05 & 59 & 136& 54& IRFM\\
Valentin \& Fischer (2005) & 91 & 118 & 59 & Fitting with synthetic spectra\\
Kovtyukh et al. (2003, 2004) & 57 & 94 & 36& Line-depth ratios\\
Casagrande et al. (2011) &131 & 127 & 69 &IRFM or T$_{\rm eff}$--colour relations\tnote{*}\\
 \hline
 \multicolumn{5}{c}{Giant stars} \\
 Alonso et al. (1999) & -4 & 77 & 15& IRFM\\
RM05 & -16& 76 & 14& IRFM\\
Kovtyukh et al. (2006, 2007) & 82 & 89 & 14& Line-depth ratios\\
 \hline
\end{tabular}
\begin{tablenotes}\small
\item[*]The relations are taken from Casagrande et al. (2010).
 \end{tablenotes}
\end{threeparttable}
\end{table*}

\subsection{Spectroscopic effective temperatures}
Santos et al. (2004) derive values of $T_{\rm eff}$ for 98 planet-hosting stars and 41 stars without known planets based on the excitation equilibrium of Fe~{\sc i} lines from high resolution spectroscopy.
A total of 24 stars are found in common with our current  sample of dwarfs. 
Their effective temperatures are on average 127 K hotter than the direct measurements, with a standard deviation s.d. = 129\,K.

A uniform compilation of stellar properties for 1,040 nearby F/G/K dwarf stars derived from high resolution spectra is presented by Valenti \& Fisher (2005).
In their study, $T_{\rm eff}$, log\,$g$ and [Fe/H] are estimated simultaneously by fitting the observed spectra with synthetic ones.
The typical uncertainties of $T_{\rm eff}$ from this method is reported to be about 44\,K.
A comparison of common objects with the current dwarf sample yields an average difference $\Delta T_{\rm eff} = 91$\,K and a standard deviation s.d. = 129\,K ($N = 59$).

\begin{figure*}
\centering
\includegraphics[scale=0.50,angle=0]{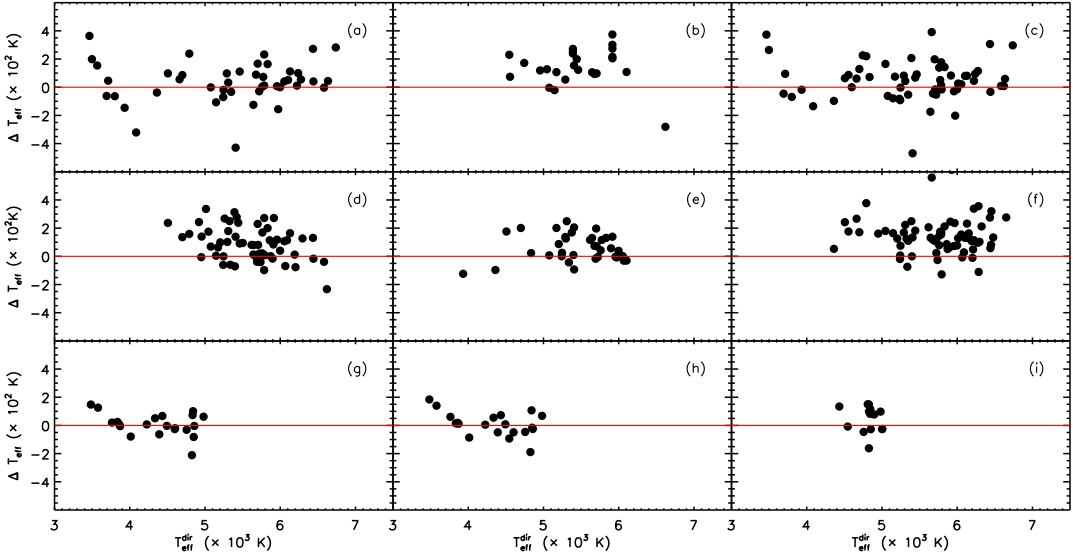}
\caption{ Differences of temperature between the alternative scales and the direct scale from  interferometry are plotted against that from the direct scale, $T_{\rm eff}^{\rm dir}$.
The top two panels are dwarfs for: (a) Alonso et al. (1996); (b) Santos et al. (2004); (${\rm c}$) RM05; (d) Valentin \& Fischer (2005); (e) Kovtyukh et al. (2003, 2004, 2006) and (f) for Casagrande et al. (2011).
The bottom panel are giants for: (g) Alonso et al. (1999); (h) RM05 and (i) Kovtyukh et al. (2006, 2007).}
\end{figure*}

\begin{figure*}
\centering
\includegraphics[scale=0.49,angle=0]{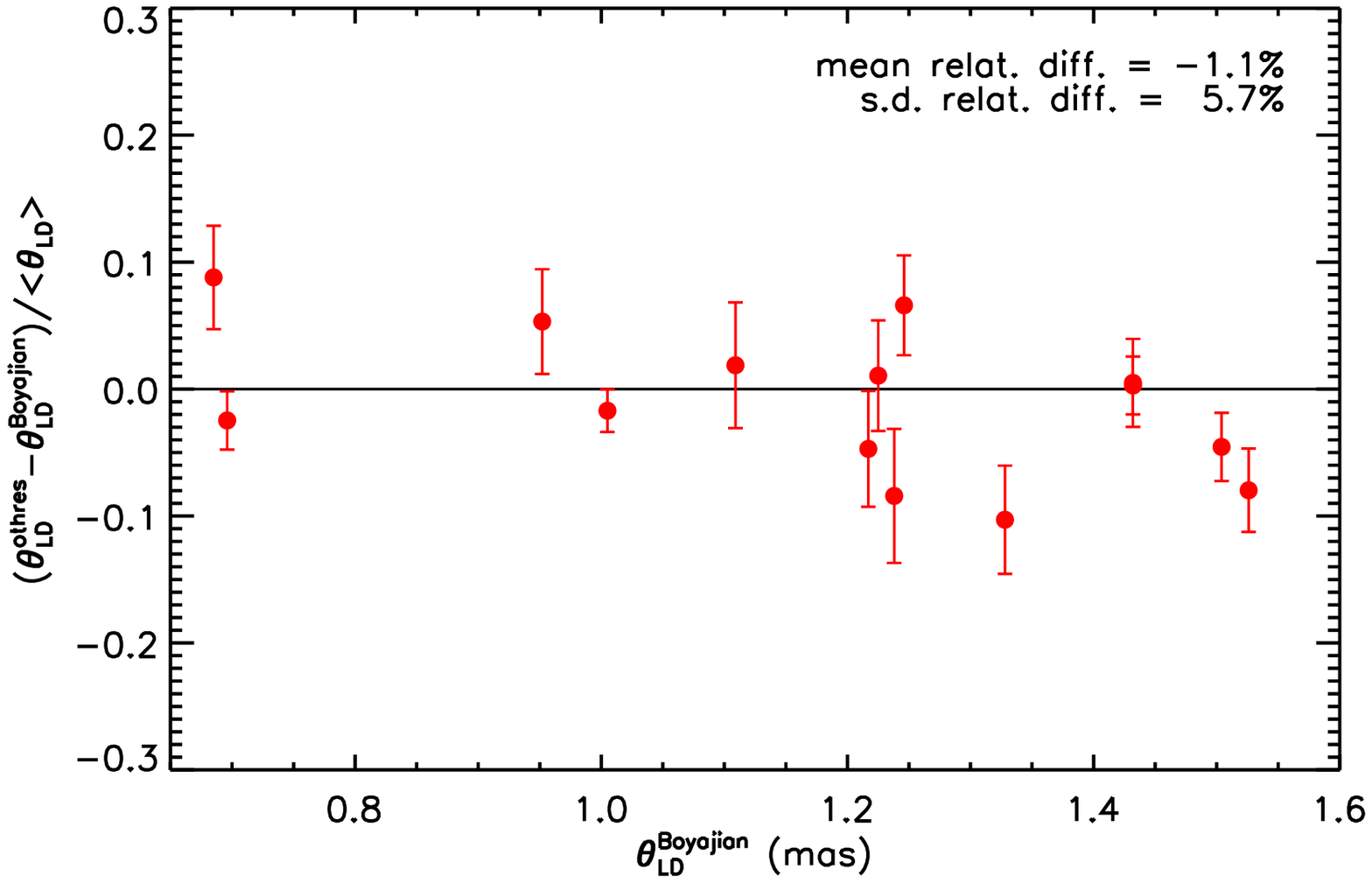}
\includegraphics[scale=0.49,angle=0]{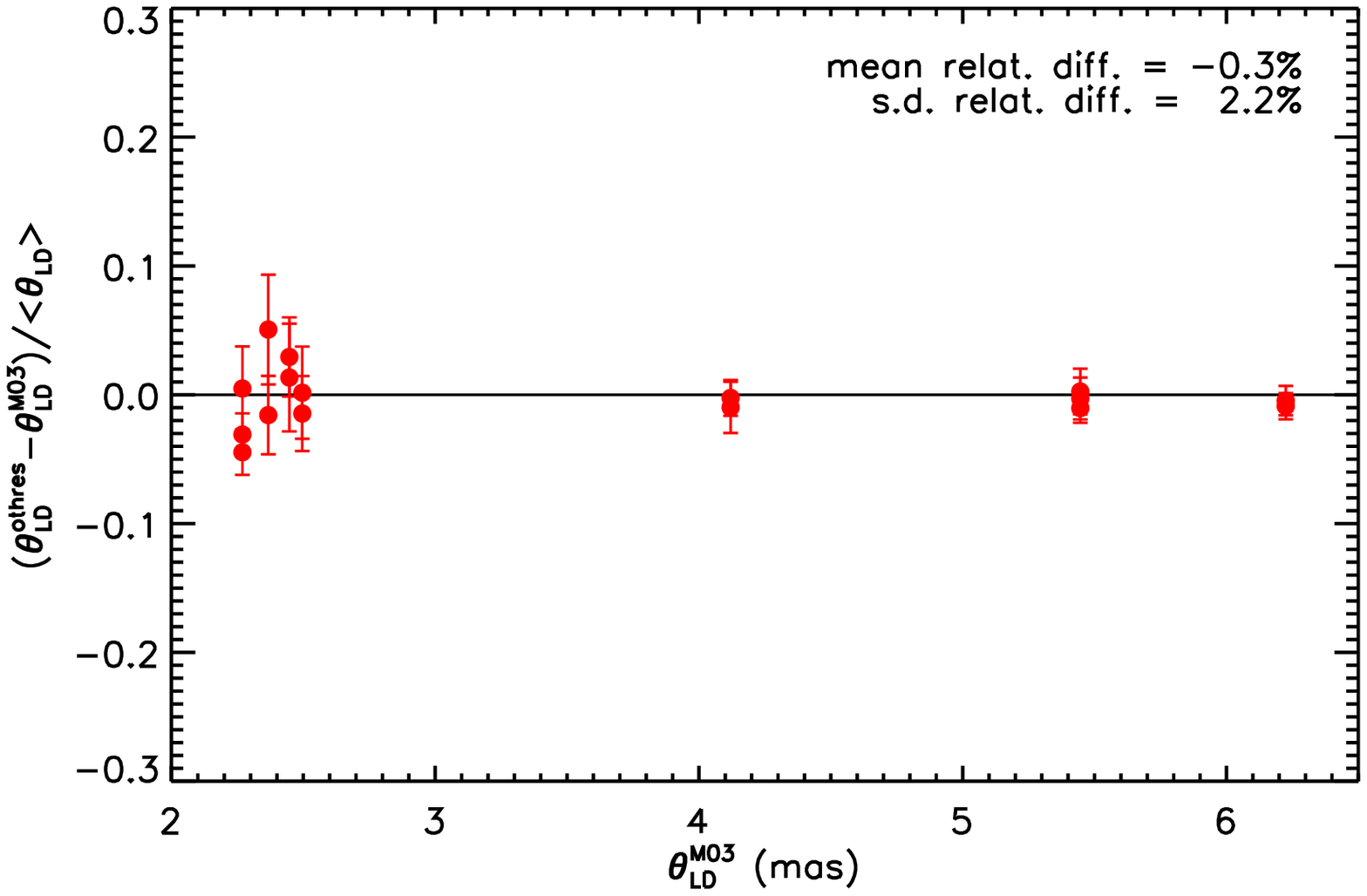}
\caption{Relative differences of  limb-darkened diameters as measured by the Boyajian's group and by others (left), and as determined by the M03 and by others (right) are plotted against the values of the Boyajian's group ($\theta_{\rm LD}^{\rm Boyajian}$) and those of M03 ($\theta_{\rm LD}^{\rm M03}$), respectively.}
\end{figure*}

On the whole, values of $T_{\rm eff}$ derived from high resolution spectra are on average $\sim$ 100\,K hotter than the direct measurements.

\subsection{Effective temperatures deduced from line-depth ratios}
To obtain effective temperatures of high precision, Kovtyukh et al. (2003, 2004, 2006, 2007) develop a method based on the observed line-depth ratios and apply the method to several hundreds dwarfs and giants stars with high resolution and high signal-to-noise ratio spectra.
The temperatures derived have an  internal error smaller than a few tens Kelvin.
The advantage of this method is that the line-depth ratios employed are insensitive to the interstellar reddening, the spectral resolution and the rotational and micro-turbulence broadening (e.g. Kovtyukh et al. 2003).
As Table\,5 and Fig.\,11  show,  effective temperatures of dwarfs yielded by the line-depth ratio technique are in general good agreement with direct measurements.
The mean difference is $\Delta T_{\rm eff} = 57$\,K, with a standard deviation s.d. = 94\,K ($N = 36$).
For giants, effective temperatures derived by this method are slightly hotter, by an average $\Delta T_{\rm eff} = 82$\,K, along with a standard deviation s.d. = 89\,K ($N = 14$).
Note that the values of s.d. are quite small for both dwarfs and giants corroborating that this method based on line-depth ratios does seem to have a very high internal precision.

\subsection{Remarks on the direct temperature scale}
 The systematic differences revealed by the above comparisons are probably caused by potential systematics in the model atmospheres or biases in the zero-point calibrations of other $T_{\rm eff}$ determination methods. 
Alternatively, they could also come from the calibrations of interferometry measurements themselves.
In what follows, we discuss the potential uncertainties of direct temperature scale  based on the current  compiled measurements of interferometric angular diameters.

\textbf{1) Interferometric measurements:}
In the current work, direct measurements of $T_{\rm eff}$ are  compiled from various studies using different telescopes.
Thus  the consistency between the different measurements needs to be checked.
As described in Section 2, most of the angular diameter measurements for dwarfs and giants used in the current work come from the work of the  Boyajian's group (B12a,b; B13) and from M03, respectively. 
Thus we compare respectively the angular diameters measured by Boyajian's group and by M03 with those yielded by other independent work.
 To do the comparisons, we present stars in our compiled catalog with multiple measurements  in Table A1.
As Fig.\,12 shows, no systematic errors are correlated with the size of the star (i.e. $\theta_{\rm LD}$) in both cases.
The mean and standard deviation of relative differences between the diameters  measured by the Boyajian's group and by others  are $-1.1$ and $5.7$ per cent, respectively.
The corresponding values of  differences between   M03 and other work are $-0.3$ and $2.2$ per cent.
In both cases  the means are quite small, much smaller than the standard deviations, implying that the measurements of Boyajian's group and of M03 are consistent with the results of other independent  work.

In addition, we have also checked the differences of diameters obtained with CHARA and PTI. 
As reported in van Belle \& von Braun (2009) and B12a, diameters obtained with CHARA are $\sim$\,(5--6)\,$\pm$6 per cent systematically  larger than those yielded by PTI.
However, we note that most of the PTI diameters included in the comparisons of the above two studies have uncertainties larger than 5 per cent.
In the current work, only the diameter measurements of uncertainties less than 5 per cent are used. 
Here, we have done a similar comparison for stars in our sample (see Table~A1) with both CHARA and PTI measurements available. 
We find that the CHARA  diameters are on average only $\sim\,1.8\,\pm\,6.2$ per cent larger than those yielded by PTI.
Thus for sample stars adopted in the current work, the discrepancies between the CHARA and PTI diameters are much less than those reported in van Belle \& von Braun (2009) and B12a, and show no significant systematics with measured diameter.

\textbf{2) 1D versus 3D model  atmospheres:} 
In direct diameter measurements, corrections for  limb-darkening effects are important and need to be properly applied.
It is possibly true that 3D model atmospheres may present a more realistic description of the limb-darkening effects (Pereria 2013).
Diameters derived using 1D model atmospheres (used for most of our sample stars) are in general  $0.5-1.0$ per cent larger than those yielded using 3D model atmospheres (e.g. Allen Prieto et al. 2002; Bigot et al. 2006).
This is confirmed by stars HD128621 and HD061421 in our sample  that have diameters determined using both1D and 3D model atmospheres (see Table A1).
Thus  the adoption of 1D model atmosphere measurements in the current work may induce a small systematic error compared to 3D model atmosphere measurements.
However, any such systematics are likely to be quite small, on the level of  $0.5-1.0$ per cent, far less than the measurements uncertainties (5 per cent).  

\begin{figure*}
\centering
\includegraphics[scale=0.49,angle=0]{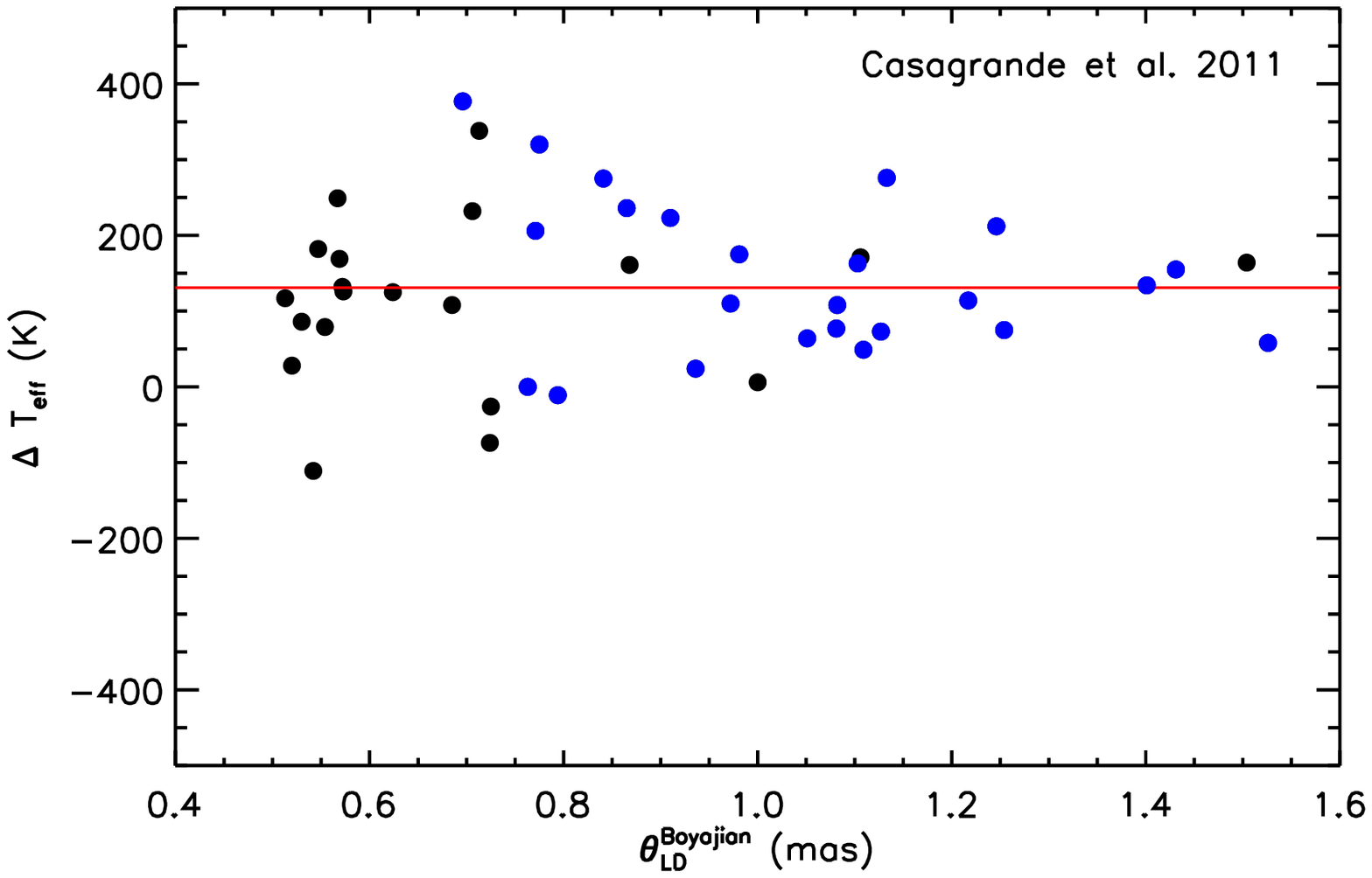}
\includegraphics[scale=0.49,angle=0]{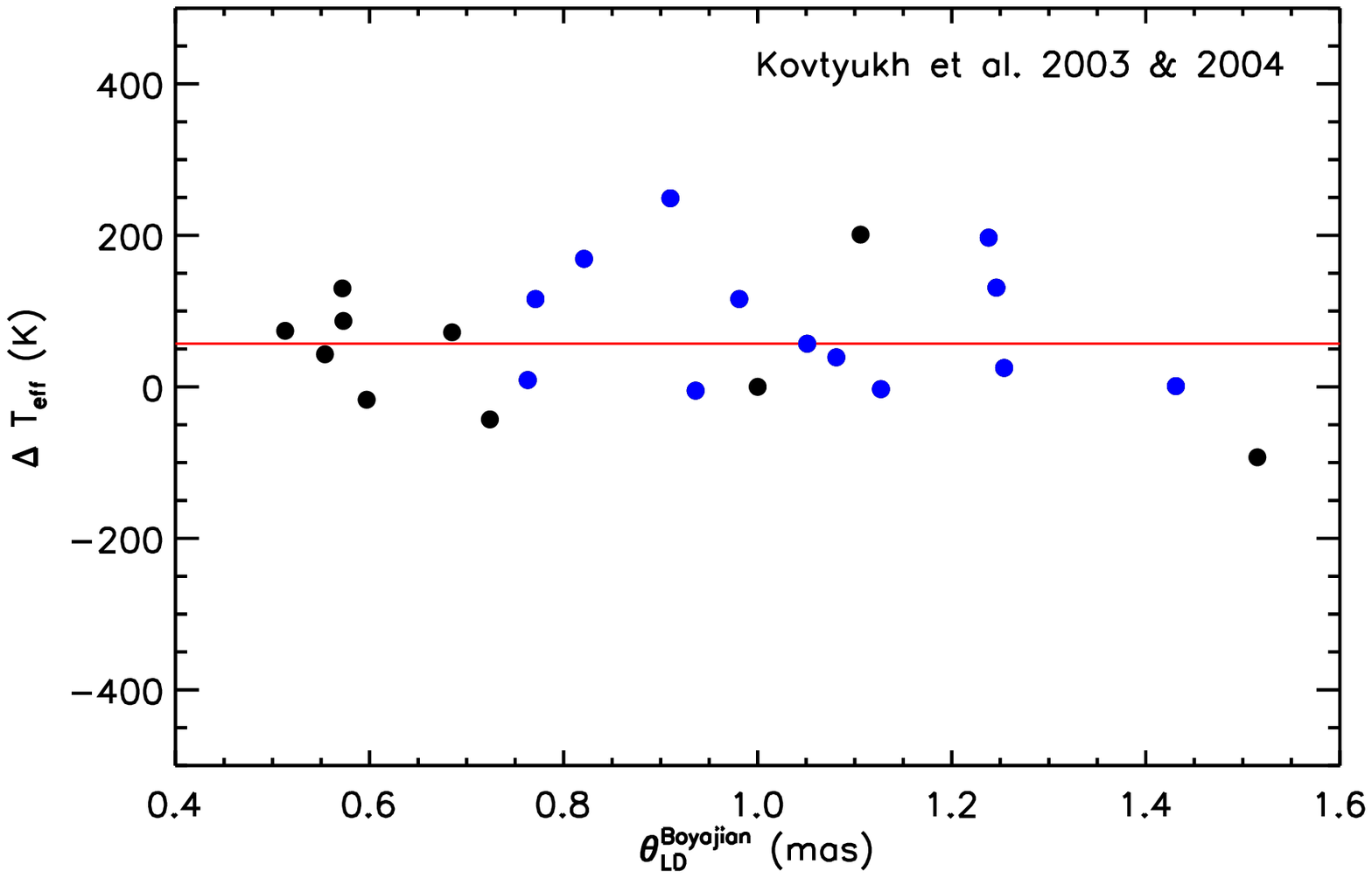}
\caption{Differences of temperatures from Casagrande et al. (2011, left) and from Kovtyukh et al. (2003, 2004, right) with respective to the interferometric measurements of the Boyajian's group (B12a,b, B13), plotted as a function of interferometric angular diameters  measured by the Boyajian's group. Blue dots represent interferometric measurements from B12a. Red lines delineate the mean temperature  differences presented in Table 5.}
\end{figure*}

\textbf{3) Systematic errors related to the stellar  angular diameters:}
For interferometric measurements, the smaller the size (angular diameter) of the star the more difficult the observations are, and thus the larger the (random plus systematic) errors of the measured diameters are.
It is thus important to check whether there are any systematic errors that correlated with the size of star of concern. 
As reported in Casagrande et al. (2014, hereafter C14), the differences of temperatures yielded by  two photometric scales, that of Casagrande et al. (2011) and Holmberg et al. (2009), and that of B12a, show a clear trend of variations (especially for stars of $\theta_{\rm LD} \le 1$ mas) as a function of interferometric diameter. 
C14 suggest that the systematic trend originates from the measurements of B12a given the nearly constant differences between temperatures from the two photometric temperature scales. 
To better understand the origin of this systematic trend, we have re-checked the differences between  temperatures of Casagrande et al. (2011) and those from the Boyajian's group (B12a,b, B13) using 46 common stars.
As described in Section\,4.1, the temperatures of Casagrande et al. (2011) are derived from the IFRM or from the $T_{\rm eff}$-colour relations calibrated with the IRFM.
For the very bright stars (e.g. like those in  our sample), no reliable near-infrared photometric measurements are available and, as a consequence, the temperatures are derived from the $T_{\rm eff}$-colour relations (mostly based on the St{\"omberg} photometry).
As the blue dots in Fig. 13 show, the differences of temperatures between those of Casagrande et al. (2011) and of B12a are indeed  essentially what C14 have seen.
However, when the black dots, from B12b and B13, are also taken into consideration, the temperature differences no longer  show any significant systematic trend with stellar diameter except for a nearly constant displacement ($\sim 130$\,K, see Table\,5)  for a wide range of stellar interferometric diameter from 0.5 to 1.5 mas.
It seems that the systematic trend noted by C14 is possibly only an artifact, caused by the limited sampling and range of angular diameters of stars included in their comparison.
To further check for any systematic errors that may be correlated with stellar angular size in interferometric measurements, we have carried out a similar comparison, comparing temperatures from Kovtyukh et al. (2003, 2004) with those from the Boyajian's group (B12a,b, B13) for 24 common stars.
As described in Section\,4.3,  temperatures from Kovtyukh et al. (2003, 2004) are obtained with line-depth ratios.
Temperatures derived with this method are of a very high internal precision (to a few tens Kelvin) and are insensitive to the interstellar reddening, the spectral resolution and the rotational and micro turbulence broadening.
Hence they are quite suitable to test if there are any systematic errors in the interferometric measurements that correlate with the size of the star.
Again, as Fig. 13 shows, no such systematic trend is found, except for a nearly constant displacement ($\sim 57$\,K, see Table\,5) for a wide diameter from 0.5 to 1.5 mas.
Similarly, there are also no significant systematic errors in diameter measurements from other work for dwarfs in our sample, considering the almost zero displacement between the diameters yielded by the Boyajian's group and by other studies,  as Fig.\,12 shows. 
For giant stars, any errors of diameter measurements caused by the differing stellar angular size are unlikely to be significant, given their large angular diameters (generally larger than 2 mas).

We conclude that: 1) The current set of diameter measurements compiled from the literature are all self-consistent; 2) The diameter measurements adopted in the current work, mostly based on 1D model atmospheres, could be possibly slightly overestimated, but only at the level of  $0.5-1.0$ per cent; 3) The current set of measurements show no significant systematic errors that correlate with the stellar angular size.

\begin{figure*}
\centering
\includegraphics[scale=0.45,angle=0]{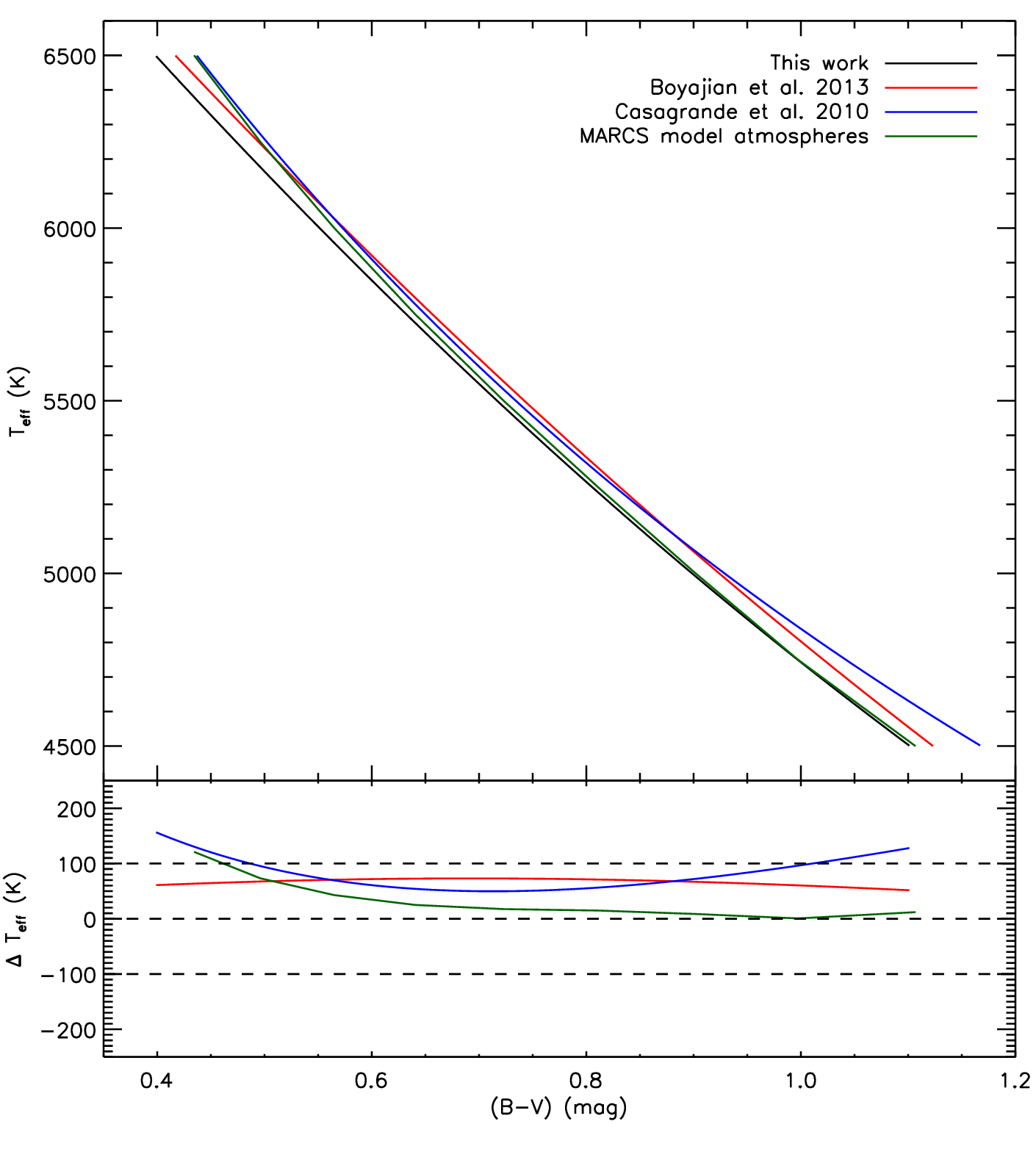}
\includegraphics[scale=0.45,angle=0]{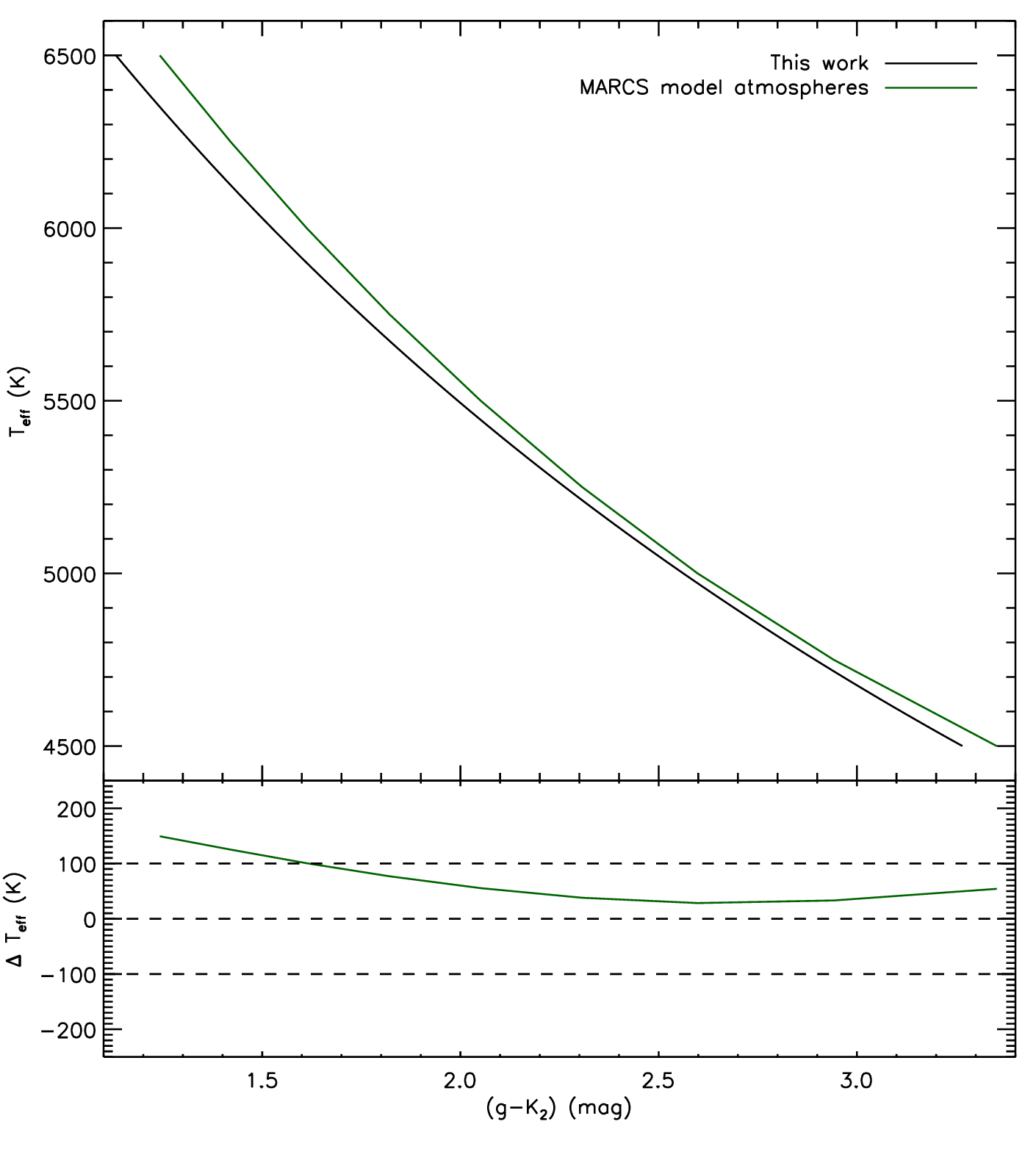}
\caption{Comparisons of the $T_{\rm eff}$--colour relations for dwarfs of solar metallicity from the current work (black) with that from B13 (red), Casagrande et al. 2010 (blue) and MARCS model atmospheres (green; assuming log\,$g$\,=\,4.0) for colour ($B-V$) (left panel) and colour ($g-K_{2}$) (right panel). 
The differences of predicted temperatures, $\Delta T_{\rm eff}$ (K), as a function of colours are plotted in the lower part of each panel.}
\end{figure*}

\section{Comparisons with other  $T_{\rm \sc \textnormal{eff}}$ -- colour relations}

\begin{table*}
\begin{center}
\caption{Colours of the Sun}
\begin{threeparttable}
\begin{tabular}{ccccccccccc}
\hline
Colour&This Work&CBC96\tnote{*a}&AAM96\tnote{*}&SF00\tnote{*}&RM05\tnote{*}&HFP06\tnote{*}&C10\tnote{*}&R12\tnote{*}&CV15\tnote{*b}\\
\hline
$  U-V $ &0.791 $\pm$ 0.070 & 0.770& 0.770 $\pm$ 0.036 &--&--&0.815$\pm$0.066&--&0.819$\pm$0.023&0.770 \\
$  B-V $ &0.623 $\pm$ 0.037& 0.630& 0.615 $\pm$ 0.020&0.626$\pm$0.018&0.619&0.642$\pm$0.016&0.641$\pm$0.024&0.653$\pm$0.005&0.621\\
$  V-R_{\rm J} $ &0.526 $\pm$ 0.036&--&0.525 $\pm$ 0.020&--&--&--&--&--&--\\
$  R_{\rm J}-I_{\rm J} $ &0.323  $\pm$ 0.026 &--&0.325  $\pm$ 0.020&--&--&--&--&--&--\\
$  V-I_{\rm J} $ &0.849 $\pm$ 0.055&--&0.850  $\pm$ 0.020&--&--&--&--&--&--\\
$ V-R_{\rm C} $ &0.348$\pm$0.025&--&--&--&0.351&0.354$\pm$0.010&0.359$\pm$0.010&0.356$\pm$0.003&0.361\\
$ V-I_{\rm C} $ &0.665$\pm$0.045&--&--&--&0.682&0.688$\pm$0.014&0.690$\pm$0.016&0.701$\pm$0.003&0.688\\
$ R_{\rm C}-I_{\rm C} $ &0.318$\pm$0.024&--&--&--&0.330&0.332$\pm$0.08&0.333$\pm$0.010&--&0.327\\
$  V-J $ &1.068  $\pm$ 0.060&--&--&--&--&--&--&--&--\\
$  V-H $ &1.362 $\pm$ 0.083 &--&--&--&--&--&--&--&--\\
$  V-K $ &1.437  $\pm$ 0.080&--&--&--&--&--&--&--&--\\
$ V-J_2 $ &1.125$\pm$0.060 &--&--&--&1.141&1.151$\pm$0.035&1.180$\pm$0.021&--&1.164\\
$ V-H_2 $ &1.366 $\pm$0.081 &--&--&--&1.396&1.409$\pm$0.035&1.460$\pm$0.023&--&1.452\\
$ V-K_2 $ &1.472  $\pm$0.080&--&--&--&1.495&1.505$\pm$0.041&1.544$\pm$0.018&--&1.537\\
$  g-r $ &0.419  $\pm$ 0.036&--&--&--&--&0.450$\pm$0.020&--&--&0.426\\
$  g-J $ &1.313  $\pm$ 0.075&--&--&--&--&--&--&--&--\\
$  g-H $ &1.611  $\pm$ 0.102&--&--&--&--&--&--&--&--\\
$  g-K $ &1.689   $\pm$0.095 &--&--&--&--&--&--&--&--\\
$ g-J_2 $ &1.370  $\pm$0.075&--&--&--&--&--&--&--&1.425\\
$ g-H_2 $ &1.618 $\pm$0.093&--&--&--&--&--&--&--&1.713\\
$ g-K_2 $ &1.723  $\pm$ 0.096&--&--&--&--&--&--&--&1.798\\
  \hline
\end{tabular}
\begin{tablenotes}
\item[*]References: CBC96 -- Colina, Bohlin \& Castelli (1996); AAM96 -- Alonso, Arribas \& Mart\'inez-Roger (1996); SF00 -- Sekiguchi \& Fukugita (2000); HFP06 -- Holmberg, Flynn \& Portinari (2006); C10 -- Casagrande et al. (2010) and R12 -- Ram\'irez et al. (2012); CV15 -- Casagrande \& VandenBerg (2014).
\item[a]Those colours of the Sun are obtained from the real solar spectrum directly.
\item[b]Those colours of the Sun are obtained using the MARCS synthetic solar spectrum (for microturbulence velocity $\xi\,=1\,$\,km\,s$^{-1}$, $T_{\rm eff}\,=\,5777$\,K, log\,$g\,=\,4.44$ and [Fe/H]\,=0). 
 \end{tablenotes}
\end{threeparttable}
\end{center}
\end{table*}
In this section, we compare the metallicity-dependent  $T_{\rm eff}$--colour relations presented in the current work to those from the previous studies.
Fig.\,16 compares the $T_{\rm eff}$ versus colour (B-V) relation for dwarfs of solar metallicity (i.e. [Fe/H]\,=\,0) obtained in  the current calibrations with those from Casagrande et al. (2010), Boyajian et al. (2013) and MARCS model atmospheres (for log\,$g$ = 4.0; Gustafsson et al. 2008), as calculated by Casagrande \& VandenBerg (2014).
Generally, $T_{\rm eff}$ predicted by the relations of Casagrande et al. (2010) is $\sim$100\,K hotter than that of the current work,  consistent with the findings in Section 4.1.
The discrepancies become larger for both blue (e.g. $B-V < 0.5$\,mag) and red (e.g. $B-V > 1.0$\,mag) colours. 
The relation based on MARCS model atmospheres is in good agreement  with our within a few tens of degrees Kelvin except for the blue (e.g. $B-V < 0.6$\,mag, $T_{\rm eff} > 5,800$\,K). 
We also compare our $T_{\rm eff}$ versus colour ($g-K_{2}$) relation for dwarfs of solar metallicity with that from MARCS model atmospheres (of log\,$g$ = 4.0) in Fig.\, 14.
The results are quite similar to those in colour $(B-V)$.

As described above, B13 consider the metallicity effects for colour ($B-V$) only when deriving the $T_{\rm eff}$--colour relation.
 To our surprise, although the samples used by B13 and the current work are quite similar (the 125 dwarfs consisting the B13 sample constitute the bulk of the current sample), temperatures predicted by the empirical relation of B13 are about 50--100\,K hotter than the values calculated from the relation of the current work, for stars hotter than $T_{\rm eff} \gtrsim 4300$\,K (or $B-V \lesssim$ 1.20\,mag), as shown in Figs.\,14 and 16 for the case of solar metallicity.
 We find that the residuals of the fit obtained by B13 show a systematic trend as a function of colour, as shown in Fig.\,16.
The systematics in the fit of B13  is fully responsible for the discrepancy between the $T_{\rm eff}$--(B-V) relations of B13 and of the current work.

For the giants, the $T_{\rm eff}$ versus colour ($V-K_{2}$) relation of solar metallicity of current work is compared with that from RM05 as well as from MARCS model atmospheres (for log\,$g$ = 2.5), as shown in Fig.\,15.
Unlike the dwarfs, the relation for giants  from MARCS model atmospheres is hotter than ours by $\sim 100$\,K.
The relation of RM05 is in excellent agreement with ours in their applicable range ($V-K_{2} <$\,3.29\,mag).

\begin{figure}
\centering
\includegraphics[scale=0.45,angle=0]{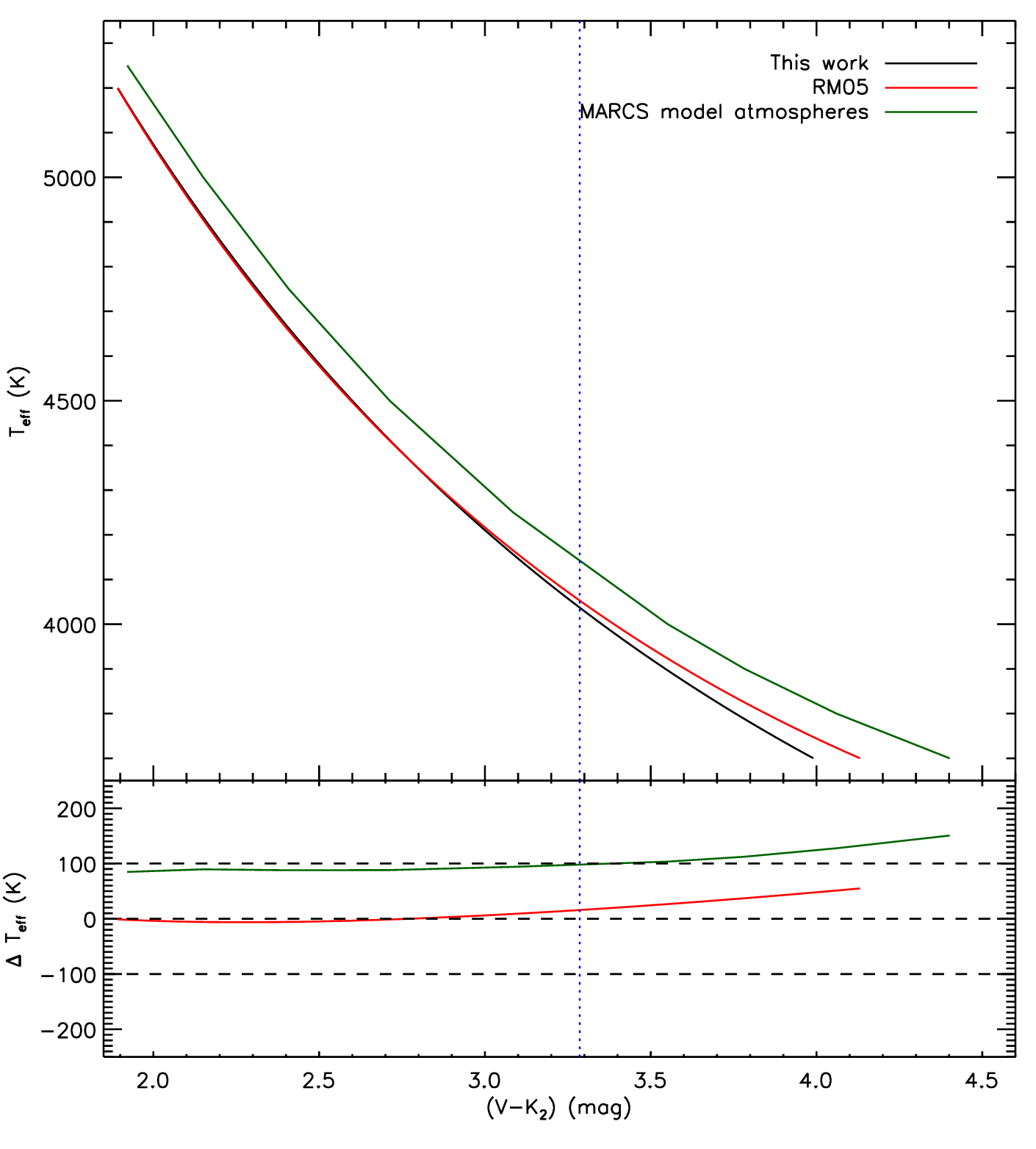}
\caption{Comparisons of the $T_{\rm eff}$--($V-K_{2}$) relation for giants of solar metallicity presented in the current work (black) with that from RM05 (red) and MARCS model atmospheres (green; log\,$g$\,=\,2.5). 
The lower part of the panel shows the differences of $T_{\rm eff}$, $\Delta\,T_{\rm eff}\,(\rm K)$, between the values predicted by relations from other studies and by ours as a function of colour.
The dashed blue line indicates the upper applicable range of colour range of the relation of RM05 $V-K_{2}\,=\,3.29$.}
\end{figure}

\section{Applications}
\subsection{Colours of the Sun}
It is not an easy task to measure colours of the Sun, being such a bright and extended source.
Most measurements of the solar colours are made indirectly, either using the \textit{solar twins} or the interpolating empirical relations (e.g. the $T_{\rm eff}$-colour-metallicity relations) to the solar values based on samples of stars with well determined physical properties (e.g. Alonso et al. 1996; Sekiguchi \& Fukugia 2000; RM05; Holmberg et al. 2006; Casagrande et al. 2010; Ram\'irez et al. 2012).
However, essentially all the effective temperature scales of the aforementioned work are based on temperatures deduced by indirect methods(e.g. the IRFM or spectroscopy). 
Consequently, there could be a systematic errors in the estimated colours of the Sun.
Using the empirical calibration based on direct effective temperature measurements from interferometry presented in the current work, we have derived the colours of the Sun assuming a solar effective temperature $T_{\rm eff} = 5,777$ K and metallicity [Fe/H] = 0\footnote{Here we use the empirical relations deduced excluding the early-type stars.}.
The results are presented in Table 6 and compared to literature values.

Table\,6 shows that, as expected, our newly deduced colours of the Sun are systematic bluer than the recent determinations of Casagrande et al. (2010), deduced with the same method but based on an IRFM effective temperature scale, which is about 130\,K higher than the direct scale derived here, as shown in Fig. 11\,(f) and Table\,5.
 As shown in Table\,6, our newly deduced colour ($B-V$) of the Sun agrees well with the direct measurement of Colina, Bohlin \& Castelli (1996).
Using a similar approach as ours, Sekiguchi \& Fukugita (2000) also find similar result.
We warn that the colours of the Sun presented in the current work that involves bands transformed from measurements in other bands, e.g. those involving the SDSS\,$gr$ or the 2MASS $J_{2}H_{2}K_{2}$ bands, may suffer from large errors (both random and systematic), considering the possible errors propagated by those transformations.
Of course, as discussed above, the interferometry measurements themselves are not entirely free of potential systematics. 
This should be kept in mind when using the colours of the Sun presented in the current work.

\subsection{Calibration of temperature scales for stellar spectroscopic surveys}

\begin{table*}
\caption{Comparisons with the effective temperature yielded by the pipelines of SDSS/SEGUE and LSS-GAC}
\begin{threeparttable}
\begin{tabular}{lccccc}
\hline
Source&$\Delta T_{\rm eff}$ (K)&s.d. (K)&$<$ SNR$>$&$N$& Method\\
\hline
\multicolumn{5}{c}{Dwarf stars} \\
SDSS DR9 (Ahn et al. 2012) & 147 & 75 &58& 10682& Multiple methods\tnote{a}\\
LSP3 (Xiang et al. 2014) & 12 & 120 & 37&26953 & Template matching\tnote{b}\\
LASP (Wu et al. 2014) & -6 & 100& 37&26562& Template matching\tnote{c}\\
 \hline
 \multicolumn{5}{c}{Giant stars} \\
SDSS DR9 (Ahn et al. 2012) & 156 & 84 &57& 280& Multiple methods\tnote{a}\\
LSP3 (Xiang et al. 2014) & 44 & 95 & 36&3747 & Template matching\tnote{b}\\
LASP (Wu et al. 2014) & -15 & 76& 36&3413& Template matching\tnote{c}\\
 \hline
\end{tabular}
\begin{tablenotes}\small
\item[a] Effective temperatures yielded by individual methods are scaled to match those given by the IRFM effective temperature scale of Casagrande et al. (2010).
\item[b] Using the MILES spectral library as templates.
\item[c] Using the ELODIE spectral library as templates.
 \end{tablenotes}
\end{threeparttable}
\end{table*}

In large scale medium- to low-resolution stellar spectroscopic surveys, such as the SDSS/SEGUE and LSS-GAC, effective temperature $T_{\rm eff}$ is generally estimated simultaneously  with other stellar atmospheric parameters including metallicity [Fe/H] and surface gravity log $g$, by, for example, template matching with either empirical or synthetic spectra (e.g. Lee et al. 2008; Luo et 2015; Xiang et al. 2014). 
As shown in Section 4, effective temperatures thus determined could deviate from the true values by hundreds Kelvin.
In this subsection, we attempt to calibrate the temperature scales of two recent  spectroscopic surveys: the SDSS and LSS-GAC, using the empirical $T_{\rm eff}$-colour relations presented in the current work and examine if there are any systematic errors in the effective temperatures yielded by those surveys.
The results are presented in Table\,7, including the average  difference $\Delta T_{\rm eff}$  between the values of $T_{\rm eff}$ as yielded by the pipelines of those surveys and the values deduced from the empirical relations  presented in the current work, along with the standard deviation of the difference, s.d., the median spectral signal-to-noise ratios (SNRs) of the stars used in the comparison, the number of stars $N$ used, and the method used by the pipeline to estimate effective temperatures.

\begin{figure}
\centering
\includegraphics[scale=0.325,angle=0]{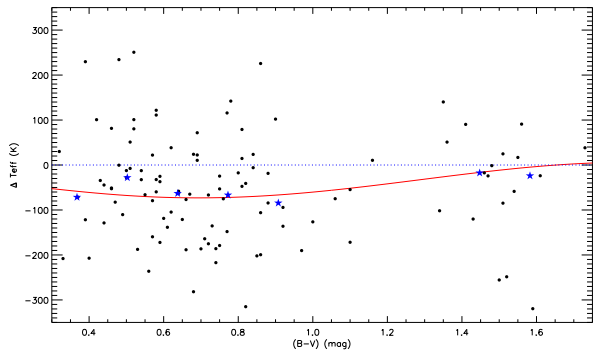}
\caption{ Differences of effective temperatures deduced from the $T_{\rm eff}$--$(B-V)$ relation of the current work and those from the relation of B13 for the solar metallicity, plotted against $(B - V)$ (red line).
Black dots represent the residuals ($T_{\rm eff} - T_{\rm eff}^{fit}$) of fit in the calibration of B13 and blue stars are the median residuals in the individual colour bins.}
\end{figure}

$\bullet$\,\textbf{SDSS:} The SDSS/SEGUE Stellar Parameters Pipeline (SSPP; Lee et al. 2008a, 2008b; Allende Prieto et al. 2008; Smolinski et al. 2011) estimates effective temperatures utilizing multiple approaches.
Each method has its favoured  applicable range in colour ($g-r$) and SNR.
The final adopted $T_{\rm eff}$ is the average of results from all methods.
In order to calibrate the SSPP effective temperatures, we select stars of spectral SNRs $>\,30$ and reddening  $E(B-V)< 0.05$\,mag, as given by the extinction map of SFD98, to minimize the effects of uncertainties of reddening corrections, from the ninth data release of SDSS (hereafter DR9; Ahn et al. 2012).
Following the results presented in Section 3, we use the $T_{\rm eff}$--($g-K_{2}$) relation to derive photometric temperatures of those selected SDSS stars (with metallicities yielded by the SSPP).
To obtain the $K_{2}$ magnitudes and reliable intrinsic colours of those stars, they are cross-matched with the 2MASS photometric catalog. 
For stars with a match, only those with a 2MASS flag  \textit{ph\_qual}  flagged by `A' in $K_{\rm s}$-band and with a $K_{\rm s}$-band photometric error less than 0.05\,mag, and at the same time with a SDSS $g$-band magnitude greater than $14.0$ mag (to avoid  potential saturation),  and a $g$-band  photometric error smaller than 0.02 mag, are retained.
Finally we select  10,682 dwarf stars with log $g$\,$\ge 3.5$ and 280 giant stars with log\,$g$\,$< 3.5$.
The photometric temperatures of those stars are then calculated using the $T_{\rm eff}$--$(g - K_2)$ relation presented in Section\,3, subject to the applicability ranges in colour and metalicity as shown in Tables 3 and 4.
A comparison of  the $T_{\rm eff}$ between the SDSS DR9 adopted values  and the photometric values yielded by our direct empirical calibration is presented in Fig.\,17.
On average, the DR9 adopted  effective temperatures are systematically hotter than our values by 147\ and 158\,K for dwarfs and giants, respectively (cf. Table 7).
The systematic offset of $\sim$ 150\,K  found here is easily understood considering the fact that the SSPP re-scales temperatures yielded by each method  to match with the IRFM scale of Casagrande et al. (2010), which has been shown to be too hot by 131\,K compared to the direct effective temperature scale (cf. Section\,4, Fig. 11\,(f) and Table\,5).
The standard deviations of the differences are smaller than 80\,K, for both dwarfs and giants, suggesting that the uncertainties of $\sim$50\,K reported by Ahn et al. (2012)  for stars of spectral  SNRs $> 30$ are reasonable estimated. 
Finally, we caution that the systematic offset, $\Delta T_{\rm eff}$, is not a constant but has some weak dependence on temperature.

\begin{figure*}
\centering
\includegraphics[scale=0.65,angle=0]{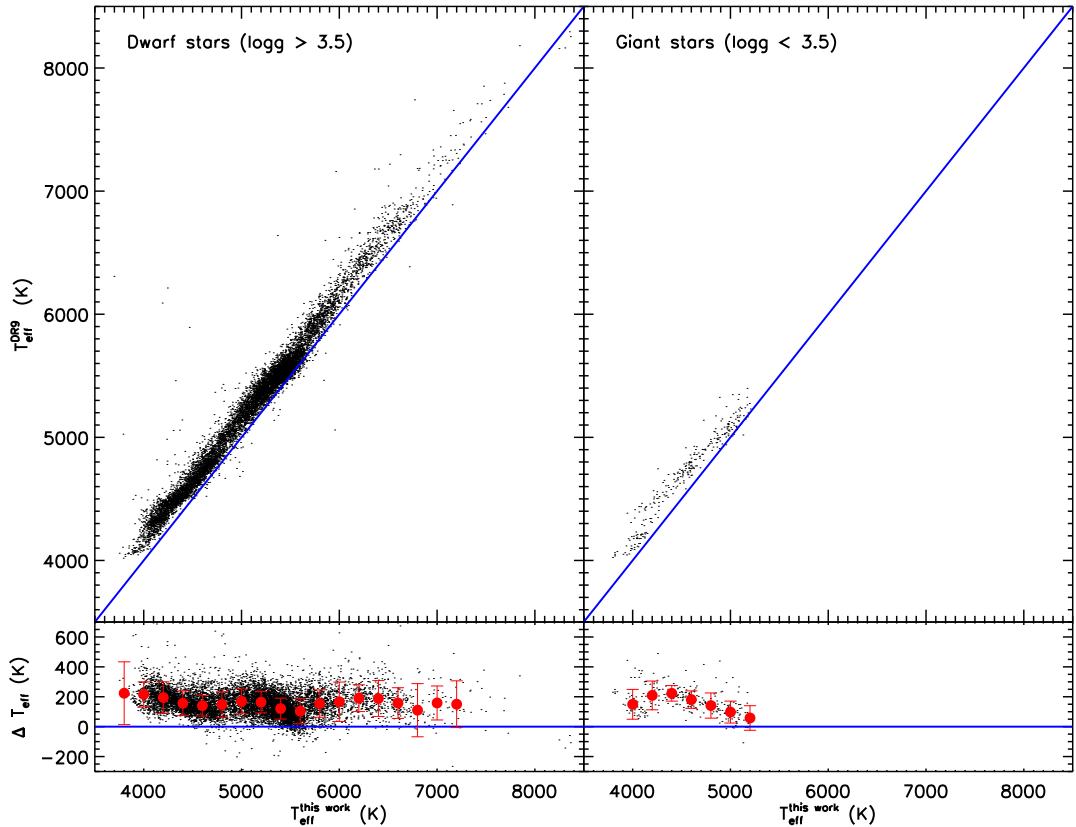}
\caption{Comparison of effective temperatures adopted by the SDSS DR9 and those derive from our $T_{\rm eff}$--($g-K_{2}$) relation. 
                  The differences, $\Delta T_{\rm eff}$ = ($T_{\rm eff}^{\rm DR9}\,-\,T_{\rm eff}^{\rm this\,work}$), are plotted at the bottom (with red dots and error bars representing the means and standard deviations of differences in the individual temperature bins).
                  The left panel is for dwarfs of log\,$g\,>\,3.5$ and the right for giants of with log\,$g\,<\,3.5$.}
\end{figure*}

\begin{figure*}
\centering
\includegraphics[scale=0.65,angle=0]{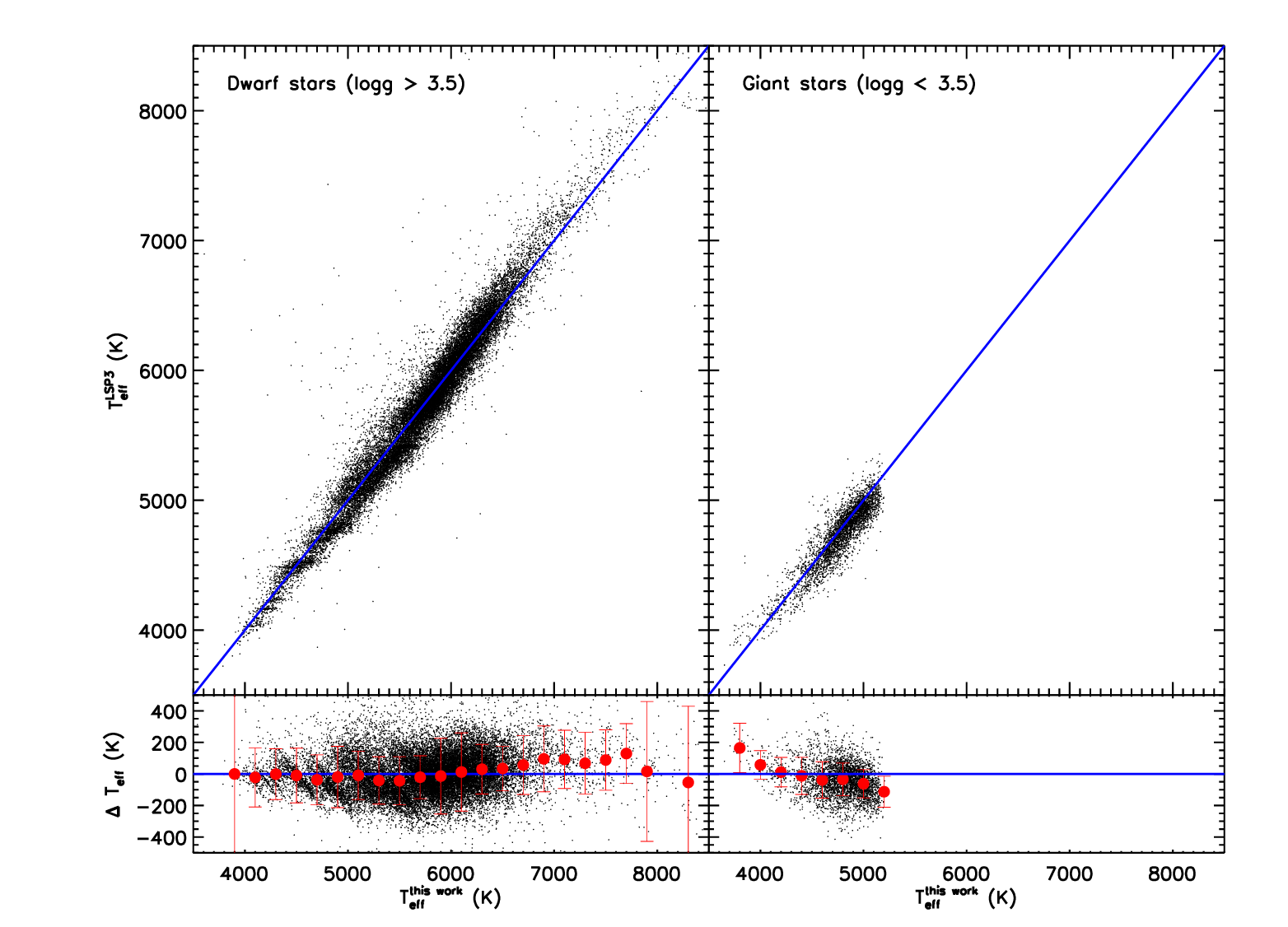}
\caption{Similar to Fig. 15 but for the LSP3.}
\end{figure*}
\begin{figure*}
\centering
\includegraphics[scale=0.65,angle=0]{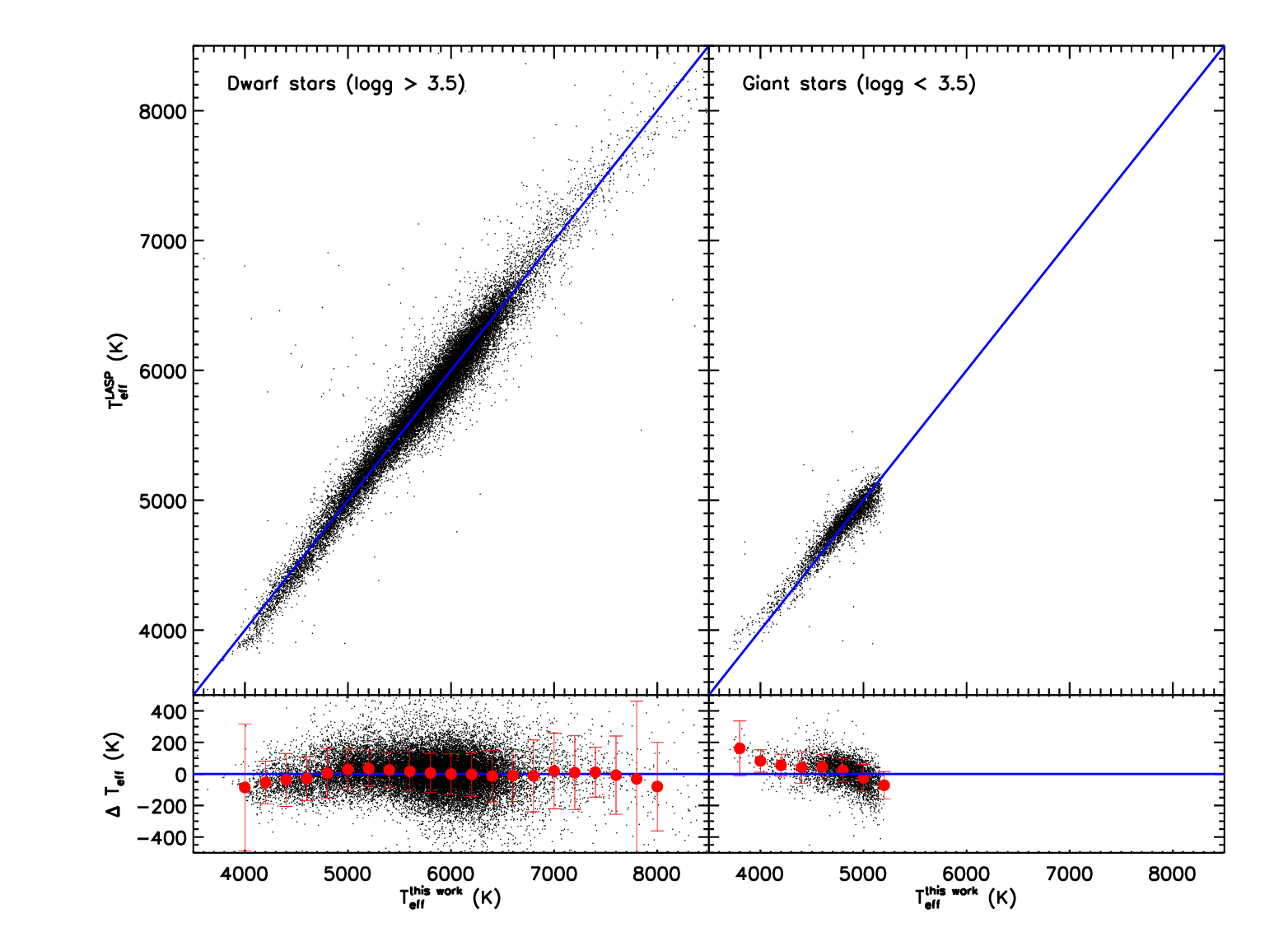}
\caption{Similar to Fig. 15 but for the LASP.}
\end{figure*}

$\bullet$\,\textbf{LSS-GAC:} As a major component of the on-going LAMOST Galactic surveys, the LSS-GAC (Liu et al. 2014) aims to collect optical ($\lambda\lambda$3700-9000), low resolution ($R \sim 1, 800$) spectra for a statistically complete sample of over a million stars of all colours in the magnitude range $14.0 \leq r < 17.8$\,mag (18.5\,mag for limited fields), distributed in a continuous sky area of $\sim$ 3, 400 sq.deg, covering Galactic longitudes $150\,<l <\,210^{\circ}$ and latitudes $|b|\,<30^{\circ}$ under good observing conditions (dark or grey lunar nights). 
In addition, with a similar target selection algorithm, over 1.5 million very bright stars brighter than 14.0\,mag (and normally fainter than 9.0\,mag), will be  observed utilizing bright lunar nights.
At present, two stellar parameter pipelines, the LAMOST Stellar Parameter Pipeline (LASP; Wu et al. 2014; Luo et al. 2015) and the LAMOST Stellar Parameter at Peking University (LSP3; Xiang et al. 2014), have been  developed that deliver spectral classifications and deduce stellar radial velocities and atmospheric parameters from  collected spectra.
 Both pipelines apply a template matching technique to estimate the atmospheric parameters, except that the  LASP uses the ELODIE library (Prugniel \& Soubiran 2001; Prugniel et al. 2007) spectra of very high resolution ($R\sim 4,200$),  obtained with an echelle spectrograph, after degrading the spectral resolution to match that of LAMOST spectra,  while the LSP3 adopts the MILES spectral library (S\'anchez-Bl\'azquez et al. 2006) consisting of accurately flux-calibrated spectra of spectral resolution comparable to that of LAMOST spectra. 
 Both libraries contain a similar number of stars ($\sim 1,000$) that cover similar ranges in stellar atmospheric parameters .
The Pilot and Regular Surveys of LSS-GAC were initiated in September 2011and 2012, respectively. 
By the summer of 2013, over one million spectra of good quality, SNR\,$(7450{\rm \AA})\,\ge\,10$, have been collected (Yuan et al. 2014). 
The Regular Survey is expected to last for 5 years.

As in the case above for the  SDSS DR9 data, we select the stars from the LSS-GAC Value-added Data Release 1 (DR1; Yuan et al. 2014) of  good LAMOST spectral SNRs (4650\AA) $>15$ and low extinction [$E(B-V)\,<\,0.05$\,mag] that also have  high precision photometry in $K_{s}$ band from the 2MASS and $g$-band from the XSTPS-GAC (Zhang et al. 2013, 2014; Liu et al. 2014).
Finally, we derive values of $T_{\rm eff}$ based on our empirical relation in colour $(g-K_{2})$ for $\sim$ 26,000 dwarfs and 3,000 giants selected from the LSS-GAC DR1(using metallicities yielded by LSP3). 
Effective temperatures yielded by the LSP3 and LASP are compared respectively in Figs.\,18 and 19 to those derived from our empirical relation.
On the whole, for both dwarfs and giants, values of $T_{\rm eff}$ yielded by the  LSP3 and LASP are consistent within a few tens Kelvin with the results given by our empirical calibration (cf.  Table\,7).
However, the agreement is less satisfactory for very hot or cool stars.
For the LSP3, it overestimates the effective temperature about by $100-200$ K for stars hotter than $6500$\,K  or those cooler than $4000$\,K,  dwarfs or giants likewise.
In Xiang et al. (2014), they correct those systematics in their final released results using the empirical calibration presented  here.
For the LASP,  the deviations for hot stars are insignificant, but systematics on the level of $\sim$\,100-200\,K are detected for stars cooler than 4500\,K, for both dwarfs and giants but with opposite trends.
For dwarfs, the LASP underestimates their effective temperatures, while for giants it does the opposite.
Most of the systematics seen in LSP3 or LASP parameters are propagated from the uncertainties of effective temperature scales of template  libraries employed by those pipelines  (MILES or ELODIE).
Ideally, in future, one can correct the systematics in effective temperature scale of those template  libraries directly  using the  empirical calibration presented here. 

To conclude, the empirical metallicity-dependent $T_{\rm eff}$--colour relations presented in the current work can be used to provide accurate, unbiased estimates of $T_{\rm eff}$ for millions of stars targeted by modern large stellar spectroscopic surveys with high efficiency, especially relation in the colour ($g-K_{2}$), given its weak dependence on both metallicity and luminosity.
To apply those photometric calibrations,  accurate estimates of the interstellar extinction are essential especially for disk stars.
Fortunately, with the multi-band photometric data now available, one can obtain highly accurate estimates of extinction towards individual stars using a variety of techniques, such as the spectral energy (SED) fitting  method (e.g.Chen et al. 2014) or the `star-pair' technique (e.g. Yuan et al. 2013).
For example, using the SED fitting technique, Chen et al. (2014) estimate  extinction values towards  more than 13 million stars, covering the entire footprint of the XSTPS-GAC area (over 6000\,\,sq. deg.), based on the multi-band photometries from the XSTPS-GAC in the optical, and the 2MASS and WISE (Wright et al. 2010) in the infrared.

\section{Conclusions}
Based on nearly two hundred dwarf (luminosity classes: IV/V) and giant (luminosity classes: II/III) stars with direct effective temperature measurements of better than 2.5 per cent collected from the literature, we have derived  metallicity-dependent $T_{\rm eff}$--colour relations in twenty-one colours for dwarfs and eighteen colours for giants  in four photometric systems (the Johnson, Cousins, SDSS and 2MASS).
The calibrations have typical percentage residuals of 2.0 and 1.5 per cents for dwarf and giant stars, respectively.
Restricted by the available calibration sample stars, at present, the calibrations are limited to the metal-rich stellar populations,  although a couple of dwarfs of metallicities [Fe/H]$ \sim -2.0$ included by our sample providing some constraints on the relations at such low metallicities. 
We expect more metal-poor stars (both dwarfs and giants) will be observed with the LBOI in the future and thus improve the calibrations at low metallicities.

The effects of metallicity  on $T_{\rm eff}$ can be well understood, which has well explained by R05 using synthetic spectra.
We explore quantitatively the effects of  metallicity and luminosity on effective temperatures derived from colour ($g-K_{2}$) or $(V - K_2)$.
Generally, for disk stars ([Fe/H] $> -1.0$), both colours show only weak dependence on both metallicity and luminosity. However, for halo stars ([Fe/H]\,$\sim\,-2.0$), the metallicity effects are significant.
 
A detailed comparison  between the empirical, direct temperature scale presented in the current work with those in the literature is presented.
We find that the IRFM effective temperature scales of Alonso et al. (1996, 1999) and RM05 match the current one within few tens Kelvin, while that of Casagrande et al. (2010, 2011) is about 130\,K hotter.
This discrepancies amongst different IRFM scales are likely caused by the different zero-points adopted. 
We also find that spectroscopic effective temperature scales (such as those based on the `excitation balance of iron lines', spectral template matching or line-depth ratios) are 50--130\,K hotter than the direct effective temperature scale. 
  The differences between the direct effective temperature scale and other indirect effective temperature scales could be due to systematics in model atmospheres or zero-point calibrations of those indirect scales, although potential biases in the calibrations of interferometry measurements themselves cannot be ruled out entirely.
 In addition, we find that the metallicity-dependent $T_{\rm eff}$--($B-V$) relation derived by B13 is inconsistent with the direct effective temperature scale presented here.
 We find that this is likely caused by the systematics in the fit of B13.

As an example, we present twenty-one colours of the Sun, deduced  from the current calibrations assuming a solar  $T_{\rm eff}\,=\,5777$\,K and [Fe/H]\,=\,0.
Also using the calibration in colour ($g-K_{2}$), we investigate possible systematics in effective temperatures yielded by two of the currently on-going large scale stellar spectroscopic surveys: the SDSS and LSS-GAC.
The effective temperatures delivered by the SSPP pipeline of SDSS seems to have systematically overestimated the effective temperatures by approximately 150\,K. 
The cause of this discrepancy can be attributed to the fact that the SSPP  calibrate all effective temperature determinations  to match the IRFM scale of Casagrande (2010). 
For the LSS-GAC, effective temperatures given by both the LSP3 and LASP agree well with the photometric values yielded by  our empirical calibration, although some deviations are seen in the results of very hot or cool stars.

With high precision photometry now available from the SDSS, XSTPS-GAC, PanSTARRS in optical and from 2MASS in the infrared, we expect that the $T_{\rm eff}$--($g-K_{2}$) calibration presented in the current work  can be an invaluable tool for the determinations of effective temperature for large numbers of stars targeted by the currently on-going or upcoming large scale stellar spectroscopic surveys. 

 \section*{Acknowledgements} 
  We thank the referee, Dr. M. Bessell, for constructive suggestions that improve the manuscript significantly. 
 This work is supported by National Key Basic Research Program of China 2014CB845700.  
 
 The Guoshoujing Telescope (the Large Sky Area Multi-Object Fiber Spectroscopic Telescope, LAMOST) is a National Major Scientific Project built by the Chinese Academy of Sciences. Funding for the project has been provided by the National Development and Reform Commission. LAMOST is operated and managed by the National Astronomical Observatories, Chinese Academy of Sciences.

This work has made use of data products from the Sloan Digital Sky Survey (SDSS),  Two Micron All Sky Survey (2MASS), Wide-field Infrared Survey Explorer (WISE) and SIMBAD database, operated at CDS, Strasbourg, France.

\appendix
\section{}
 Table A1 presents stars in our sample with multiple angular measurements.
It lists angular diameters from different references and the instruments used.
The ratios of differences to combined measurement uncertainties, as well as ratios of measurements are also provided.
\begin{table*}
\centering
\scriptsize
  \caption{Sample stars with multiple angular diameter measurements}
  \begin{threeparttable}
  \begin{tabular}{cccccccc}
  \hline
Star&$\theta_{\rm LD} \pm \sigma$&Reference&Instrument&$(\theta_{{\rm LD,}i}-\theta_{{\rm LD,}1})/\sigma_{\rm C}^{a}$&$\theta_{{\rm LD,}i}/\theta_{{\rm LD,}1}$\\
&(mas)&&&&\\
\hline
\multicolumn{6}{c}{Dwarf stars} \\
HIP087937&$0.952 \pm 0.005$ & B12b & CHARA& $0.0$&$1.00 \pm 0.00$\\
&$1.004 \pm 0.040$ & Lane et al. (2001) & PTI & $1.3$&$1.05 \pm 0.04$\\

HD126660&$1.109 \pm 0.007$&B12a&CHARA&$0.0$&$1.00 \pm 0.00$\\
&$1.130 \pm 0.055$&van Belle \& von Braun (2009)&PTI&$0.4$&$1.02 \pm 0.05$\\

HD142860&$1.217 \pm 0.005$&B12a&CHARA&$0.0$&$1.00 \pm 0.00$\\
&$1.161 \pm 0.054$&van Belle \& von Braun (2009)&PTI&$-1.0$&$0.95 \pm 0.04$\\

HD088230&$1.225 \pm 0.008$&B12b&CHARA&$0.0$&$1.00 \pm 0.00$\\
&$1.238 \pm 0.053$&van Belle \& von Braun (2009)&PTI&$0.2$&$1.01 \pm 0.04$\\

HD109358&$1.238 \pm 0.030$&B12a&CHARA&$0.0$&$1.00 \pm 0.00$\\
&$1.138 \pm 0.055$&van Belle \& von Braun (2009)&PTI&$-1.6$&$0.92 \pm 0.05$\\

HD019373&$1.246 \pm 0.008$ &B12a & CHARA& $0.0$&$1.00 \pm 0.00$\\
&$1.331 \pm 0.050$&van Belle \& von Braun (2009)&PTI&$1.7$&$1.07 \pm 0.04$\\

HD097603&$1.328 \pm 0.009$ &B12a & CHARA& $0.0$&$1.00 \pm 0.00$\\
&$1.198 \pm 0.053$&van Belle \& von Braun (2009)&PTI&$-2.4$&$0.90 \pm 0.04$\\

HD095735&$1.432 \pm 0.013$&B12b&CHARA&$0.0$&$1.00 \pm 0.00$\\
&$1.436 \pm 0.030$ & Lane et al. (2001) &PTI&$0.1$&$1.00 \pm 0.02$\\
&$1.439 \pm 0.048$ & van Belle \& von Braun (2009)&PTI&$0.1$&$1.00 \pm 0.03$\\

HD030652&$1.526 \pm 0.004$ &B12a & CHARA& $0.0$&$1.00 \pm 0.00$\\
&$1.409 \pm 0.048$&van Belle \& von Braun (2009)&PTI&$-2.4$&$0.92 \pm 0.03$\\

HD201092&$1.581 \pm 0.022$ &Kervella et al. (2008) & CHARA& $0.0$&$1.00 \pm 0.00$\\
&$1.666 \pm 0.046$&van Belle \& von Braun (2009)&PTI&$1.7$&$1.05 \pm 0.03$\\

HD201091&$1.775 \pm 0.013$ &Kervella et al. (2008) & CHARA& $0.0$&$1.00 \pm 0.00$\\
&$1.628 \pm 0.046$&van Belle \& von Braun (2009)&PTI&$-3.1$&$0.92 \pm 0.03$\\

HD026965&$1.504 \pm 0.006$&B12b&CHARA&$0.0$&$1.00 \pm 0.00$\\
&$1.437 \pm 0.039$ & Demory et al. (2009)&VLTI&$-1.7$&$0.96 \pm 0.03$\\

HD217014 & $0.685 \pm 0.011$ &B13&CHARA&$0.0$&$1.00 \pm 0.00$\\
&$0.748 \pm 0.027$& Baines et al. (2008) & CHARA&$2.2$&$1.09 \pm 0.04$\\

HD103095 & $0.696 \pm 0.005$ &B12a&CHARA&$0.0$&$1.00 \pm 0.00$\\
&$0.679 \pm 0.015$& Creevey et al. (2012) & CHARA&$-1.1$&$0.98 \pm 0.02$\\

HD001326 & $1.005 \pm 0.005$ &B12b&CHARA&$0.0$&$1.00 \pm 0.00$\\
&$0.988 \pm 0.016$& Berger et al. (2006) & CHARA&$-1.0$&$0.98 \pm 0.02$\\

HD009826&$1.114 \pm 0.009$ & Baines et al. (2008) & CHARA& $0.0$&$1.00 \pm 0.00$\\
&$1.180 \pm 0.010$ & Ligi et al. (2012) & CHARA & $4.9$&$1.06 \pm 0.01$\\

HD217987&$1.304 \pm 0.032$ & Demory et al. (2009) & VLTI & $0.0$&$1.00 \pm 0.00$\\
&$1.388 \pm 0.040$& S{\'egransan} et al. (2003) & VLTI &$1.6$&$1.06 \pm 0.04$\\

HD128621&$6.000 \pm 0.021^{b}$ & Bigot et al. (2006) & VLTI & $0.0$&$1.00 \pm 0.00$\\
&$6.001 \pm 0.034$& Kervella et al. (2003) & VLTI &$0.0$&$1.00 \pm 0.01$\\

HD121370&$2.269 \pm 0.025$ & M03 & Mark III & $0.0$&$1.00 \pm 0.00$\\
&$2.170 \pm 0.030$& Nordgren et al. (2001) & Mark III &$-2.5$&$0.96 \pm 0.02$\\
&$2.280 \pm 0.070$& Nordgren et al. (2001) & NPOI &$0.1$&$1.00 \pm 0.03$\\
&$2.200 \pm 0.027$& Th{\'e}venin et al. (2005)& VLTI &$-1.9$&$0.97 \pm 0.02$\\

HD150680&$2.367 \pm 0.051$ & M03 & Mark III & $0.0$&$1.00 \pm 0.00$\\
&$2.330 \pm 0.050$& Nordgren et al. (2001) & Mark III &$-0.5$&$0.98 \pm 0.03$\\
&$2.490 \pm 0.090$& Nordgren et al. (2001) & NPOI &$1.2$&$1.05 \pm 0.04$\\

HD061421&$5.446 \pm 0.054$ & M03 & Mark III & $0.0$&$1.00 \pm 0.00$\\
&$5.460 \pm 0.080$& Nordgren et al. (2001) & Mark III &$0.1$&$1.00 \pm 0.02$\\
&$5.430 \pm 0.070$& Nordgren et al. (2001) & NPOI &$-0.2$&$1.00 \pm 0.02$\\
&$5.390 \pm 0.030 ^{b}$& Chiavassa et al. (2012)& VLTI &$-0.9$&$0.99 \pm 0.01$\\

\hline
\multicolumn{6}{c}{Giant stars} \\
HD133208&$2.477 \pm 0.065$ & M03 & Mark III & $0.0$&$1.00 \pm 0.00$\\
&$2.520 \pm 0.040$&Baines et al. (2010) & PTI &$0.6$&$1.02 \pm 0.03$\\
&$2.480 \pm 0.080$&Baines et al. (2010) & NPOI &$0.0$&$1.00 \pm 0.04$\\

HD216131&$2.496 \pm 0.040$ & M03 & Mark III & $0.0$&$1.00 \pm 0.00$\\
&$2.460 \pm 0.060$&Baines et al. (2010) & PTI &$-0.5$&$0.96 \pm 0.03$\\
&$2.500 \pm 0.080$&Baines et al. (2010) & NPOI &$0.0$&$1.00 \pm 0.04$\\

HD096833&$4.120 \pm 0.041$ & M03 & Mark III & $0.0$&$1.00 \pm 0.00$\\
&$4.110 \pm 0.040$&Baines et al. (2010) & PTI &$-0.2$&$1.00 \pm 0.01$\\
&$4.080 \pm 0.070$&Baines et al. (2010) & NPOI &$-0.5$&$0.99 \pm 0.02$\\

HD189319&$6.225 \pm 0.062$ & M03 & Mark III & $0.0$&$1.00 \pm 0.00$\\
&$6.197 \pm 0.035$&Wittkowski et al. (2006) & NPOI &$-0.4$&$1.00 \pm 0.01$\\
&$6.170 \pm 0.012$&Wittkowski et al. (2006) & VLTI &$-0.9$&$0.99 \pm 0.01$\\
  \hline
\end{tabular}
\begin{tablenotes}
\item[a]We define the combined errors as $\sigma_{\rm C} = (\sigma_{\rm LD,1}^{2} + \sigma_{\rm LD,i}^{2})^{1/2}$
\item[b]Values of $\theta_{\rm LD}$ are derived with 3D model atmospheres. Others are all derived with 1D model atmospheres.
 \end{tablenotes}
\end{threeparttable}
\end{table*}

\end{document}